# Interactions between Knowledge and Time in a First-Order Logic for Multi-Agent Systems: Completeness Results

**F. Belardinelli**                                    F.BELARDINELLI@IMPERIAL.AC.UK
**A. Lomuscio**                                        A.LOMUSCIO@IMPERIAL.AC.UK
*Department of Computing*
*Imperial College London, UK*

## Abstract

We investigate a class of first-order temporal-epistemic logics for reasoning about multi-agent systems. We encode typical properties of systems including *perfect recall*, *synchronicity*, *no learning*, and having a *unique initial state* in terms of variants of quantified interpreted systems, a first-order extension of interpreted systems. We identify several monodic fragments of first-order temporal-epistemic logic and show their completeness with respect to their corresponding classes of quantified interpreted systems.

## 1. Introduction

While reactive systems (Pnueli, 1977) are traditionally specified using plain temporal logic, there is a well-established tradition in Artificial Intelligence (AI) and, in particular, Multi-Agent Systems (MAS) research to adopt more expressive languages. Much of this tradition is inspired by the earlier, seminal work in AI by McCarthy (1979, 1990) and others aimed at adopting an "intentional stance" (Dennett, 1987) when reasoning about intelligent systems. Specifically, logics for knowledge (Fagin, Halpern, Moses, & Vardi, 1995), beliefs, desires, intentions, obligations, etc., have been put forward to represent the *informational* and *motivational* attitudes of agents in the system. Theoretical explorations have focused on the soundness and completeness of a number of axiomatisations as well as the decidability and computational complexity of the corresponding logics.

The great majority of work in these lines focuses on propositional languages. Yet, specifications supporting quantification are increasingly required in applications. For example, it is often necessary to refer to different individuals at different instances of time.

Quantified modal languages (Garson, 2001) have long attracted considerable attention. Early work included analysing the philosophical and logical implications of different setups for the quantification domains, particularly in combination with temporal concepts. More recently, considerable attention has been given to identifying suitable fragments that preserve completeness and decidability, and then studying the resulting computational complexity of the satisfiability problem. This article follows this direction.

In more detail, we investigate the meta-theoretical properties of monodic fragments of quantified temporal-epistemic logic where interactions between quantifiers, time, and knowledge of the agents are present. There is a deep-rooted interest (Fagin et al., 1995; Meyden, 1994) in understanding the implications of interaction axioms in this context, as they often express interesting properties of MAS, including "perfect recall", "synchronicity", and "no learning". These features are well-understood at the propositional level (Fagin, Halpern, & Vardi, 1992; Halpern, van der Meyden, & Vardi, 2004) and are commonly used in several application areas. The technical question this paper aims to resolve is whether a similar range of results can be provided in the presence of (limited forms of) quantification. As we shall demonstrate, the answer to this question is largely positive.

## 1.1 State of the Art

The analysis and application of temporal-epistemic logic in a first-order setting has an established tradition in AI. One of the early contributions is the work of Moore (1990), which presents a theory of action that takes into consideration the epistemic preconditions to actions and their effects on knowledge. More recently, a number of first-order temporal-epistemic logics for reasoning about MAS were introduced by Wooldridge et al. (2002, 2006, 1999), often in the context of the MABLE programming language for agents. The same authors introduced a first-order branching time temporal logic for MAS (Wooldridge & Fisher, 1992), and developed it in a series of papers (Wooldridge et al., 2002, 2006). First-order multi-modal logics also constitute the conceptual base of a number of other agent theories, such as BDI logics (Rao & Georgeff, 1991), the KQML framework (Cohen & Levesque, 1995), and the $\mathcal{LORA}$ framework (Wooldridge, 2000b). All of these include operators for mental attitudes (e.g., knowledge, belief, intention, desire, etc.), as well as temporal and dynamic operators with some form of quantification. However, most of the current literature has so far fallen short of a systematic analysis of the formal properties of these frameworks. Some of the frameworks above are so rich that they are unlikely to be finitely axiomatisable, let alone decidable. Still, these earlier contributions are an inspiration to the present investigation, as they are among the few to have explicitly addressed the subject of first-order temporal-epistemic languages in a MAS setting.

At a purely theoretical level, first-order temporal and epistemic logics have also received increasing attention with a range of contributions on axiomatisability (Degtyarev et al., 2003; Sturm et al., 2000; Wolter & Zakharyaschev, 2002), decidability (Degtyarev et al., 2002; Hodkinson et al., 2000; Wolter & Zakharyaschev, 2001), and complexity (Hodkinson, 2006; Hodkinson et al., 2003). Wolter and Zakharyaschev (2001) introduced the monodic fragment of quantified modal logic, where the modal operators are restricted to formulas with at most one free variable, and they proved the decidability of various fragments. Similar results have been obtained for monodic fragments of first-order temporal logic (Hodkinson, 2002; Hodkinson et al., 2000), and the computational complexity of these formalisms have been analysed (Hodkinson, 2006; Hodkinson et al., 2003). Further, Wolter and Zakharyaschev (2002) provided a complete axiomatisation of the monodic first-order validities on the natural numbers. The monodic fragment of first-order epistemic logic has also been explored (Sturm et al., 2000; Sturm, Wolter, & Zakharyaschev, 2002), and an axiomatisation including common knowledge has been provided. These lines of research constitute the theoretical background against which this research is set.

The contributions discussed previously used plain Kripke models as the underlying semantics. However, it has been argued though that in applications a "computationally-grounded" semantics (Wooldridge, 2000a) is preferable, as this enables systems to be modelled directly. We introduced *quantified interpreted systems* (QIS) to fill this gap (Belar-dinelli & Lomuscio, 2009). This enabled us to provide a complete axiomatisation of the monodic fragment of quantified temporal-epistemic logic on linear time (Belardinelli & Lomuscio, 2011). However, no interaction between temporal and epistemic modalities was studied. Preliminary investigations into the interactions between temporal and epistemic operators in a first-order setting have already appeared (Belardinelli & Lomuscio, 2010). In this paper we extend the previous results and also consider epistemic languages containing the common knowledge operator.

## 1.2 The Present Contribution

This paper extends the current state of the art in first-order temporal-epistemic logic by introducing a family of provably complete calculi for a variety of quantified interpreted systems characterising a range of properties including perfect recall, no learning, synchronicity, and having a unique initial state. We prove the completeness of the presented first-order temporal-epistemic logics via a *quasimodel* construction, which has previously been used (Hodkinson, Wolter, & Zakharyaschev, 2002; Hodkinson et al., 2000) to prove decidability for monodic fragments of first-order temporal logic (FoTL). Quasimodels have also been applied to first-order temporal as well as epistemic logic (Sturm et al., 2000; Wolter & Zakharyaschev, 2002). Wolter et al. (2002) present a complete axiomatisation for the monodic fragment of FoTL on the natural numbers; a similar result for a variety of first-order epistemic logics with common knowledge has also appeared (Sturm et al., 2000). However, the interaction between temporal and epistemic modalities in a first order setting has not been taken into account yet, nor has the interpreted systems semantics. Nonetheless, both of these features are essential for applications to multi-agent systems and are the subject of analysis here.

### 1.2.1 STRUCTURE OF THE PAPER.

In Section 2 we first introduce the first-order temporal-epistemic languages $\mathcal{L}_m$ and $\mathcal{L}C_m$ with common knowledge for a set $Ag = \{1, \ldots, m\}$ of agents. We then present the relevant classes of QIS as well as the monodic fragments of $\mathcal{L}_m$ and $\mathcal{L}C_m$. In Sections 3 we introduce the axiomatisations for these classes of QIS, while the details of the completeness proofs are presented in Sections 4 and 5. Finally, in Section 6 we elaborate on the results obtained and discuss possible extensions and future work.

## 2. First-Order Temporal-Epistemic Logics

Interpreted systems are a standard semantics for interpreting temporal-epistemic logics in a multi-agent setting (Fagin et al., 1995; Parikh & Ramanujam, 1985). We extend interpreted systems to the first-order case by enriching these structures with a domain of individuals. We first investigated "static" quantified interpreted systems, where no account for the evolution of the system is given (Belardinelli & Lomuscio, 2008, 2009). Then, fully-fledged QIS on a language with also temporal modalities were introduced (Belardinelli & Lomuscio, 2010, 2011). We follow the definition of QIS provided in the references.

## 2.1 First-Order Temporal-Epistemic Languages

Given a set $Ag = \{1, \ldots, m\}$ of agents, the first-order temporal-epistemic language $\mathcal{L}_m$ contains individual variables $x_1, x_2, \ldots$, individual constants $c_1, c_2, \ldots$, $n$-ary predicate constants $P_1^n, P_2^n, \ldots$, for $n \in \mathbb{N}$, the propositional connectives $\neg$ and $\rightarrow$, the quantifier $\forall$, the linear time operators $\bigcirc$ and $\mathcal{U}$, and the epistemic operator $K_i$ for each agent $i \in Ag$. The language $\mathcal{L}C_m$ also contains the common knowledge operator $C$ (Fagin et al., 1992). For simplicity we consider only one group of agents for the common knowledge modality, that is, the whole of $Ag$; $C$ is really tantamount to $C_{Ag}$. The extension to proper non-empty subsets of $Ag$ is not problematic.

The languages $\mathcal{L}_m$ and $\mathcal{L}C_m$ contain no symbol for functions; so all terms $t_1, t_2, \ldots$ in these languages are either individual variables or constants.

**Definition 1.** *Formulas in $\mathcal{L}_m$ are defined in the Backus-Naur form as follows:*

$$\phi \quad ::= \quad P^k(t_1, \ldots, t_k) \mid \neg \psi \mid \psi \rightarrow \psi' \mid \forall x \psi \mid \bigcirc \psi \mid \psi \mathcal{U} \psi' \mid K_i \psi$$

*The language $\mathcal{L}C_m$ extends $\mathcal{L}_m$ with the following clause:*

- *if $\phi$ is a formula in $\mathcal{L}C_m$, then also $C\phi$ is a formula in $\mathcal{L}C_m$.*

The formulas $\bigcirc \phi$ and $\phi \mathcal{U} \phi'$ are read as "*at the next step $\phi$*" and "*eventually $\phi'$ and until then $\phi$*" respectively. The formula $K_i \phi$ represents "*agent $i$ knows $\phi$*", while $C\phi$ stands for "*$\phi$ is common knowledge*" in the set $Ag$ of agents.

We define the symbols $\wedge$, $\vee$, $\leftrightarrow$, $\exists$, $G$ (always in the future), $F$ (some time in the future) as standard. Further, we introduce some abbreviations. The operator $\bar{K}_i$ is dual to $K_i$, that is, $\bar{K}_i \phi$ is defined as $\neg K_i \neg \phi$, while $E\phi$ is a shorthand for $\bigwedge_{i \in Ag} K_i \phi$. For $k \in \mathbb{N}$, $E^k \phi$ is defined as follows: $E^0 \phi = \phi$ and $E^{k+1} \phi = EE^k \phi$. The formulas $\bar{K}_i \phi$ and $E\phi$ are read as "*agent $i$ considers $\phi$ possible*" and "*every agent knows $\phi$*" respectively.

Free and bound variables are defined as standard. By $\phi[\vec{y}]$ we mean that $\vec{y} = y_1, \ldots, y_n$ are all the free variables in $\phi$. Additionally, $\phi[\vec{y}/\vec{t}]$ is the formula obtained by substituting simultaneously some, possibly all, free occurrences of $\vec{y}$ in $\phi$ with $\vec{t} = t_1, \ldots, t_n$ while renaming bound variables. A sentence is a formula with no free variables.

## 2.2 Quantified Interpreted Systems

To introduce quantified interpreted systems we assume a set $L_i$ of local states $l_i, l_i', \ldots$ for each agent $i \in Ag$ in a multi-agent system. We consider a set $L_e$ of local states for the environment $e$ as well. The set $\mathcal{S} \subseteq L_e \times L_1 \times \ldots \times L_m$ contains all and only the global states of the multi-agent system. To represent the temporal evolution of the MAS we consider the flow of time $\mathbb{N}$ of natural numbers; a *run* is a function $r : \mathbb{N} \rightarrow \mathcal{S}$. Intuitively, a run represents one possible evolution of the MAS assuming $\mathbb{N}$ as the flow of time. Given the above, we define quantified interpreted systems for the languages $\mathcal{L}_m$ and $\mathcal{L}C_m$ as follows:

**Definition 2** (QIS). *A* quantified interpreted system *is a triple $\mathcal{P} = \langle \mathcal{R}, \mathcal{D}, I \rangle$ where:*

- *$\mathcal{R}$ is a non-empty set of runs;*

- *$\mathcal{D}$ is a non-empty set of individuals;*

- *for $r \in \mathcal{R}$, $n \in \mathbb{N}$, $I$ is a first-order interpretation, that is, a function such that*

    - *for every constant $c$, $I(c, r(n)) \in \mathcal{D}$,*
    - *for every predicate constant $P^k$, $I(P^k, r(n))$ is a $k$-ary relation on $\mathcal{D}$.*

  *Further, for every $r, r' \in \mathcal{R}$, $n, n' \in \mathbb{N}$, $I(c, r(n)) = I(c, r'(n'))$.*

Notice that we assume a *unique domain of interpretation*, as well as a *fixed interpretation* for individual constants; so we simply write $I(c)$. Following standard notation (Fagin et al., 1995), for $r \in \mathcal{R}$ and $n \in \mathbb{N}$, a pair $(r, n)$ is a *point* in $\mathcal{P}$. If $r(n) = \langle l_e, l_1, \dots, l_m \rangle$ is the global state at point $(r, n)$ then $r_e(n) = l_e$ and $r_i(n) = l_i$ are the environment's and agent $i$'s local state at $(r, n)$ respectively. Further, for $i \in Ag$ the epistemic equivalence relation $\sim_i$ is defined such that $(r, n) \sim_i (r', n')$ iff $r_i(n) = r'_i(n')$. Clearly, each $\sim_i$ is an equivalence relation. Two points $(r, n)$ and $(r', n')$ are said to be *epistemically reachable*, or simply reachable, if $(r, n) \sim (r', n')$ where $\sim$ is the transitive closure of $\bigcup_{i \in Ag} \sim_i$.

In this paper we consider the following classes of QIS.

**Definition 3.** *A quantified interpreted system $\mathcal{P}$ satisfies*

| | | |
|---|---|---|
| synchronicity | *iff* | *for every $i \in Ag$, for all points $(r, n)$, $(r', n')$,* |
| | | *$(r, n) \sim_i (r', n')$ implies $n = n'$* |
| | | |
| perfect recall for agent $i$ | *iff* | *for all points $(r, n)$, $(r', n')$, if $(r, n) \sim_i (r', n')$ and $n > 0$* |
| | | *then either $(r, n-1) \sim_i (r', n')$ or* |
| | | *there is $k < n'$ such that $(r, n-1) \sim_i (r', k)$ and* |
| | | *for all $k'$, $k < k' \leq n'$ implies $(r, n) \sim_i (r', k')$* |
| | | |
| no learning for agent $i$ | *iff* | *for all points $(r, n)$, $(r', n')$, if $(r, n) \sim_i (r', n')$* |
| | | *then either $(r, n+1) \sim_i (r', n')$ or* |
| | | *there is $k > n'$ such that $(r, n+1) \sim_i (r', k)$ and* |
| | | *for all $k'$, $k > k' \geq n'$ implies $(r, n) \sim_i (r', k')$* |
| | | |
| unique initial state | *iff* | *for all $r$, $r' \in \mathcal{R}$, $r(0) = r'(0)$* |

These conditions have extensively been discussed in the literature (Halpern et al., 2004) together with equivalent formulations. Intuitively, a QIS is synchronous if time is part of the local state of each agent. A QIS satisfies perfect recall for agent $i$ if $i$'s local state records everything that has happened to him (from the agent's point of view) so far in the run. No learning is dual to perfect recall: agent $i$ does not acquire any new knowledge during a run. Finally, a QIS has a unique initial state if all runs start from the same global state.

A QIS $\mathcal{P}$ satisfies perfect recall (resp. no learning) if $\mathcal{P}$ satisfies perfect recall (resp. no learning) for all agents. We denote the class of QIS with $m$ agents as $\mathcal{QIS}_m$; the superscripts $pr$, $nl$, $sync$, $uis$ denote the subclasses of $\mathcal{QIS}_m$ satisfying perfect recall, no learning, synchronicity, and having a unique initial state respectively. For instance, $\mathcal{QIS}_m^{sync, uis}$ is the class of all synchronous QIS with $m$ agents and having a unique initial state.

We now assign an interpretation to the formulas in $\mathcal{L}_m$ and $\mathcal{LC}_m$ by means of quantified interpreted systems. Let $\sigma$ be an assignment from variables to individuals in $\mathcal{D}$, the valuation $I^\sigma(t)$ of a term $t$ is defined as $\sigma(y)$ for $t = y$, and $I^\sigma(t) = I(c)$ for $t = c$. A variant $\sigma_a^x$ of an assignment $\sigma$ assigns $a \in \mathcal{D}$ to $x$ and agrees with $\sigma$ on all other variables.

**Definition 4.** *The satisfaction of a formula $\phi \in \mathcal{L}_m$ at point $(r, n) \in \mathcal{P}$ under an assignment $\sigma$, denoted $(\mathcal{P}^\sigma, r, n) \models \phi$, is defined inductively as follows:*

$$
\begin{aligned}
&(\mathcal{P}^\sigma, r, n) \models P^k(t_1, \ldots, t_k) &&\text{iff} &&\langle I^\sigma(t_1), \ldots, I^\sigma(t_k) \rangle \in I(P^k, r(n)) \\
&(\mathcal{P}^\sigma, r, n) \models \neg\psi &&\text{iff} &&(\mathcal{P}^\sigma, r, n) \not\models \psi \\
&(\mathcal{P}^\sigma, r, n) \models \psi \rightarrow \psi' &&\text{iff} &&(\mathcal{P}^\sigma, r, n) \not\models \psi \text{ or } (\mathcal{P}^\sigma, r, n) \models \psi' \\
&(\mathcal{P}^\sigma, r, n) \models \forall x \psi &&\text{iff} &&\text{for all } a \in \mathcal{D},\ (\mathcal{P}^{\sigma_a^x}, r, n) \models \psi \\
&(\mathcal{P}^\sigma, r, n) \models \bigcirc \psi &&\text{iff} &&(\mathcal{P}^\sigma, r, n+1) \models \psi \\
&(\mathcal{P}^\sigma, r, n) \models \psi \mathcal{U} \psi' &&\text{iff} &&\text{there is } n' \geq n \text{ such that } (\mathcal{P}^\sigma, r, n') \models \psi' \\
& && &&\text{and } (\mathcal{P}^\sigma, r, n'') \models \psi \text{ for all } n \leq n'' < n' \\
&(\mathcal{P}^\sigma, r, n) \models K_i \psi &&\text{iff} &&\text{for all } r',\ n',\ (r, n) \sim_i (r', n') \text{ implies } (\mathcal{P}^\sigma, r', n') \models \psi
\end{aligned}
$$

For $\phi \in \mathcal{LC}_m$ we have to consider also the case for the common knowledge operator:

$$
(\mathcal{P}^\sigma, r, n) \models C\psi \qquad\qquad \text{iff} \qquad \text{for all } k \in \mathbb{N}, (\mathcal{P}^\sigma, r, n) \models E^k \psi
$$

The truth conditions for $\wedge$, $\vee$, $\leftrightarrow$, $\exists$, $G$ and $F$ are defined from those above. From the definition above it follows that $(\mathcal{P}^\sigma, r, n) \models C\psi$ iff for all $(r', n')$ reachable from $(r, n)$, $(\mathcal{P}^\sigma, r', n') \models \psi$.

A formula $\phi$ is *true at a point* $(r, n)$ if it is satisfied at $(r, n)$ by every assignment $\sigma$; $\phi$ is *true on a QIS $\mathcal{P}$* if it is true at every point in $\mathcal{P}$; $\phi$ is *valid on a class $\mathcal{C}$ of QIS* if it is true on every QIS in $\mathcal{C}$. Further, a formula $\phi$ is *satisfiable on a QIS $\mathcal{P}$* if it is satisfied at some point in $\mathcal{P}$, for some assignment $\sigma$; $\phi$ is *satisfiable on a class $\mathcal{C}$ of QIS* if it is satisfiable on some QIS in $\mathcal{C}$.

By considering all combinations of $pr$, $nl$, $sync$ and $uis$ we obtain 16 subclasses of $\mathcal{QIS}_m$ for any $m \in \mathbb{N}$. Not all of them are independent, nor axiomatisable. Indeed, some of these are not axiomatisable even at the propositional level (Halpern & Moses, 1992; Halpern & Vardi, 1989). In the first column of Table 1 we group together the classes of QIS that share the same set of validities on the languages $\mathcal{L}_m$ and $\mathcal{LC}_m$. The proofs of these equivalences are similar to those of the propositional case (Halpern et al., 2004) and are not reported here. Further, we define the languages $\mathcal{PL}_m$ and $\mathcal{PLC}_m$ as the propositional fragments of $\mathcal{L}_m$ and $\mathcal{LC}_m$ respectively (formally, $\mathcal{PL}_m$ and $\mathcal{PLC}_m$ are obtained by restricting atomic formulas to 0-ary predicate constants $p_1, p_2, \ldots$). Table 1 summarises the results by Halpern et al. (2004) concerning the axiomatisability of propositional validities in $\mathcal{PL}_m$ and $\mathcal{PLC}_m$. Observe that, as regards the language $\mathcal{PL}_m$, for $m = 1$ all sets of validities on the various classes of QIS are axiomatisable, while for $m \geq 2$ no axiomatisation can be given for $\mathcal{QIS}_m^{nl,uis}$ and $\mathcal{QIS}_m^{nl,pr,uis}$ (Halpern & Vardi, 1986, 1989). As to the language $\mathcal{PLC}_m$, we restrict to the case for $m \geq 2$, as for $m = 1$ $\mathcal{PLC}_m$ has the same expressive power as $\mathcal{PL}_m$. For $m \geq 2$ no class of validities on $\mathcal{PLC}_m$ has a recursive axiomatisation but $\mathcal{QIS}_m$, $\mathcal{QIS}_m^{sync}$, $\mathcal{QIS}_m^{uis}$, $\mathcal{QIS}_m^{sync,uis}$ and $\mathcal{QIS}_m^{nl,sync,uis}$, $\mathcal{QIS}_m^{nl,pr,sync,uis}$.

In the next section we show that the axiomatisability results at the propositional level can be lifted to the *monodic* fragment of the languages $\mathcal{L}_m$ and $\mathcal{LC}_m$.

| $QIS$ | $\mathcal{PL}_1$ | $\mathcal{PL}_m,\ m \geq 2$ | $\mathcal{PLC}_m,\ m \geq 2$ |
|---|---|---|---|
| $\mathcal{QIS}_m,\ \mathcal{QIS}_m^{sync},\ \mathcal{QIS}_m^{uis},\ \mathcal{QIS}_m^{sync,uis}$ | ✓ | ✓ | ✓ |
| $\mathcal{QIS}_m^{pr},\ \mathcal{QIS}_m^{pr,uis}$ | ✓ | ✓ | ✗ |
| $\mathcal{QIS}_m^{pr,sync},\ \mathcal{QIS}_m^{pr,sync,uis}$ | ✓ | ✓ | ✗ |
| $\mathcal{QIS}_m^{nl}$ | ✓ | ✓ | ✗ |
| $\mathcal{QIS}_m^{nl,sync}$ | ✓ | ✓ | ✗ |
| $\mathcal{QIS}_m^{nl,pr}$ | ✓ | ✓ | ✗ |
| $\mathcal{QIS}_m^{nl,pr,sync}$ | ✓ | ✓ | ✗ |
| $\mathcal{QIS}_m^{nl,uis}$ | ✓ | ✗ | ✗ |
| $\mathcal{QIS}_m^{nl,pr,uis}$ | ✓ | ✗ | ✗ |
| $\mathcal{QIS}_m^{nl,sync,uis},\ \mathcal{QIS}_m^{nl,pr,sync,uis}$ | ✓ | ✓ | ✓ |

Table 1: Equivalences between classes of QIS and axiomatisability results for the propositional fragments $\mathcal{PL}_m$ and $\mathcal{PLC}_m$. The sign ✓ indicates that the set of validities on a specific class is axiomatisable; while ✗ indicates that it is not.

## 2.3 The Monodic Fragment

In the rest of the paper we will show that a sufficient condition for lifting the results in Table 1 to the first-order case is to restrict the languages $\mathcal{L}_m$ and $\mathcal{LC}_m$ to their monodic fragments.

**Definition 5.** *The monodic fragment $\mathcal{L}_m^1$ is the set of formulas $\phi \in \mathcal{L}_m$ such that any subformula of $\phi$ of the form $K_i\psi$, $\bigcirc\psi$, or $\psi_1 \mathcal{U} \psi_2$ contains at most one free variable. Similarly, the monodic fragment $\mathcal{LC}_m^1$ is the set of formulas $\phi \in \mathcal{LC}_m$ such that any subformula of $\phi$ of the form $K_i\psi$, $C\psi$, $\bigcirc\psi$, or $\psi_1 \mathcal{U} \psi_2$ contains at most one free variable.*

The monodic fragments of a number of first-order modal logics have been thoroughly investigated in the literature (Hodkinson et al., 2000, 2003; Wolter & Zakharyaschev, 2001, 2002). In the case of $\mathcal{L}_m$ and $\mathcal{LC}_m$ these fragments are quite expressive as they contain formulas like the following:

$$\forall y \ C(\forall z Available(y,z)\mathcal{U}\exists x Request(x,y)) \tag{1}$$

$$K_i \bigcirc \forall xyz(Request(x,y) \to Available(y,z)) \to$$
$$\to \bigcirc K_i \forall xyz(Request(x,y) \to Available(y,z)) \tag{2}$$

Formula (1) intuitively states that it is common knowledge that every resource $y$ will eventually be requested by somebody, but until that time the resource remains available to everybody. Notice that $y$ is the only free variable within the scope of modal operators $\mathcal{U}$ and $C$. Formula (2) represents that if agent $i$ knows that at the next step every resource is available whenever it is requested, then at the next step agent $i$ knows that this is indeed the case. However, note that the formula

$$\forall x K_i(Process(x) \to \forall y FAccess(x,y))$$

which intuitively means that agent $i$ knows that every process will eventually try to access every resource, is not in $\mathcal{L}_m^1$ as both $x$ and $y$ occur free within the scope of modal operator $F$. Still, the monodic fragments of $\mathcal{L}_m$ and $\mathcal{LC}_m$ are quite expressive as they contain all *de dicto* formulas, i.e., formulas where no free variable appears in the scope of any modal operator, as in (2). So, the limitation is really only on *de re* formulas.

We stress the fact that the formulas above have no propositional equivalent in the case that they are interepreted on quantified interpreted systems in which the domain of quantification is infinite, or its cardinality cannot be bounded in advance.

Finally, observe that the Barcan formulas $\bigcirc \forall x \phi \leftrightarrow \forall x \bigcirc \phi$ and $K_i \forall x \phi \leftrightarrow \forall x K_i \phi$ are both true in all quantified interpreted systems, as each QIS includes a unique domain of quantification. This implies that the universal quantifier commutes with the temporal modality $\bigcirc$ and the epistemic modality $K_i$. Thus, it can be the case that for some formulas $\psi, \psi' \in \mathcal{L}_m$, we have that $\psi \leftrightarrow \psi'$ is a validity, but $\psi \in \mathcal{L}_m^1$ and $\psi' \notin \mathcal{L}_m^1$. For instance, consider $\psi = \bigcirc \forall x P(x,y)$ and $\psi' = \forall x \bigcirc P(x,y)$. We will see that this remark does not interfere with our results.

## 3. Axiomatisations

In this section we present sound and complete axiomatisations of the sets of monodic validities for the classes of quantified interpreted systems in Section 2. First, we introduce the basic system $\text{QKT}_m$ that extends to the first-order case the multi-modal epistemic logic $\text{S5}_m$ combined with the linear temporal logic LTL.

**Definition 6.** *The system $QKT_m$ contains the following schemes of axioms and rules, where $\phi$, $\psi$ and $\chi$ are formulas in $\mathcal{L}_m^1$ and $\Longrightarrow$ is the inference relation.*

| First-order logic | Taut | classical propositional tautologies |
|---|---|---|
| | MP | $\phi \rightarrow \psi, \phi \Longrightarrow \psi$ |
| | Ex | $\forall x \phi \rightarrow \phi[x/t]$ |
| | Gen | $\phi \rightarrow \psi[x/t] \Longrightarrow \phi \rightarrow \forall x \psi$, where $x$ is not free in $\phi$ |
| Temporal logic | K | $\bigcirc(\phi \rightarrow \psi) \rightarrow (\bigcirc \phi \rightarrow \bigcirc \psi)$ |
| | T1 | $\bigcirc \neg \phi \leftrightarrow \neg \bigcirc \phi$ |
| | T2 | $\phi \mathcal{U} \psi \leftrightarrow \psi \vee (\phi \wedge \bigcirc(\phi \mathcal{U} \psi))$ |
| | Nec | $\phi \Longrightarrow \bigcirc \phi$ |
| | T3 | $\chi \rightarrow \neg \psi \wedge \bigcirc \chi \Longrightarrow \chi \rightarrow \neg(\phi \mathcal{U} \psi)$ |
| Epistemic logic | K | $K_i(\phi \rightarrow \psi) \rightarrow (K_i \phi \rightarrow K_i \psi)$ |
| | T | $K_i \phi \rightarrow \phi$ |
| | 4 | $K_i \phi \rightarrow K_i K_i \phi$ |
| | 5 | $\neg K_i \phi \rightarrow K_i \neg K_i \phi$ |
| | Nec | $\phi \Longrightarrow K_i \phi$ |

The operator $K_i$ is an S5 modality, while the next $\bigcirc$ and until $\mathcal{U}$ operators are axiomatised as linear-time modalities (Fagin et al., 1995). To this we add the classical postulates $Ex$ and $Gen$ for quantification, which are both sound as we consider a unique domain of individuals in the quantified interpreted systems.

**Definition 7.** *The system $QKTC_m$ extends $QKT_m$ with the following schemes of axioms for common knowledge, where $\phi$, $\psi$ and $\chi$ are formulas in $\mathcal{LC}_m^1$ and $\Longrightarrow$ is the inference relation.*

| C1 | $C\phi \leftrightarrow (\phi \wedge EC\phi)$ |
|---|---|
| C2 | $\phi \rightarrow (\psi \wedge E\phi) \Longrightarrow \phi \rightarrow C\psi$ |

We consider the standard definitions of *proof* and *theorem*; $\vdash_S \phi$ means that the formula $\phi$ is a theorem in the formal system $S$. We remark that the Barcan formula $(BF)$ $\Box \forall x \phi \leftrightarrow \forall x \Box \phi$ is provable for any unary modal operator $\Box$ by the axioms $K$ and $Ex$, and the rules $MP$ and $Gen$. The notions of soundness and completeness of a system $S$ with respect to a class $\mathcal{C}$ of QIS are defined as standard: $S$ is *sound* w.r.t. $\mathcal{C}$ if for all $\phi$, $S \vdash \phi$ implies $\mathcal{C} \models \phi$. Similarly, $S$ is *complete* w.r.t. $\mathcal{C}$ if for all $\phi$, $\mathcal{C} \models \phi$ implies $S \vdash \phi$.

In this paper we focus on the schemes of axioms in Table 2 that specify key interactions between time and knowledge (Halpern et al., 2004). We use $1, \ldots, 5$ as superscripts to denote the systems obtained by adding to $QKT_m$ or $QKTC_m$ any combination of KT1-KT5. For instance, the system $QKTC_m^{2,3}$ extends $QKTC_m$ with the axioms KT2 and KT3.

| KT1 | $K_i\phi \wedge \bigcirc(K_i\psi \wedge \neg K_i\chi) \rightarrow K_i((K_i\phi)\mathcal{U}((K_i\psi)\mathcal{U}\neg\chi))$ |
|---|---|
| KT2 | $K_i \bigcirc \phi \rightarrow \bigcirc K_i\phi$ |
| KT3 | $(K_i\phi)\mathcal{U}K_i\psi \rightarrow K_i((K_i\phi)\mathcal{U}K_i\psi)$ |
| KT4 | $\bigcirc K_i\phi \rightarrow K_i \bigcirc \phi$ |
| KT5 | $K_i\phi \leftrightarrow K_j\phi$ |

Table 2: the axioms KT1-KT5.

It is straightforward to check that the axioms of $QKT_m$ and $QKTC_m$ are valid on every QIS and the inference rules preserve validity. However, the axioms KT1-KT5 are valid only on specific classes of QIS as stated in the following Remark.

**Remark 1.** *A QIS $\mathcal{P}$ satisfies any of the axioms KT1-KT5 in the first column if $\mathcal{P}$ satisfies the corresponding semantical condition in the second column.*

| Axiom | Condition on QIS |
|---|---|
| KT1 | *perfect recall* |
| KT2 | *perfect recall, synchronicity* |
| KT3 | *no learning* |
| KT4 | *no learning, synchronicity* |
| KT5 | *all agents share the same knowledge, i.e.,* |
| | *for all $i, j \in Ag$, $(r, n) \sim_i (r', n')$ iff $(r, n) \sim_j (r', n')$.* |

These results can be shown in a similar way to the propositional case (Halpern et al., 2004); so the proofs are omitted.

By using Remark 1 we can prove soundness results for all our first-order temporal-epistemic systems.

**Theorem 1** (Soundness). *The systems reported in the first and second column of the following table are sound w.r.t. the corresponding classes of QIS in the third column.*

| Systems | | QIS |
|---------|---------|------|
| $QKT_m$ | $QKTC_m$ | $\mathcal{QIS}_m$, $\mathcal{QIS}_m^{sync}$, $\mathcal{QIS}_m^{uis}$, $\mathcal{QIS}_m^{sync,uis}$ |
| $QKT_m^1$ | $QKTC_m^1$ | $\mathcal{QIS}_m^{pr}$, $\mathcal{QIS}_m^{pr,uis}$ |
| $QKT_m^2$ | $QKTC_m^2$ | $\mathcal{QIS}_m^{pr,sync}$, $\mathcal{QIS}_m^{pr,sync,uis}$ |
| $QKT_m^3$ | $QKTC_m^3$ | $\mathcal{QIS}_m^{nl}$, $\mathcal{QIS}_m^{nl,uis}$ |
| $QKT_m^4$ | $QKTC_m^4$ | $\mathcal{QIS}_m^{nl,sync}$ |
| $QKT_m^{2,3}$ | $QKTC_m^{2,3}$ | $\mathcal{QIS}_m^{nl,pr}$, $\mathcal{QIS}_m^{nl,pr,uis}$ |
| $QKT_m^{1,4}$ | $QKTC_m^{1,4}$ | $\mathcal{QIS}_m^{nl,pr,sync}$ |
| $QKT_m^{1,4,5}$ | $QKTC_m^{1,4,5}$ | $\mathcal{QIS}_m^{nl,sync,uis}$, $\mathcal{QIS}_m^{nl,pr,sync,uis}$ |

**Proof.** These results follow from Remark 1 by a line of reasoning similar to that used in the propositional case (Fagin et al., 1995; Halpern et al., 2004). Notice that if a quantified interpreted systems $\mathcal{P}$ satisfies no learning, synchronicity, and has a unique initial state, then $\mathcal{P}$ satisfies also perfect recall, that is, $\mathcal{P} \in \mathcal{QIS}_m^{nl,sync,uis}$ implies $\mathcal{P} \in \mathcal{QIS}_m^{nl,pr,sync,uis}$. Further, all agents share the same knowledge, therefore KT5 holds in $\mathcal{P}$. □

As we anticipated above, not all calculi are complete w.r.t. the corresponding classes of quantified interpreted systems in Theorem 1. In the next theorem we summarise the completeness results that will be proved in the rest of this paper.

**Theorem 2** (Completeness). *The systems reported in the first and second column of the following table are complete w.r.t. the corresponding classes of QIS in the third column.*

| Systems | | QIS |
|---------|---------|------|
| $QKT_m$ | $QKTC_m$ | $\mathcal{QIS}_m$, $\mathcal{QIS}_m^{sync}$, $\mathcal{QIS}_m^{uis}$, $\mathcal{QIS}_m^{sync,uis}$ |
| $QKT_m^1$ | | $\mathcal{QIS}_m^{pr}$, $\mathcal{QIS}_m^{pr,uis}$ |
| $QKT_m^2$ | | $\mathcal{QIS}_m^{pr,sync}$, $\mathcal{QIS}_m^{pr,sync,uis}$ |
| $QKT_m^3$ | | $\mathcal{QIS}_m^{nl}$ |
| $QKT_m^4$ | | $\mathcal{QIS}_m^{nl,sync}$ |
| $QKT_m^{2,3}$ | | $\mathcal{QIS}_m^{nl,pr}$ |
| $QKT_m^{1,4}$ | | $\mathcal{QIS}_m^{nl,pr,sync}$ |
| $QKT_m^{2,3}$ | | $\mathcal{QIS}_1^{nl,uis}$, $\mathcal{QIS}_1^{nl,pr,uis}$ |
| $QKT_m^{1,4,5}$ | $QKTC_m^{1,4,5}$ | $\mathcal{QIS}_m^{nl,sync,uis}$, $\mathcal{QIS}_m^{nl,pr,sync,uis}$ |

We observe that, as regards the language $\mathcal{L}_m^1$, the sets of monodic validities are axiomatisable for all classes introduced but $\mathcal{QIS}_m^{nl,uis}$ and $\mathcal{QIS}_m^{nl,pr,uis}$. However, for $\mathcal{L}_1^1$ we have that $\mathcal{QIS}_1^{nl,uis}$ and $\mathcal{QIS}_1^{nl,pr,uis}$ are equivalent to $\mathcal{QIS}_1^{nl,pr}$. Thus, the sets of monodic validities on $\mathcal{QIS}_1^{nl,pr,uis}$ and $\mathcal{QIS}_1^{nl,uis}$ are axiomatised by $QKT_1^{2,3}$.

As regards the language $\mathcal{LC}_m^1$, only the set of monodic validities on $\mathcal{QIS}_m$, $\mathcal{QIS}_m^{sync}$, $\mathcal{QIS}_m^{uis}$, $\mathcal{QIS}_m^{sync,uis}$ are axiomatisable, as well as those on $\mathcal{QIS}_m^{nl,sync,uis}$ and $\mathcal{QIS}_m^{nl,pr,sync,uis}$. All other classes are not recursively axiomatisable, as this is the case already at the propositional level (Halpern & Moses, 1992; Halpern & Vardi, 1986, 1989).

For proving the completeness results reported above we introduce Kripke models as a generalisation of quantified interpreted systems.

### 3.1 Kripke Models

To prove the completeness results in Theorem 2, we first introduce an appropriate class of Kripke models as a generalisation of QIS and prove completeness for these models. Then we apply a correspondence result between Kripke models and QIS to obtain the desired results.

**Definition 8** (Kripke model). *A Kripke model is a tuple $\mathcal{M} = \langle W, \mathcal{R}_W, \{\sim_i\}_{i \in Ag}, \mathcal{D}, I \rangle$ such that*

- *$W$ is a non-empty set of states;*

- *$\mathcal{R}_W$ is a non-empty set of functions $r : \mathbb{N} \to W$;*

- *for every agent $i \in Ag$, $\sim_i$ is an equivalence relation on $W$;*

- *$\mathcal{D}$ is a non-empty set of individuals;*

- *for every $w \in W$, $I$ is a first-order interpretation, that is, a function such that*

    - *for every constant $c$, $I(c, w) \in \mathcal{D}$,*
    - *for every predicate constant $P^k$, $I(P^k, w)$ is a $k$-ary relation on $\mathcal{D}$.*

    *Further, for every $w, w' \in W$, $I(c, w) = I(c, w')$.*

Notice that Def. 8 differs from other notions of Kripke model in that it includes the set $\mathcal{R}_W$ of functions to guarantee that the correspondence between Kripke models and QIS is one-to-one. We also assume a *unique domain of interpretation*, as well as a *fixed interpretation* for individual constants, so also in this case we simply write $I(c)$. Kripke models are a generalisation of QIS in that they do not specify the inner structure of the states in $W$. Also for Kripke models we introduce *points* as pairs $(r, n)$ for $r \in \mathcal{R}_W$ and $n \in \mathbb{N}$. A point derives its properties from the corresponding state; for instance, $(r, n) \sim_i (r', n')$ if $r(n) \sim_i r'(n')$.

We consider Kripke models satisfying synchronicity, perfect recall, no learning, and having a unique initial state. The definition of these subclasses is analogous to Def. 3.

**Definition 9.** *A Kripke model $\mathcal{M}$ satisfies*

| | | |
|---|---|---|
| synchronicity | *iff* | *for every $i \in Ag$, for all points $(r, n)$, $(r', n')$,* |
| | | *$(r, n) \sim_i (r', n')$ implies $n = n'$* |
| | | |
| perfect recall for agent $i$ | *iff* | *for all points $(r, n)$, $(r', n')$, if $(r, n) \sim_i (r', n')$ and $n > 0$* |
| | | *then either $(r, n - 1) \sim_i (r', n')$ or* |
| | | *there is $k < n'$ such that $(r, n - 1) \sim_i (r', k)$ and* |
| | | *for all $k'$, $k < k' \leq n'$ implies $(r, n) \sim_i (r', k')$.* |
| | | |
| no learning for agent $i$ | *iff* | *for all points $(r, n)$, $(r', n')$, if $(r, n) \sim_i (r', n')$* |
| | | *then either $(r, n + 1) \sim_i (r', n')$ or* |
| | | *there is $k > n'$ such that $(r, n + 1) \sim_i (r', k)$ and* |
| | | *for all $k'$, $k > k' \geq n'$ implies $(r, n) \sim_i (r', k')$.* |
| | | |
| unique initial state | *iff* | *for all $r$, $r' \in \mathcal{R}_W$, $r(0) = r'(0)$.* |

Now let $\mathcal{K}_m$ be the class of Kripke models with $m$ agents. Hereafter we adopt the same naming conventions as for QIS; for instance, $\mathcal{K}_m^{sync,uis}$ is the class of synchronous Kripke models with $m$ agents and having a unique initial state. Further, the inductive clauses for the satisfaction relation $\models$ with respect to an assignment $\sigma$ are straightforwardly defined from those for QIS, as well as the notions of truth and validity.

**Definition 10.** *The satisfaction of a formula $\phi \in \mathcal{L}_m$ (resp. $\mathcal{LC}_m$) at point $(r,n) \in \mathcal{M}$ for an assignment $\sigma$, or $(\mathcal{M}^\sigma, r, n) \models \phi$, is inductively defined as follows:*

| | | |
|---|---|---|
| $(\mathcal{M}^\sigma, r, n) \models P^k(t_1, \ldots, t_k)$ | iff | $\langle I^\sigma(t_1), \ldots, I^\sigma(t_k) \rangle \in I(P^k, r(n))$ |
| $(\mathcal{M}^\sigma, r, n) \models \neg\psi$ | iff | $(\mathcal{M}^\sigma, r, n) \not\models \psi$ |
| $(\mathcal{M}^\sigma, r, n) \models \psi \to \psi'$ | iff | $(\mathcal{M}^\sigma, r, n) \not\models \psi$ or $(\mathcal{M}^\sigma, r, n) \models \psi'$ |
| $(\mathcal{M}^\sigma, r, n) \models \forall x \psi$ | iff | for all $a \in \mathcal{D}$, $(\mathcal{M}^{\sigma^x_a}, r, n) \models \psi$ |
| $(\mathcal{M}^\sigma, r, n) \models \bigcirc\psi$ | iff | $(\mathcal{M}^\sigma, r, n+1) \models \psi$ |
| $(\mathcal{M}^\sigma, r, n) \models \psi\mathcal{U}\psi'$ | iff | there is $n' \geq n$ such that $(\mathcal{M}^\sigma, r, n') \models \psi'$ |
| | | and $n \leq n'' < n'$ implies $(\mathcal{M}^\sigma, r, n'') \models \psi$ |
| $(\mathcal{M}^\sigma, r, n) \models K_i \psi$ | iff | for all $r'$, $n'$, $(r,n) \sim_i (r', n')$ implies $(\mathcal{M}^\sigma, r', n') \models \psi$ |
| $(\mathcal{M}^\sigma, r, n) \models C\psi$ | iff | for all $k \in \mathbb{N}$, $(\mathcal{M}^\sigma, r, n) \models E^k \psi$ |

A formula $\phi$ is *true at a point* $(r,n)$ if it is satisfied at $(r,n)$ by every assignment $\sigma$; $\phi$ is *true on a Kripke model* $\mathcal{M}$ if it is true at every point in $\mathcal{M}$; $\phi$ is *valid on a class $\mathcal{C}$ of Kripke models* if it is true on every Kripke model in $\mathcal{C}$. Further, a formula $\phi$ is *satisfiable on a Kripke model* $\mathcal{M}$ if it is satisfied at some point in $\mathcal{M}$, for some assignment $\sigma$; $\phi$ is *satisfiable on a class $\mathcal{C}$ of Kripke models* if it is satisfiable on some Kripke model in $\mathcal{C}$.

We relate Kripke models and quantified interpreted systems by means of a map $g : \mathcal{K}_m \to \mathcal{QIS}_m$ (Lomuscio & Ryan, 1998). Let $\mathcal{M} = \langle W, \mathcal{R}_W, \{\sim_i\}_{i \in Ag}, \mathcal{D}, I \rangle$ be a Kripke model. For every agent $i \in Ag$, for $(r,n) \in \mathcal{M}$, let the equivalence class $[(r,n)]_i = \{(r',n') \mid (r,n) \sim_i (r',n')\}$ be a local state for agent $i$; while each $(r,n)$ is a local state for the environment. Then define $g(\mathcal{M})$ as the tuple $\langle \mathcal{R}', \mathcal{D}, I' \rangle$ where $\mathcal{R}'$ contains the runs $\mathbf{r}^r$ for $r \in \mathcal{R}_W$ such that $\mathbf{r}^r(n) = \langle (r,n), [(r,n)]_1, \ldots, [(r,n)]_m \rangle$. Further, $\mathcal{D}$ is the same as in $\mathcal{M}$, and for every constant $c$, $I'(c, \mathbf{r}^r(n)) = I(c, r(n))$, and $I'(P^k, \mathbf{r}^r(n)) = I(P^k, r(n))$. The structure $g(\mathcal{M})$ is a QIS that satisfies the following result:

**Lemma 1.** *For every $\phi$ in $\mathcal{L}_m$ (resp. $\mathcal{LC}_m$),*

$$(\mathcal{M}^\sigma, r, n) \models \phi \quad iff \quad (g(\mathcal{M})^\sigma, \mathbf{r}^r, n) \models \phi$$

**Proof.** The proof is by induction on the structure of $\phi$. If $\phi$ is an atomic formula $P^k(t_1, \ldots, t_k)$, then $(\mathcal{M}^\sigma, r, n) \models \phi$ iff $\langle I^\sigma(t_1), \ldots, I^\sigma(t_k) \rangle \in I(P^k, r(n))$, iff we have $\langle I'^\sigma(t_1), \ldots, I'^\sigma(t_k) \rangle \in I'(P^k, \mathbf{r}^r(n))$, iff $(g(\mathcal{M})^\sigma, \mathbf{r}^r, n) \models \phi$. The inductive cases for the propositional connectives and quantifiers are straightforward, as well as those for the temporal operators $\bigcirc$ and $\mathcal{U}$. As to $\phi = K_i\psi$, we have that $(\mathcal{M}^\sigma, r, n) \models \phi$ iff $(r,n) \sim_i (r', n')$ implies $(\mathcal{M}^\sigma, r', n') \models \psi$, but $(r,n) \sim_i (r',n')$ iff $\mathbf{r}_i^r(n) = \mathbf{r}_i^{r'}(n')$. Thus, $(\mathcal{M}^\sigma, r, n) \models \phi$ iff $(\mathbf{r}^r, n) \sim_i' (\mathbf{r}^{r'}, n')$ implies $(\mathcal{M}^\sigma, r', n') \models \psi$. Again, by the induction hypothesis $(\mathcal{M}^\sigma, r, n) \models \phi$ iff $(\mathbf{r}^r, n) \sim_i' (\mathbf{r}^{r'}, n')$ implies $(g(\mathcal{M})^\sigma, \mathbf{r}^{r'}, n') \models \psi$, i.e., iff $(g(\mathcal{M})^\sigma, \mathbf{r}^r, n) \models \phi$. The case for $\phi = C\psi$ is treated similarly by considering the epistemic reachability relation. $\square$

Notice that if $\mathcal{M}$ satisfies synchronicity, or perfect recall, or no learning, or has a unique initial state, then also $g(\mathcal{M})$ satisfies the same property. This follows from the fact that $(r, n) \sim_i (r', n')$ iff $(\mathbf{r}_i^r, n) \sim_i' (\mathbf{r}^{r'}, n')$. Thus, $g$ defines a map from each of the 16 subclasses of $\mathcal{K}_m$ outlined in Def. 9 to the corresponding subclass of $\mathcal{QIS}_m$ and we obtain the following corollary to Lemma 1.

**Corollary 1.** *Let $X$ be any subset of $\{pr, nl, sync, uis\}$. For every monodic formula $\phi \in \mathcal{L}_m^1$ (resp. $\mathcal{L}C_m^1$), if $\phi$ is satisfiable in $\mathcal{K}_m^X$, then $\phi$ is satisfiable in $\mathcal{QIS}_m^X$.*

For reasoning about the monodic fragments of $\mathcal{L}_m$ and $\mathcal{L}C_m$ when we are dealing with no learning and perfect recall, we introduce the following class of "monodic friendly" Kripke models. These structures are motivated by the fact that KT1 and KT3 are too weak to enforce either perfect recall or no learning on Kripke models when these axioms are restricted to monodic formulas. However, they suffice for monodic friendly structures. In the following, we also prove that satisfiability in Kripke models is equivalent to satisfiability in monodic friendly structures when we restrict our languages to monodic formulas.

**Definition 11** (mf-model). *A monodic friendly Kripke model is a tuple $\mathcal{M}_{mf} = \langle W, \mathcal{R}_W, \{\sim_{i,a}\}_{i \in Ag, a \in \mathcal{D}}, \mathcal{D}, I \rangle$ such that*

- *$W$, $\mathcal{R}_W$, $\mathcal{D}$ and $I$ are defined as for Kripke models;*

- *for $i \in Ag$, $a \in \mathcal{D}$, $\sim_{i,a}$ is an equivalence relation on $W$.*

We can define synchronicity, perfect recall, no learning, and having a unique initial state also for mf-models by parametrising Def. 9 to each relation $\sim_{i,a}$. For instance, an mf-model satisfies perfect recall for agent $i$ if for all points $(r, n)$, $(r', n')$, for all $a \in \mathcal{D}$, whenever $(r, n) \sim_{i,a} (r', n')$ and $n > 0$ then either $(r, n-1) \sim_{i,a} (r', n')$ or there is $k < n'$ such that $(r, n-1) \sim_{i,a} (r', k)$ and for all $k'$, $k < k' \leq n'$ implies $(r, n) \sim_{i,a} (r', k')$. As regards the subclasses of the class $\mathcal{MF}_m$ of all mf-models with $m$ agents, we adopt the same naming conventions as for QIS and Kripke models. Notice that Kripke models can be seen as mf-models such that for all $i \in Ag$, $a, b \in \mathcal{D}$, $\sim_{i,a}$ is equal to $\sim_{i,b}$.

Finally, the satisfaction relation $\models$ for $\phi \in \mathcal{L}_m^1$ (resp. $\mathcal{L}C_m^1$) in a mf-model $\mathcal{M}_{mf}$ is defined in the same way as in Kripke models, except for the epistemic operators:

$$(\mathcal{M}_{mf}^\sigma, r, n) \models K_i \psi[y] \quad \text{iff} \quad \text{for all } r', n', (r, n) \sim_{i, \sigma(y)} (r', n') \text{ implies } (\mathcal{M}_{mf}^\sigma, r', n') \models \psi[y]$$

where at most $y$ appears free in $\psi$. Notice that if $\psi$ is a sentence, then $(\mathcal{M}_{mf}^\sigma, r, n) \models K_i \psi$ iff $(r, n) \sim_{i,a} (r', n')$ implies $(\mathcal{M}_{mf}^\sigma, r', n') \models \psi$ for all $a \in \mathcal{D}$. The case for the common knowledge operator $C$ is straightforward by definition of $E^k$. In particular, two points $(r, n)$ and $(r', n')$ are *epistemically reachable* for $a \in \mathcal{D}$, or simply reachable, if $(r, n) \sim_a (r', n')$, where $\sim_a$ is the transitive closure of $\bigcup_{i \in Ag} \sim_{i,a}$.

We remark that the converse of the Barcan formula, or $CBF$, $K_i \forall x \psi \rightarrow \forall x K_i \psi$ holds in all mf-models; while the Barcan formula, or $BF$, $\forall x K_i \psi \rightarrow K_i \forall x \psi$ does not. To check this consider the mf-model $\mathcal{M} = \langle W, \mathcal{R}_W, \{\sim_{i,a}\}_{i \in Ag, a \in \mathcal{D}}, \mathcal{D}, I \rangle$ in Fig.1(a) such that

- $W = \{w, w', w''\}$

- $\mathcal{R}_W = \{r, r', r''\}$ and $r(0) = w$, $r'(0) = w'$, and $r''(0) = w''$

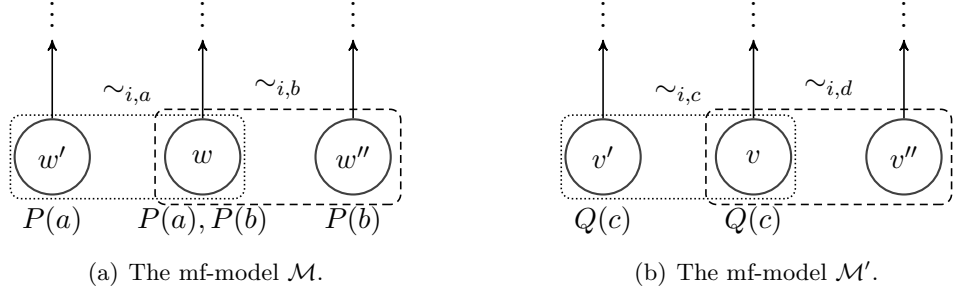

(a) The mf-model $\mathcal{M}$.

(b) The mf-model $\mathcal{M}'$.

Figure 1: Arrows represent the system runs; while epistemically related states are grouped together.

- $\mathcal{D} = \{a, b\}$

- $I(P^1, r(0)) = \{a, b\}$, $I(P^1, r'(0)) = \{a\}$ and $I(P^1, r''(0)) = \{b\}$

- $\sim_{i,a}$ and $\sim_{i,b}$ are equivalence relations such that $(r, 0) \sim_{i,a} (r', 0)$ and $(r, 0) \sim_{i,b} (r'', 0)$.

We can see that $(\mathcal{M}, r, 0) \models \forall x K_i P(x)$, but $(\mathcal{M}, r, 0) \not\models K_i \forall x P(x)$ as $(r, 0) \sim_{i,a} (r', 0)$ and $(\mathcal{M}^\sigma, r', 0) \not\models P(x)$ for $\sigma(x) = b$.

Furthermore, the $K$ axiom $K_i(\psi \rightarrow \psi') \rightarrow (K_i \psi \rightarrow K_i \psi')$ is not valid on mf-models either. In fact, consider the mf-model $\mathcal{M}' = \langle W', \mathcal{R}'_{W'}, \{\sim_{i,a}\}_{i \in Ag, a \in \mathcal{D}'}, \mathcal{D}', I' \rangle$ in Fig.1(b) such that

- $W' = \{v, v', v''\}$

- $\mathcal{R}'_{W'} = \{q, q', q''\}$ and $q(0) = v$, $q'(0) = v'$, and $q''(0) = v''$

- $\mathcal{D}' = \{c, d\}$

- $I'(Q^1, q(0)) = \{c\}$, $I'(Q^1, q'(0)) = \{c\}$ and $I'(Q^1, q''(0)) = \emptyset$

- $\sim_{i,c}$ and $\sim_{i,d}$ are equivalence relations such that $(q, 0) \sim_{i,c} (q', 0)$ and $(q, 0) \sim_{i,d} (q'', 0)$.

Finally, let $\sigma(x) = c$. We can check that $(\mathcal{M}^\sigma, q, 0) \models (Q(x) \rightarrow \exists x Q(x)) \wedge Q(x)$ and $(\mathcal{M}^\sigma, q', 0) \models (Q(x) \rightarrow \exists x Q(x)) \wedge Q(x)$. Thus, $(\mathcal{M}^\sigma, q, 0) \models K_i(Q(x) \rightarrow \exists x Q(x)) \wedge K_i Q(x)$. But $(\mathcal{M}, q'', 0) \not\models \exists x Q(x)$, so $(\mathcal{M}^\sigma, q, 0) \not\models K_i \exists x Q(x)$.

We now prove the following lemma, which will be used in the completeness proof for systems satisfying perfect recall or no learning. The lemma states that, when we deal with satisfiability of monodic formulas, mf-models suffice.

**Lemma 2.** Let $\mathcal{MF}^{K,BF}_m$ be the class of all mf-models validating the formulas $K$ and $BF$. For every monodic formula $\phi \in \mathcal{L}^1_m$ (resp. $\mathcal{LC}^1_m$),

$$\mathcal{K}_m \models \phi \quad iff \quad \mathcal{MF}^{K,BF}_m \models \phi$$

**Proof.** The implication from right to left follows from the fact that the class $\mathcal{K}_m$ of all Kripke models is isomorphic to the subclass of monodic friendly Kripke models such that for all $i \in Ag$, $a, b \in \mathcal{D}$, $\sim_{i,a}$ is equal to $\sim_{i,b}$. In other words, given a Kripke model $\mathcal{M} = \langle W, R_W, \{\sim_i\}_{i \in Ag}, \mathcal{D}, I \rangle$ we can define the mf-model $\mathcal{M}' = \langle W, R_W, \{\sim_{i,a}\}_{i \in Ag, a \in \mathcal{D}}, \mathcal{D}, I \rangle$, where for every $a \in \mathcal{D}$, $\sim_{i,a}$ is equal to $\sim_i$. It is straightforward to see that $\mathcal{M}'$ validates both $K$ and $BF$ (in particular, the counterexamples in Fig. 1 are ruled out). Further, if $\mathcal{M} \not\models \phi$ then $\mathcal{M}' \not\models \phi$. Thus, if $\mathcal{MF}_m^{K,BF} \models \phi$, then $\mathcal{K}_m \models \phi$.

For the implication from left to right, assume that $\mathcal{M}_{mf} = \langle W, \mathcal{R}_W, \{\sim_{i,a}\}_{i \in Ag, a \in \mathcal{D}}, \mathcal{D}, I \rangle$ is an mf-model validating $K$ and $BF$ such that $(\mathcal{M}_{mf}^\sigma, r, n) \not\models \phi$ for some point $(r, n)$ and some assignment $\sigma$. We can then build a Kripke model $\mathcal{M}' = \langle W', \mathcal{R}'_{W'}, \{\sim'_i\}_{i \in Ag}, \mathcal{D}', I' \rangle$ from $\mathcal{M}_{mf}$ such that $(\mathcal{M}'^\sigma, r, n) \not\models \phi$ as follows. We start by assuming $W' = W$, $\mathcal{R}' = \mathcal{R}$ and $\mathcal{D}' = \mathcal{D}$. Further, for each $i \in Ag$, define $\sim'_i$ as the transitive closure of $\bigcup_{a \in \mathcal{D}} \sim_{i,a}$. Finally, set $I' = I$. We now have to check that the Kripke model $\mathcal{M}'$ is well defined and does not validate $\phi$.

First of all, we point out the following issue associated with the construction above: it can be the case that for some point $(q, k)$ and some monodic formula $\psi[x]$, it happens that $(\mathcal{M}_{mf}^\sigma, q, k) \models K_i \psi[x]$, $(q, k) \sim_{i,\sigma(x)} (q', k')$ and $(q, k) \sim_{i,\sigma(y)} (q'', k'')$ for some $\sigma(x) \neq \sigma(y)$. Further, suppose that $(\mathcal{M}_{mf}^\sigma, q'', k'') \not\models \psi[x]$, while we obviously have that $(\mathcal{M}_{mf}^\sigma, q', k') \models \psi[x]$. Now by the definition of $\sim'_i$ we have that $(\mathcal{M}'^\sigma, q, k) \not\models K_i \psi[x]$; so the two models do not satisfy the same formulas. We can solve this problem by modifying the interpretation $I$ according to the structure of the monodic formula $\psi[x]$, while keeping the same truth value for $\psi[x]$ at point $(q, k)$. We consider the relevant cases according to the structure of $\psi[x]$; the induction hypothesis consists of the fact that we are able to find such an interpretation $I$ for all subformulas of $\psi[x]$.

For $\psi[x] = P(x)$ we simply assume that $\sigma(x) \in I(P, q''(k''))$, so that $(\mathcal{M}'^\sigma, q'', k'') \models \psi[x]$ and $(\mathcal{M}'^\sigma, q, k) \models K_i \psi[x]$. Note that this does not change the truth value of any epistemic formula in $(q, k)$ as we assumed that $(q, k) \not\sim_{i,\sigma(x)} (q'', k'')$ (otherwise $\psi[x]$ would be satisfied in $(q'', k'')$). The cases for propositional connectives and modal operators are similarly dealt with by applying the induction hypothesis. For $\psi[x] = \forall y \theta[x, y]$ we have that $(\mathcal{M}_{mf}^\sigma, q'', k'') \not\models \psi[x]$, therefore there exists $b \in \mathcal{D}$ such that $(\mathcal{M}_{b^y}^\sigma, q'', k'') \not\models \theta[x, y]$. Now we have to consider 4 different cases depending on whether $(q, k)$ satisfies any of these 4 formulas:

$$K_i \ \forall x \ \neg\theta[x, y] \tag{3}$$

$$\bar{K}_i \ \forall x \ \neg\theta[x, y] \tag{4}$$

$$K_i \ \exists x \ \neg\theta[x, y] \tag{5}$$

$$\bar{K}_i \ \exists x \ \neg\theta[x, y] \tag{6}$$

By using the axioms and inference rules in $\mathrm{QKT}_m$ for Formula (3) we can show what follows (where $\Rightarrow$ is used for entailment):

$(\mathcal{M}_{mf}^\sigma, q, k) \models K_i \forall y \theta[x, y] \wedge K_i \forall x \neg\theta[x, y] \Rightarrow$
$\quad \Rightarrow (\mathcal{M}_{mf}^\sigma, q, k) \models \exists x K_i \forall y \theta[x, y] \wedge \exists y K_i \forall x \neg\theta[x, y] \ \text{ by } Ex$
$\quad \Rightarrow (\mathcal{M}_{mf}^\sigma, q, k) \models K_i \exists x \forall y \theta[x, y] \wedge K_i \exists y \forall x \neg\theta[x, y] \ \text{ by } \exists z K_i \phi \rightarrow K_i \exists z \phi$
$\quad \Rightarrow (\mathcal{M}_{mf}^\sigma, q, k) \models K_i (\exists x \forall y \theta[x, y] \wedge \exists y \forall x \neg\theta[x, y]) \ \text{ by } K_i (\phi \wedge \psi) \leftrightarrow K_i \phi \wedge K_i \psi$

$$\Rightarrow (\mathcal{M}^{\sigma}_{mf}, q, k) \models K_i(\exists x \forall u \theta[x, u] \wedge \exists y \forall v \neg \theta[v, y]) \quad \text{by change of variables}$$
$$\Rightarrow (\mathcal{M}^{\sigma}_{mf}, q, k) \models K_i \exists x \exists y \forall u \forall v (\theta[x, u] \wedge \neg \theta[v, y]) \quad \text{by prefixing}$$
$$\Rightarrow (\mathcal{M}^{\sigma}_{mf}, q, k) \models K_i \exists x \exists y (\theta[x, y] \wedge \neg \theta[x, y]) \quad \text{by } Ex$$

but the last formula is a contradiction; so (3) cannot hold in $(q, k)$. Similarly, Formula (4) cannot hold in $(q, k)$ either because:

$$(\mathcal{M}^{\sigma}_{mf}, q, k) \models K_i \forall y \theta[x, y] \wedge \bar{K}_i \forall x \neg \theta[x, y]$$
$$\Rightarrow (\mathcal{M}^{\sigma}_{mf}, q, k) \models \exists x K_i \forall y \theta[x, y] \wedge \exists y \bar{K}_i \forall x \neg \theta[x, y] \quad \text{by } Ex$$
$$\Rightarrow (\mathcal{M}^{\sigma}_{mf}, q, k) \models K_i \exists x \forall y \theta[x, y] \wedge \exists y \bar{K}_i \forall x \neg \theta[x, y] \quad \text{by } \exists z K_i \phi \rightarrow K_i \exists z \phi$$
$$\Rightarrow (\mathcal{M}^{\sigma}_{mf}, q, k) \models K_i \exists x \forall y \theta[x, y] \wedge \bar{K}_i \exists y \forall x \neg \theta[x, y] \quad \text{by } \exists z \bar{K}_i \phi \leftrightarrow \bar{K}_i \exists z \phi$$
$$\Rightarrow (\mathcal{M}^{\sigma}_{mf}, q, k) \models \bar{K}_i (\exists x \forall y \theta[x, y] \wedge \exists y \forall x \neg \theta[x, y]) \quad \text{by } K_i \phi \wedge \bar{K}_i \psi \rightarrow \bar{K}_i (\phi \wedge \psi)$$
$$\Rightarrow (\mathcal{M}^{\sigma}_{mf}, q, k) \models \bar{K}_i \exists x \exists y (\theta[x, y] \wedge \neg \theta[x, y]) \quad \text{similarly to above}$$

Note that in both derivations we make use of the formulas $K$ and $BF$ (for instance, to prove the theorems $K_i(\phi \wedge \psi) \leftrightarrow K_i \phi \wedge K_i \psi$ and $\exists z K_i \phi \rightarrow K_i \exists z \phi$). Finally, to satify the Formulas (5) and (6) in $(q, k)$, we have to guarantee the existence of an individual $x$ while avoiding the clash with $\sigma(x)$. So, we introduce a new individual $a'$ in the domain $\mathcal{D}'$ such that $a'$ and $\sigma(x)$ satisfy the same formulas at all points. Thus, $a'$ can be seen as a copy of $\sigma(x)$. Finally, by the induction hypothesis we can modify the interpretation $I'$ so that $(\mathcal{M}'^{\sigma}, q'', k'') \models \theta[x, y]$.

The case for the common knowledge operator derives from the one for $K_i$. As a result, we obtain a Kripke model $\mathcal{M}'$ such that $(\mathcal{M}'^{\sigma}, r, n) \not\models \phi$. □

Moreover, by the procedure described above, if $\mathcal{M}_{mf}$ satisfies perfect recall, or no learning, or synchronicity, or has unique initial state, then also $\mathcal{M}'$ satisfies the same property. Thus, by Lemma 2 we can prove the following result.

**Corollary 2.** *Let $X$ be any subset of $\{pr, nl, sync, uis\}$. For every monodic formula $\phi \in \mathcal{L}^1_m$ (resp. $\mathcal{LC}^1_m$), if $\phi$ is satisfiable in $\mathcal{MF}^X_m$ also validating the formulas $K$ and $BF$, then $\phi$ is satisfiable in $\mathcal{K}^X_m$.*

**Proof.** It is easy to see that if $\mathcal{M}_{mf}$ satisfies either synchronicity or has a unique initial state, then $\mathcal{M}'$ does as well by the way it is defined. Further, suppose that $\mathcal{M}_{mf}$ satisfies perfect recall, and $(r, n) \sim'_i (r', n')$ for $n > 0$. This means that there is a sequence $a_1, \ldots, a_k$ of individuals in $\mathcal{D}$ and a sequence $(q_1, m_1), \ldots, (q_k, m_k)$ of points such that (i) $(r, n) = (q_1, m_1)$ and $(r', n') = (q_k, m_k)$; and (ii) $(q_j, m_j) \sim_{i, a_j} (q_{j+1}, m_{j+1})$ for $j < k$. We show the result for $k = 3$, the case for $k > 3$ follows by a straightforward generalisation.

If $(q_1, m_1 - 1) \sim_{i, a_1} (q_2, m_2)$, then by the definition of $\sim'_i$ we have that $(q_1, m_1 - 1) \sim'_i (q_3, m_3)$ as well. Hence, $\mathcal{M}_{mf}$ satisfies perfect recall. Otherwise, suppose that by perfect recall there is $l_2 < m_2$ such that $(q_1, m_1 - 1) \sim_{i, a_1} (q_2, l_2)$, and for all $l'_2$, $l_2 < l'_2 \leq m_2$ implies $(q_1, m_1) \sim_{i, a_1} (q_2, l'_2)$. Now consider each $l'_2(h) = m_2 - h$, for $0 \leq h < m_2 - l_2$. By perfect recall, either (i) there exists $p_3(h)$ such that $(q_2, l'_2(h)) \sim_{i, a_2} (q_3, p_3(h))$, and for all $p'_3(h)$, $p_3(h) < p'_3(h) \leq p_3(h - 1)$ implies $(q_2, l_2(h)) \sim_{i, a_2} (q_3, p'_3(h))$, or (ii) $(q_2, l'_2(h) - 1) \sim_{i, a_2} (q_3, p_3(h - 1))$, where $p_3(-1) = m_3$. Notice that in both cases, by the definition of $\sim'_i$, we have that $(q_1, m_1) \sim'_i (q_3, p'_3(h))$ for all $0 \leq h < m_2 - l_2$, that is, $(q_1, m_1) \sim'_i (q_3, p'_3)$ for all $p_3[l_2 + 1] < p'_3 \leq m_3$. Further, for $l_2$, either (i) there exists $l_3$ such that $(q_2, l_2 - 1) \sim_{i, a_2} (q_3, l_3)$, and for all $l_3'$, $l_3 < l_3' \leq p_3[l_2 + 1]$ implies $(q_2, l_2) \sim_{i, a_2} (q_3, l_3')$, or (ii) $(q_2, l_2 - 1) \sim_{i, a_2} (q_3, p_3[l_2 + 1])$. In the first case, if some $l_3'$ is strictly less than $m_3$, then there is $l_3'$ such that $(q_1, m_1 - 1) \sim_i' (q_3, l_3')$ and for all $l_3''$, $l_3' < l_3'' \leq m_3$ implies $(q_1, m_1) \sim_i' (q_3, l_3'')$. Otherwise, we have that $(q_1, m_1 - 1) \sim_i' (q_3, m_3)$. Hence, $\mathcal{M}_{mf}$ satisfies perfect recall.

The proof for no learning is similar. $\qquad\square$

Finally, by combining Corollaries 1 and 2 we immediately obtain the following result.

**Corollary 3.** *Let $X$ be any subset of $\{pr, nl, sync, uis\}$. For every monodic formula $\phi \in \mathcal{L}_m^1$ (resp. $\mathcal{LC}_m^1$), if $\phi$ is satisfiable in $\mathcal{MF}_m^X$ also validating the formulas $K$ and $BF$, then $\phi$ is satisfiable in $\mathcal{QIS}_m^X$.*

In the next section we show that it is indeed possible to build such an mf-model.

## 4. The Completeness Proof

In this section we outline the main steps of the completeness proof, which is based on a *quasimodel* construction (Gabbay, Kurucz, Wolter, & Zakharyaschev, 2003; Hodkinson et al., 2000). Differently from these contributions, here we explicitly take into account the interaction between temporal and epistemic modalities. Intuitively, a quasimodel for a monodic formula $\phi$ is a relational structure whose points are sets of sets of subformulas of $\phi$. Each set of sets of subformulas describes a "possible state of affairs", and contains sets of subformulas defining the individuals in that state. More formally, given a formula $\phi \in \mathcal{LC}_n^1$ we define

$$sub_C \phi = sub\phi \cup \{EC\psi \mid C\psi \in sub\phi\} \cup \{K_i C\psi \mid C\psi \in sub\phi, i \in Ag\}$$

where $sub\phi$ is the set of subformulas of $\phi$. For $\phi \in \mathcal{L}_n^1$, $sub_C \phi$ is simply $sub\phi$. Further, we define

$$sub_{C \bigcirc \neg} \phi = sub_C \phi \cup \{\neg\psi \mid \psi \in sub_C \phi\} \cup \{\bigcirc\psi \mid \psi \in sub_C \phi\} \cup \{\bigcirc\neg\psi \mid \psi \in sub_C \phi\}$$

Observe that $sub_{C \bigcirc \neg} \phi$ is closed under negation modulo equivalences $\phi \leftrightarrow \neg\neg\phi$ and $T1$, that is, for all $\psi \in sub_{C \bigcirc \neg} \phi$, if $\psi$ is not of the form $\neg\theta$ then $\neg\psi \in sub_{C \bigcirc \neg} \phi$; otherwise, $\theta \in sub_{C \bigcirc \neg} \phi$. Finally, let $sub_n \phi$ be the subset of $sub_{C \bigcirc \neg} \phi$ containing formulas with at most $n$ free variables. So, $sub_0 \phi$ is the set of sentences in $sub_n \phi$. If $x$ is a variable not occurring in $\phi$, we define $sub_x \phi = \{\psi[y/x] \mid \psi[y] \in sub_1 \phi\}$. Clearly, $x$ is the only free variable in the formulas in $sub_x \phi$. By $con\phi$ we denote the set of all constants occurring in $\phi$. In Table 3 we report the set $sub_y \phi$ for $\phi$ equal to Formula (1) thus abbreviated:

$$\forall y \ C(\forall z Av(y, z)\mathcal{U} \exists x Req(x, y))$$

Further, for $k \in \mathbb{N}$ we define the closures $cl_k \phi$ and $cl_{k,i} \phi$ by mutual recursion.

**Definition 12.** *Let $cl_0 \phi = sub_x \phi$ and for $k \geq 0$, $cl_{k+1} \phi = \bigcup_{i \in Ag} cl_{k,i} \phi$. For $k \geq 0$, $i \in Ag$, $cl_{k,i} \phi = cl_k \phi \cup \{K_i(\psi_1 \vee \ldots \vee \psi_n), \neg K_i(\psi_1 \vee \ldots \vee \psi_n) \mid \psi_1, \ldots, \psi_n \in cl_k \phi\}$.*

| $sub_y\phi$ | $\{\phi,\ C(\forall z Av(y,z)\mathcal{U}\exists x Req(x,y)),\ EC(\forall z Av(y,z)\mathcal{U}\exists x Req(x,y)),$ |
|---|---|
| | $\{K_i C(\forall z Av(y,z)\mathcal{U}\exists x Req(x,y))\}_{i\in Ag},\ \forall z Av(y,z)\mathcal{U}\exists x Req(x,y),\ \forall z Av(y,z),\ \exists x Req(x,y),$ |
| | $\neg\phi,\ \neg C(\forall z Av(y,z)\mathcal{U}\exists x Req(x,y)),\ \neg EC(\forall z Av(y,z)\mathcal{U}\exists x Req(x,y)),$ |
| | $\{\neg K_i C(\forall z Av(y,z)\mathcal{U}\exists x Req(x,y))\}_{i\in Ag},\ \neg\forall z Av(y,z)\mathcal{U}\exists x Req(x,y),\ \neg\forall z Av(y,z),$ |
| | $\neg\exists x Req(x,y),$ |
| | $\bigcirc\phi,\ \bigcirc C(\forall z Av(y,z)\mathcal{U}\exists x Req(x,y)),\ \bigcirc EC(\forall z Av(y,z)\mathcal{U}\exists x Req(x,y)),$ |
| | $\{\bigcirc K_i C(\forall z Av(y,z)\mathcal{U}\exists x Req(x,y))\}_{i\in Ag},\ \bigcirc\forall z Av(y,z)\mathcal{U}\exists x Req(x,y),\ \bigcirc\forall z Av(y,z),$ |
| | $\bigcirc\exists x Req(x,y),$ |
| | $\bigcirc\neg\phi,\ \bigcirc\neg C(\forall z Av(y,z)\mathcal{U}\exists x Req(x,y)),\ \bigcirc\neg EC(\forall z Av(y,z)\mathcal{U}\exists x Req(x,y)),$ |
| | $\{\bigcirc\neg K_i C(\forall z Av(y,z)\mathcal{U}\exists x Req(x,y))\}_{i\in Ag},\ \bigcirc\neg\forall z Av(y,z)\mathcal{U}\exists x Req(x,y),\ \bigcirc\neg\forall z Av(y,z),$ |
| | $\bigcirc\neg\exists x Req(x,y)\}$ |

Table 3: the set $sub_y\phi$ for $\phi$ equal to Formula (1).

| t | $\{\phi,\ C(\forall z Av(y,z)\mathcal{U}\exists x Req(x,y)),\ EC(\forall z Av(y,z)\mathcal{U}\exists x Req(x,y)),$ |
|---|---|
| | $\{K_i C(\forall z Av(y,z)\mathcal{U}\exists x Req(x,y))\}_{i\in Ag},\ \forall z Av(y,z)\mathcal{U}\exists x Req(x,y),\ \forall z Av(y,z),\ \exists x Req(x,y),$ |
| | $\bigcirc\neg\phi,\ \bigcirc\neg C(\forall z Av(y,z)\mathcal{U}\exists x Req(x,y)),\ \bigcirc\neg EC(\forall z Av(y,z)\mathcal{U}\exists x Req(x,y)),$ |
| | $\{\bigcirc\neg K_i C(\forall z Av(y,z)\mathcal{U}\exists x Req(x,y))\}_{i\in Ag},\ \bigcirc\neg\forall z Av(y,z)\mathcal{U}\exists x Req(x,y),\ \bigcirc\forall z Av(y,z),$ |
| | $\bigcirc\neg\exists x Req(x,y)\}$ |

Table 4: a type t in $cl_0\phi$, for $\phi$ equal to Formula (1).

We define $ad(\phi)$ as the greatest number of alternations of distinct $K_i$'s along any branch in $\phi$'s parse tree (Halpern et al., 2004). Further, an *index* is any finite sequence $\iota = i_1, \ldots, i_k$ of agents such that $i_n \neq i_{n+1}$, for $1 \leq n < k$; the length of $\iota$ is denoted by $|\iota|$. Also, $\iota\sharp i$ is the absorptive concatenation of indexes $\iota$ and $i$ such that $\iota\sharp i = \iota$ if $i_k = i$. Finally, $K_\iota\psi$ is a shorthand for $K_{i_1}\ldots K_{i_k}\psi$. Now let $\iota$ be an index such that $|\iota| \leq ad(\phi)$. If $\iota$ is the empty sequence $\epsilon$ then $cl_\iota\phi = cl_{ad(\phi)}\phi$. If $\iota = \iota'\sharp i$, then $cl_\iota\phi = cl_{k,i}\phi$ for $k = ad(\phi) - |\iota|$. We now introduce *types* for quasimodels, which intuitively can be seen as individuals described by maximal and consistent sets of formulas.

**Definition 13** (Type). *A $\iota$-type t for $\phi$ is any maximal and consistent subset of $cl_\iota\phi$, i.e., for every monodic formulas $\psi$ and $\psi'$ in $cl_\iota\phi$,*

*(i)* $\neg\psi \in$ t *iff* $\psi \notin$ t*;*

*(ii)* $\psi \wedge \psi' \in$ t *iff* $\psi,\ \psi' \in$ t*.*

Two $\iota$-types t, t' are said to *agree on $sub_0\phi$* if t $\cap\ sub_0\phi =$ t' $\cap\ sub_0\phi$, i.e., if they share the same sentences. Given a $\iota$-type t for $\phi$ and a constant $c \in con\phi$, $\langle$t$,c\rangle$ is an *indexed type* for $\phi$, abbreviated as t$^c$. In Table 4 we report a type t in $cl_0\phi$, for $\phi$ equal to Formula (1).

We now introduce *state candidates*, which intuitively represent the states of a quasimodel.

**Definition 14** (State candidate). *A $\iota$-state candidate for $\phi$ is a pair $\mathfrak{C} = \langle T, T^{con}\rangle$ such that*

*(i)* $T$ *is a set of $\iota$-types for $\phi$ that agree on $sub_0\phi$;*

*(ii)* $T^{con}$ *is a set containing for each $c \in con\phi$ an indexed type t$^c$ such that* t $\in T$.

We also introduce the notion of *point*, which describes a state candidate from the perspective of a particular type.

**Definition 15** (Point). *A $\iota$-point for $\phi$ is a pair $\mathfrak{P} = \langle \mathfrak{C}, \mathfrak{t} \rangle$ such that*

*(i) $\mathfrak{C} = \langle T, T^{con} \rangle$ is a $\iota$-state candidate for $\phi$;*

*(ii) $\mathfrak{t} \in T$ is a $\iota$-type.*

Note that, by a slight abuse of notation, we call *points* both the pairs $(r, n)$ in QIS and the pairs $\mathfrak{P} = \langle \mathfrak{C}, \mathfrak{t} \rangle$. This is to be consistent with previous work (Fagin et al., 1995; Halpern et al., 2004); the context will disambiguate. Also, we write $\mathfrak{t} \in \mathfrak{C}$ for $\mathfrak{C} = \langle T, T^{con} \rangle$ and $\mathfrak{t} \in T$. Similarly for $\mathfrak{t} \in \mathfrak{P}$. Given a $\iota$-state candidate $\mathfrak{C} = \langle T, T^{con} \rangle$ and a point $\mathfrak{P} = \langle \mathfrak{C}, \mathfrak{t} \rangle$ we define the formulas $\alpha_{\mathfrak{C}}$ and $\beta_{\mathfrak{P}}$ as follows:

$$\alpha_{\mathfrak{C}} := \bigwedge_{\mathfrak{t} \in T} \exists x \mathfrak{t}[x] \wedge \forall x \bigvee_{\mathfrak{t} \in T} \mathfrak{t}[x] \wedge \bigwedge_{\mathfrak{t}^c \in T^{con}} \mathfrak{t}[x/c]$$
$$\beta_{\mathfrak{P}} := \alpha_{\mathfrak{C}} \wedge \mathfrak{t}$$

where we do not distinguish between a type $\mathfrak{t}$ and the conjuction of formulas it contains.

A $\iota$-state candidate $\mathfrak{C}$ is *S-consistent* if the formula $\alpha_{\mathfrak{C}}$ is consistent w.r.t. the system $S$, i.e., if $\nvdash_S \neg \alpha_{\mathfrak{C}}$. Similarly, a $\iota$-point $\mathfrak{P}$ is *S-consistent* if the formula $\beta_{\mathfrak{P}}$ is consistent w.r.t. $S$. We refer to plain *consistency* whenever the system $S$ of reference is understood. Consistent state candidates represent the states of our quasimodels. We now define the relations of *suitability* that constitute the relational part of quasimodels.

**Definition 16.**
- *A $\iota_1$-type $\mathfrak{t}_1$ and a $\iota_2$-type $\mathfrak{t}_2$ are $\bigcirc$-suitable, or $\mathfrak{t}_1 \Rightarrow \mathfrak{t}_2$, iff $\iota_1 = \iota_2$ and $\mathfrak{t}_1 \wedge \bigcirc \mathfrak{t}_2$ is consistent. They are i-suitable, or $\mathfrak{t}_1 \approx_i \mathfrak{t}_2$, iff $\iota_1 \sharp i = \iota_2 \sharp i$ and $\mathfrak{t}_1 \wedge \bar{K}_i \mathfrak{t}_2$ is consistent.*

- *A $\iota_1$-state candidate $\mathfrak{C}_1$ and a $\iota_2$-state candidate $\mathfrak{C}_2$ are $\bigcirc$-suitable, or $\mathfrak{C}_1 \Rightarrow \mathfrak{C}_2$, iff $\iota_1 = \iota_2$ and $\alpha_{\mathfrak{C}_1} \wedge \bigcirc \alpha_{\mathfrak{C}_2}$ is consistent. They are i-suitable, or $\mathfrak{C}_1 \approx_i \mathfrak{C}_2$, iff $\iota_1 \sharp i = \iota_2 \sharp i$ and $\alpha_{\mathfrak{C}_1} \wedge \bar{K}_i \alpha_{\mathfrak{C}_2}$ is consistent.*

- *A $\iota_1$-point $\mathfrak{P}_1$ and a $\iota_2$-point $\mathfrak{P}_2$ are $\bigcirc$-suitable, or $\mathfrak{P}_1 \Rightarrow \mathfrak{P}_2$, iff $\iota_1 = \iota_2$ and $\beta_{\mathfrak{P}_1} \wedge \bigcirc \beta_{\mathfrak{P}_2}$ is consistent. They are i-suitable, or $\mathfrak{P}_1 \approx_i \mathfrak{P}_2$, iff $\iota_1 \sharp i = \iota_2 \sharp i$ and $\beta_{\mathfrak{P}_1} \wedge \bar{K}_i \beta_{\mathfrak{P}_2}$ is consistent.*

- *A $\iota_1$-point $\mathfrak{P}_1 = \langle \mathfrak{C}_1, \mathfrak{t}_1 \rangle$ and a $\iota_2$-point $\mathfrak{P}_2 = \langle \mathfrak{C}_2, \mathfrak{t}_2 \rangle$ are $\bigcirc$-suitable for a constant $c \in con\phi$, or $\mathfrak{P}_1 \Rightarrow^c \mathfrak{P}_2$, iff $\mathfrak{P}_1 \Rightarrow \mathfrak{P}_2$, $\mathfrak{t}_1^c \in T_1^{con}$ and $\mathfrak{t}_2^c \in T_2^{con}$. They are i-suitable for c, or $\mathfrak{P}_1 \approx_i^c \mathfrak{P}_2$, iff $\mathfrak{P}_1 \approx_i \mathfrak{P}_2$, $\mathfrak{t}_1^c \in T_1^{con}$ and $\mathfrak{t}_2^c \in T_2^{con}$.*

By using the axioms $T$, $4$ and $5$ it can be shown that the relation $\approx_i$ is reflexive, transitive and symmetric, that is, an equivalence relation. Also, the relation $\Rightarrow$ is serial. In the following lemma we list some properties of relations $\Rightarrow$ and $\approx_i$ that will be useful in what follows.

**Lemma 3.** *(i) Let $\bigcirc\psi \in sub_x\phi$, if $\mathfrak{t}_1 \Rightarrow \mathfrak{t}_2$ then $\bigcirc\psi \in \mathfrak{t}_1$ iff $\psi \in \mathfrak{t}_2$.*

(ii) Let $K_i\psi \in sub_x\phi$ and let $\mathfrak{t}$ be a $\iota$-type, $K_i\psi \in \mathfrak{t}$ iff for all $\iota$-types $\mathfrak{t}'$, $\mathfrak{t} \approx_i \mathfrak{t}'$ implies $\psi \in \mathfrak{t}'$. Moreover, let $|\iota\sharp i| \leq ad(\phi)$, then $K_i\psi \in \mathfrak{t}$ iff for all $\iota\sharp i$-types $\mathfrak{t}'$, $\mathfrak{t} \approx_i \mathfrak{t}'$ implies $\psi \in \mathfrak{t}'$.

**Proof.**

(i) The proof is similar to the one for Lemma 9(i) in the work of Wolter et al. (2002). If $\bigcirc\psi \in \mathfrak{t}_1$ and $\psi \notin \mathfrak{t}_2$ then $\neg\psi \in \mathfrak{t}_2$ and since $\mathfrak{t}_1 \wedge \bigcirc\mathfrak{t}_2$ is consistent, then also $\bigcirc\psi \wedge \bigcirc\neg\psi$ is consistent, which is a contradiction. From right to left, if $\psi \in \mathfrak{t}_2$ and $\bigcirc\psi \notin \mathfrak{t}_1$ then $\neg\bigcirc\psi \in \mathfrak{t}_1$. Since $\mathfrak{t}_1 \wedge \bigcirc\mathfrak{t}_2$ is consistent, then also $\neg\bigcirc\psi \wedge \bigcirc\psi$ is consistent, which is a contradiction.

(ii) From left to right, if $K_i\psi \in \mathfrak{t}$ and $\psi \notin \mathfrak{t}'$ then $\neg\psi \in \mathfrak{t}'$ and since $\mathfrak{t} \wedge \bar{K}_i\mathfrak{t}'$ is consistent, then also $K_i\psi \wedge \bar{K}_i\neg\psi$ is consistent, which is a contradiction. From right to left, if $K_i\psi \notin \mathfrak{t}$ then we can extend the set $\{\neg\psi\} \cup \{\theta \mid K_i\theta \in \mathfrak{t}\}$ to a $\iota$-type $\mathfrak{t}'$. In particular, $\mathfrak{t} \approx_i \mathfrak{t}'$ and $\neg\psi \in \mathfrak{t}'$. Moreover, if $|\iota\sharp i| \leq ad(\phi)$ then we can similarly prove that $K_i\psi \in \mathfrak{t}$ iff for all $\iota\sharp i$-types $\mathfrak{t}'$, $\mathfrak{t} \approx_i \mathfrak{t}'$ implies $\psi \in \mathfrak{t}'$. $\qquad\square$

We now present the frame underlying the quasimodel for $\phi$.

**Definition 17** (Frame). *A frame $\mathcal{F}$ is a tuple $\langle \mathcal{R}, \mathcal{D}, \{\sim_{i,a}\}_{i \in Ag, a \in \mathcal{D}}, \mathfrak{f} \rangle$ where*

(i) *$\mathcal{R}$ is a non-empty set of indexes $r, r', \ldots$;*

(ii) *$\mathcal{D}$ is a non-empty set of individuals;*

(iii) *for every $i \in Ag$, $a \in \mathcal{D}$, $\sim_{i,a}$ is an equivalence relation on the set of points $(r, n)$ for $r \in \mathcal{R}$ and $n \in \mathbb{N}$;*

(iv) *$\mathfrak{f}$ is a partial function associating to each point $(r, n)$ a consistent state candidate $\mathfrak{f}(r, n) = \mathfrak{C}_{r,n}$ such that*

    (a) *the domain of $\mathfrak{f}$ is not empty;*

    (b) *if $\mathfrak{f}$ is defined on $(r, n)$, then it is defined on $(r, n + 1)$;*

    (c) *if $\mathfrak{f}$ is defined on $(r, n)$ and $(r, n) \sim_{i,a} (r', n')$, then $\mathfrak{f}$ is defined on $(r', n')$.*

The function $\mathfrak{f}$ is partial to take into consideration the case of synchronous systems. Also, it is straightforward to introduce frames satisfying perfect recall, no learning, synchronicity, or having a unique initial state, by following the same definitions given for mf-models. Next, we provide the definition of *objects*, which correspond to the runs of Gabbay et al. (2003). We choose this terminology to avoid confusion with the runs in QIS.

**Definition 18** (Object). *Given an individual $a \in \mathcal{D}$, an object in the frame $\mathcal{F}$ is a map $\rho_a$ associating a type $\rho_a(r, n) \in T_{r,n}$ to every $(r, n) \in Dom(\mathfrak{f})$ with $\mathfrak{f}(r, n) = \mathfrak{C}_{r,n} = \langle T_{r,n}, T_{r,n}^{con} \rangle$ such that*

1. *$\rho_a(r, n) \Rightarrow \rho_a(r, n + 1)$*

2. *if $(r, n) \sim_{i,a} (r', n')$ then $\rho_a(r, n) \approx_i \rho_a(r', n')$*

3. $\chi \mathcal{U} \psi \in \rho_a(r, n)$ *iff there is* $n' \geq n$ *such that* $\psi \in \rho_a(r, n')$ *and for all* $n''$, $n \leq n'' < n'$ *implies* $\chi \in \rho_a(r, n'')$

4. *if* $\rho_a(r, n) \approx_i \mathfrak{t}$ *are* $\iota$-*types, then for some* $(r', n')$, $(r, n) \sim_{i,a} (r', n')$ *and* $\rho_a(r', n') = \mathfrak{t}$

5. *if* $\neg C \psi \in \rho_a(r, n)$ *then there exists a point* $(r', n')$ *reachable from* $(r, n)$ *such that* $\neg \psi \in \rho_a(r', n')$

An *object*[+] satisfies (1), (2), (3), (5) above and (4′) below instead of (4).

(4′) *if* $\rho_a(r, n)$ *is a* $\iota$-type, $\mathfrak{t}$ *is a* $\iota \sharp i$-*type and* $\rho_a(r, n) \approx_i \mathfrak{t}$, *then for some* $(r', n') \sim_{i,a} (r, n)$, $\rho_a(r', n') = \mathfrak{t}$.

Intuitively, an object identifies the same individual, here represented by types, across different state candidates. Now we have all the elements to give the definition of quasimodel.

**Definition 19** (Quasimodel). *A quasimodel for* $\phi$ *is a tuple* $\mathfrak{Q} = \langle \mathcal{R}, \mathcal{O}, \{\sim_{i,\rho}\}_{i \in Ag, \rho \in \mathcal{O}}, \mathfrak{f} \rangle$ *such that* $\langle \mathcal{R}, \mathcal{O}, \{\sim_{i,\rho}\}_{i \in Ag, \rho \in \mathcal{O}}, \mathfrak{f} \rangle$ *is a frame, and*

1. $\phi \in \mathfrak{t}$ *for some* $\mathfrak{t} \in T_{r,n}$ *and* $T_{r,n} \in \mathfrak{C}_{r,n}$

2. $\mathfrak{C}_{r,n} \Rightarrow \mathfrak{C}_{r,n+1}$

3. *if* $(r, n) \sim_{i,\rho} (r', n')$ *then* $\rho(r, n) \approx_i \rho(r', n')$

4. *for every* $\mathfrak{t} \in T_{r,n}$ *there exists an object* $\rho \in \mathcal{O}$ *such that* $\rho(r, n) = \mathfrak{t}$

5. *for every* $c \in con\phi$, *the function* $\rho^c$ *such that* $\rho^c(r, n) = \mathfrak{t}^c \in T_{r,n}^{con}$ *is an object in* $\mathcal{O}$.

A *quasimodel*[+] is defined as a quasimodel but where clauses (4) and (5) refer to objects[+] rather than objects. We can define quasimodels (resp. quasimodel[+]) satisfying perfect recall, no learning, synchronicity, or having a unique initial state, by assuming the corresponding condition on the underlying frame. The difference between objects (resp. quasimodel) and objects[+] (resp. quasimodel[+]) is purely technical. In particular, the latter are needed for systems satisfying perfect recall and no learning as it will become apparent in Section 5. In the following lemma we list some properties of quasimodels that will be useful in what follows.

**Lemma 4.** *In every quasimodel* $\mathfrak{Q}$, *for every object* $\rho \in \mathcal{O}$,

(i) $K_i \chi \in \rho(r, n)$ *iff for all* $(r', n')$, $(r', n') \sim_{i,\rho} (r, n)$ *implies* $\chi \in \rho(r', n')$.

(ii) $C \chi \in \rho(r, n)$ *iff for all points* $(r', n')$ *reachable from* $(r, n)$ *we have that* $\chi \in \rho(r', n')$.

**Proof.**

(i) The implication from left to right follows from the fact that $(r', n') \sim_{i,\rho} (r, n)$ implies $\rho(r, n) \approx_i \rho(r', n')$. For the implication from right to left, if $K_i \chi \notin \rho(r, n)$ then by Lemma 3(ii) there is a $\iota$-type $\mathfrak{t}$ such that $\rho(r, n) \approx_i \mathfrak{t}$ and $\neg \chi \in \mathfrak{t}$. By Definition 18 for some $(r', n')$, $(r, n) \sim_{i,\rho} (r', n')$ and $\neg \chi \in \rho(r', n') = \mathfrak{t}$.

(ii) The implication from left to right is proved by induction on the length of the path from $(r, n)$ to $(r', n')$. Both the base case and the inductive step follow by axiom C1. The implication from right to left follows from Definition 18. □

We now state the main result of this section, that is, satisfiability in quasimodels implies satisfiability in mf-models. In what follows a quasimodel $\mathfrak{Q}$ *validates* a formula $\phi$ if $\phi$ belongs to every type in every state-candidate in $\mathfrak{Q}$.

**Theorem 3.** *If there is a quasimodel (resp. quasimodel+) $\mathfrak{Q}$ for a monodic formula $\phi$, then $\phi$ is satisfiable in a mf-model $\mathcal{M}_{mf}$. Moreover, if $\mathfrak{Q}$ validates the formulas K and BF, then so does $\mathcal{M}_{mf}$. Finally, if $\mathfrak{Q}$ satisfies perfect recall, or no learning, or synchronicity, or has a unique initial state, then so does $\mathcal{M}_{mf}$.*

**Proof.** This proof is inspired by those of Lemmas 11.72 and 12.9 in the work of Gabbay et al. (2003), but here we consider monodic friendly Kripke models rather than standard Kripke models. First, for every monodic formula $\psi$ of the form $K_i\chi$, $C\chi$, $\bigcirc\chi$ or $\chi_1\mathcal{U}\chi_2$ we introduce a $k$-ary predicate constant $P_\psi^k$ for $k$ equal to 0 or 1, depending whether there are 0 or 1 free variables in $\psi$. The formula $P_\psi^k(x)$ is called the *surrogate* of $\psi$. Given a monodic formula $\phi$ we denote by $\overline{\phi}$ the formula obtained from $\phi$ by substituting all its modal subformulas which are not within the scope of another modal operator with their surrogates. Since every state candidate $\mathfrak{C}$ in the quasimodel $\mathfrak{Q}$ is consistent and all the system $S$ of first-order temporal-epistemic logic considered in Section 3 are based on classical first-order logic, the formula $\overline{\alpha}_\mathfrak{C}$ is consistent with respect to first-order (non-modal) logic. By Gödel's completeness theorem there is a first-order structure $\mathcal{I} = \langle I, \mathcal{D} \rangle$, where $\mathcal{D}$ is a non-empty set of individuals and $I$ is a first-order interpretation on $\mathcal{D}$, that satisfies $\overline{\alpha}_\mathfrak{C}$, i.e., $\mathcal{I}^\sigma \models \overline{\alpha}_\mathfrak{C}$ for some assignment $\sigma$ of the variables to the elements in $\mathcal{D}$. We intend to build an mf-model by joining all these first-order structures. However, it is possible that these structures have different domains with different cardinalities. To solve this problem, we consider a cardinal number $\kappa \geq \aleph_0$ greater than the cardinality of the set $\mathcal{O}$ of all objects in $\mathfrak{Q}$ and define

$$\mathcal{D} = \{\langle \rho, \xi \rangle \mid \rho \in \mathcal{O}, \xi < \kappa\}$$

Then, for $(r, n) \in \mathfrak{Q}$, for any $\iota$-type $\mathfrak{t} \in T_{r,n}$ we have that

$$|\{\langle \rho, \xi \rangle \in \mathcal{D} \mid \rho(r, n) = \mathfrak{t}\}| = \kappa$$

By the method described in Claim 11.24 by Gabbay et al. (2003), we can expand each first-order structure to obtain a structure $\mathcal{I}_{r,n} = \langle I_{r,n}, \mathcal{D} \rangle$ with domain $\mathcal{D}$ such that $\mathcal{I}_{r,n}$ satisfies $\overline{\alpha}_{\mathfrak{C}_{r,n}}$ and

$$|\{a \in \mathcal{D} \mid \sigma(x) = a \text{ and } \mathcal{I}_{r,n}^\sigma \models \overline{\mathfrak{t}}[x]\}| = \kappa$$

So, we can assume without loss of generality that all first-order structures $\mathcal{I}_{r,n}$ share the same domain $\mathcal{D}$, and for every $\mathfrak{t} \in T_{r,n}$, $\langle \rho, \xi \rangle \in \mathcal{D}$, we have

$$\rho(r, n) = \mathfrak{t} \quad \text{iff} \quad \mathcal{I}_{r,n}^\sigma \models \overline{\mathfrak{t}}[x]$$

for $\sigma(x) = \langle \rho, \xi \rangle$. Equivalently, for $\mathfrak{t} \in T_{r,n}$, $\sigma(x) = \langle \rho, \xi \rangle \in \mathcal{D}$,

$$\rho(r, n) = \{\psi \in cl_\iota\phi \mid I_{r,n}^\sigma \models \overline{\psi}[x]\} \tag{7}$$

Moreover, $I_{r,n}(c) = \langle \rho^c, 0 \rangle$ for every $c \in con\phi$.

We define the mf-model $\mathcal{M}_{mf}$ as the tuple $\langle W, \mathcal{R}, \{\sim_{i,a}\}_{i \in Ag, a \in \mathcal{D}}, \mathcal{D}, I \rangle$ such that $W$ is the set of points $(r, n)$ for $r$ in $\mathcal{R} \in \mathfrak{Q}$ and $n \in \mathbb{N}$; $\mathcal{R}$ is the set of runs from $\mathbb{N}$ to $W$ such that $r(n) = (r, n)$; $\mathcal{D}$ is defined as above; for $i \in Ag$ and $\langle \rho, \xi \rangle \in \mathcal{D}$, $\sim_{i,\langle \rho, \xi \rangle}$ is defined as $\sim_{i,\rho}$; and the interpretation $I$ is obtained by joining the various first-order interpretations $I_{r,n}$, i.e., $I(P, r(n)) = I_{r,n}(P)$ for every predicate constant $P$. We can now prove the following result for $\mathcal{M}_{mf}$.

**Lemma 5.** *If the mf-model $\mathcal{M}_{mf}$ is obtained from a quasimodel $\mathfrak{Q}$ as described above, then for every $\psi \in sub_x\phi$,*

$$\mathcal{I}^\sigma_{r,n} \models \overline{\psi} \quad iff \quad (\mathcal{M}^\sigma_{mf}, r, n) \models \psi$$

*Moreover, if $\mathfrak{Q}$ is a quasimodel$^+$, $\mathfrak{f}(r, n)$ is a $\iota$-state candidate and $ad(K_\iota \psi) \leq ad(\phi)$ then*

$$\mathcal{I}^\sigma_{r,n} \models \overline{\psi} \quad iff \quad (\mathcal{M}^\sigma_{mf}, r, n) \models \psi$$

**Proof.** The proof is similar to Lemma 12.10 in the work of Gabbay et al. (2003). We begin with the first part. The base case of induction follows by definition of the interpretation $I$ in the mf-model. The step for propositional connectives and quantifiers follows by the induction hypothesis and equations $\overline{\psi_1 \rightarrow \psi_2} = \overline{\psi_1} \rightarrow \overline{\psi_2}$, $\overline{\neg\psi_1} = \neg\overline{\psi_1}$, $\overline{\forall x \psi_1} = \forall x \overline{\psi_1}$.

Now let $\psi = \bigcirc \chi[x]$ and assume that $\sigma(x) = \langle \rho, \xi \rangle$, then we have:

$$\mathcal{I}^\sigma_{r,n} \models \overline{\bigcirc \chi}[x] \quad \text{iff} \quad \bigcirc \chi[x] \in \rho(r, n) \tag{8}$$
$$\text{iff} \quad \chi[x] \in \rho(r, n+1) \tag{9}$$
$$\text{iff} \quad \mathcal{I}^\sigma_{r,n+1} \models \overline{\chi}[x] \tag{10}$$
$$\text{iff} \quad (\mathcal{M}^\sigma_{mf}, r, n+1) \models \chi[x] \tag{11}$$
$$\text{iff} \quad (\mathcal{M}^\sigma_{mf}, r, n) \models \bigcirc \chi[x]$$

Steps (8) and (10) follow by Equation (7). Step (9) is motivated by Lemma 3(i), and step (11) follows by the induction hypothesis.

Let $\psi = (\chi \mathcal{U} \chi')[x]$ and $\sigma(x) = \langle \rho, \xi \rangle$, then we have:

$$\mathcal{I}^\sigma_{r,n} \models \overline{(\chi \mathcal{U} \chi')}[x] \quad \text{iff} \quad (\chi \mathcal{U} \chi')[x] \in \rho(r, n) \tag{12}$$
$$\text{iff} \quad \text{there is } n' \geq n \text{ such that } \chi'[x] \in \rho(r, n')$$
$$\text{and } \chi[x] \in \rho(r, n'') \text{ for all } n \leq n'' < n' \tag{13}$$
$$\text{iff} \quad \text{there is } n' \geq n \text{ such that } \mathcal{I}^\sigma_{r,n'} \models \overline{\chi'}[x]$$
$$\text{and } \mathcal{I}^\sigma_{r,n''} \models \overline{\chi}[x] \text{ for all } n \leq n'' < n' \tag{14}$$
$$\text{iff} \quad \text{there is } n' \geq n \text{ such that } (\mathcal{M}^\sigma_{mf}, r, n') \models \chi'[x]$$
$$\text{and } (\mathcal{M}^\sigma_{mf}, r, n'') \models \chi[x] \text{ for all } n \leq n'' < n' \tag{15}$$
$$\text{iff} \quad (\mathcal{M}^\sigma_{mf}, r, n) \models \chi \mathcal{U} \chi'[x]$$

Steps (12) and (14) follow by Equation (7). Step (13) is motivated by Def. 18, and step (15) follows by the induction hypothesis.

Let $\psi = K_i \chi[x]$ and $\sigma(x) = \langle \rho, \xi \rangle$, then we have:

$$\mathcal{I}_{r,n}^{\sigma} \models \overline{K_i \chi}[x] \quad \text{iff} \quad K_i \chi[x] \in \rho(r,n) \tag{16}$$

$$\text{iff} \quad \text{for all } (r', n') \sim_{i,\rho} (r,n), \chi[x] \in \rho(r', n') \tag{17}$$

$$\text{iff} \quad \text{for all } (r', n') \sim_{i,\langle \rho, \xi \rangle} (r,n), \mathcal{I}_{r',n'}^{\sigma} \models \overline{\chi}[x] \tag{18}$$

$$\text{iff} \quad \text{for all } (r', n') \sim_{i,\langle \rho, \xi \rangle} (r,n), (\mathcal{M}_{mf}^{\sigma}, r', n') \models \chi[x] \tag{19}$$

$$\text{iff} \quad (\mathcal{M}_{mf}^{\sigma}, r, n) \models K_i \chi[x]$$

Steps (16) and (18) follow by Equation (7). Step (17) is motivated by Lemma 4(i), and step (19) follows by the induction hypothesis.

Let $\psi = C\chi[x]$ and $\sigma(x) = \langle \rho, \xi \rangle$, then we have:

$$\mathcal{I}_{r,n}^{\sigma} \models \overline{C\chi}[x] \quad \text{iff} \quad C\chi[x] \in \rho(r,n) \tag{20}$$

$$\text{iff} \quad \text{for all } (r', n') \text{ reachable from } (r,n), \chi[x] \in \rho(r', n') \tag{21}$$

$$\text{iff} \quad \text{for all } (r', n') \text{ reachable from } (r,n), \mathcal{I}_{r',n'}^{\sigma} \models \overline{\chi}[x] \tag{22}$$

$$\text{iff} \quad \text{for all } (r', n') \text{ reachable from } (r,n), (\mathcal{M}, r', n') \models \chi[x] \tag{23}$$

$$\text{iff} \quad (\mathcal{M}, r, n) \models C\chi[x]$$

Steps (20) and (22) follow by Equation (7). Step (21) is motivated by Lemma 4(ii), and step (23) follows by the induction hypothesis.

Now we prove the second part of the lemma. All cases are identical to the first part, except for $\psi = K_\iota \chi$. Suppose that $\mathfrak{f}(r,n)$ is a $\iota$-state candidate and $ad(K_\iota \psi) \leq ad(\phi)$. For the implication from left to right, if $(r,n) \sim_{i,\rho} (r', n')$ then $\rho(r', n')$ is a $\iota'$-type such that $\iota \sharp i = \iota' \sharp i$. Thus, $ad(K_\iota \chi) \leq ad(K_{\iota \sharp i} \chi) \leq ad(K_\iota K_i \chi) \leq ad(\phi)$. So, we can apply the induction hypothesis. For the implication from right to left, if $ad(K_\iota K_i \chi) \leq ad(\phi)$ then $|\iota \sharp i| \leq ad(\phi)$ and by Lemma 3(ii) there is some $\iota \sharp i$-type $\mathfrak{t}$ such that $\mathfrak{t} \approx_i \rho(r,n)$ and $\neg \psi \in \mathfrak{t}$. By Def. 18 there is $(r', n')$ such that $(r, n) \sim_{i,\rho} (r', n')$ and $\rho(r', n') = \mathfrak{t}$. Since $ad(K_{\iota \sharp i} \chi) = ad(K_\iota K_i \chi) \leq ad(\phi)$ we can apply the induction hypothesis. $\qquad\square$

To complete the proof of Theorem 3, by definition of quasimodel we have that $\phi \in \mathfrak{t}$ for some $\mathfrak{t} \in T_{r,n}$ and $T_{r,n} \in \mathfrak{C}_{r,n}$. Therefore, $\phi$ is satisfied in the mf-model $\mathcal{M}_{mf}$ at point $(r, n)$. We also remark that if $\mathfrak{Q}$ validates the formulas $K$ and $BF$, so does $\mathcal{M}_{mf}$. This is the case as, if $K$ and $BF$ belong to every type in every state-candidate in $\mathfrak{Q}$, then by Lemma 5 we have that $\mathcal{M}_{mf}$ validates $K$ and $BF$ as well.

Finally, if $\mathfrak{Q}$ satisfies perfect recall, or no learning, or synchronicity, or has a unique initial state, then the mf-model obtained from $\mathfrak{Q}$ satisfies the corresponding constraints by construction. We show this fact for perfect recall: if $(r,n) \sim_{i,\langle \rho, \xi \rangle} (r', n')$ and $n > 0$, then in particular $(r,n) \sim_{i,\rho} (r', n')$. Since $\mathfrak{Q}$ satisfies perfect recall, either $(r, n-1) \sim_{i,\rho} (r', n')$, or there is $k < n'$ such that $(r, n-1) \sim_{i,\rho} (r', k)$ and for all $k'$, $k < k' \leq n'$ implies $(r,n) \sim_{i,\rho} (r', k')$. By the definition of $\sim_{i,\langle \rho, \xi \rangle}$ we obtain that either $(r, n-1) \sim_{i,\langle \rho, \xi \rangle} (r', n')$, or there is $k < n'$ such that $(r, n-1) \sim_{i,\langle \rho, \xi \rangle} (r', k)$ and for all $k'$, $k < k' \leq n'$ implies $(r,n) \sim_{i,\langle \rho, \xi \rangle} (r', k')$, that is, $\mathcal{M}_{mf}$ satisfies perfect recall as well. $\qquad\square$

We next show the existence of quasimodels for any monodic $\phi$.

## 5. Dealing with each System

In this section we consider the completeness proof for each system in Theorem 2. In particular, we show that if a monadic formula $\phi$ is consistent with respect to a system $S$, then we can build a quasimodel (or a quasimodel$^+$ in specific cases) for $\phi$ based on a frame for $S$. In the following sections the symbol $\vdash$ represents provability in the appropriate system $S$. We start with some lemmas that are useful for the construction of the quasimodel for any system.

**Lemma 6.** (i) For any consistent monadic formula $\phi$ there is a consistent $\epsilon$-state candidate $\mathfrak{C} = \langle T, T^{con} \rangle$ such that $\phi \in \mathfrak{t}$ for some $\mathfrak{t} \in T$.

(ii) Let $\mathfrak{P} = \langle \mathfrak{C}, \mathfrak{t} \rangle$ be a consistent $\iota$-point for $\phi$ such that $\mathfrak{C} = \langle T, T^{con} \rangle$, and let $c \in con\phi$. Then,

   (a) if $\mathfrak{C} \Rightarrow \mathfrak{C}'$ then there exists a $\iota$-point $\mathfrak{P}' = \langle \mathfrak{C}', \mathfrak{t}' \rangle$ such that $\mathfrak{P} \Rightarrow \mathfrak{P}'$.

   (b) if $\mathfrak{t}^c \in T^{con}$ and $\mathfrak{C} \Rightarrow \mathfrak{C}'$ then there exists a $\iota$-point $\mathfrak{P}' = \langle \mathfrak{C}', \mathfrak{t}' \rangle$ such that $\mathfrak{P} \Rightarrow^c \mathfrak{P}'$.

   (c) if $\psi_1 \mathcal{U} \psi_2 \in \mathfrak{t}$ then there is a sequence of $\iota$-points $\mathfrak{P}_j = \langle \mathfrak{C}_j, \mathfrak{t}_j \rangle$ for $j \leq k$ that realises $\psi_1 \mathcal{U} \psi_2$, i.e., $\mathfrak{P} = \mathfrak{P}_0 \Rightarrow \ldots \Rightarrow \mathfrak{P}_k$, $\psi_2 \in \mathfrak{t}_k$ and $\psi_1 \in \mathfrak{t}_j$ for $j < k$.

   (d) if $\psi_1 \mathcal{U} \psi_2 \in \mathfrak{t}^c$ then there is a sequence of $\iota$-points $\mathfrak{P}_j = \langle \mathfrak{C}_j, \mathfrak{t}_j \rangle$ for $j \leq k$ that $c$-realises $\psi_1 \mathcal{U} \psi_2$, i.e., the sequence realises $\psi_1 \mathcal{U} \psi_2$ and $\mathfrak{P}_0 \Rightarrow^c \ldots \Rightarrow^c \mathfrak{P}_k$.

   (e) if $\neg K_i \psi \in \mathfrak{t}$ then there is a $\iota$-point $\mathfrak{P}' = \langle \mathfrak{C}', \mathfrak{t}' \rangle$ such that $\mathfrak{P} \approx_i \mathfrak{P}'$ and $\neg \psi \in \mathfrak{t}'$.

   (f) if $\neg K_i \psi \in \mathfrak{t}^c$ then there is a $\iota$-point $\mathfrak{P}' = \langle \mathfrak{C}', \mathfrak{t}' \rangle$ such that $\mathfrak{P} \approx_i^c \mathfrak{P}'$ and $\neg \psi \in \mathfrak{t}'$.

   (g) if $\neg C \psi \in \mathfrak{t}$ then there is a sequence of $\iota$-points $\mathfrak{P}_j = \langle \mathfrak{C}_j, \mathfrak{t}_j \rangle$ for $j \leq k$ such that $\mathfrak{P} = \mathfrak{P}_0 \approx_{i_0} \ldots \approx_{i_{k-1}} \mathfrak{P}_k$ and $\neg \psi \in \mathfrak{t}_k$.

   (h) if $\neg C \psi \in \mathfrak{t}^c$ then there is a sequence of $\iota$-points $\mathfrak{P}_j = \langle \mathfrak{C}_j, \mathfrak{t}_j \rangle$ for $j \leq k$ such that $\mathfrak{P} = \mathfrak{P}_0 \approx_{i_0}^c \ldots \approx_{i_{k-1}}^c \mathfrak{P}_k$ and $\neg \psi \in \mathfrak{t}_k$.

**Proof.** The proof is similar to the one for Claims 11.75, 11.76 and 12.13 in the work of Gabbay et al. (2003), but here we consider $\iota$-state candidates and $\iota$-points. Let $\pi_\phi$ be the disjunction of all formulas $\beta_\mathfrak{P}$ for all $\iota$-points $\mathfrak{P}$ for $\phi$. Consider the formula $\overline{\pi}_\phi$, which is obtained by substituting all subformulas of $\pi_\phi$ of the form $K_i \psi$, $C \psi$, $\bigcirc \psi$ or $\psi_1 \mathcal{U} \psi_2$ that are not within the scope of another modal operator with their surrogates. We can check that $\overline{\pi}_\phi$ is true on all (non-modal) first-order structures. Since both QKT$_m$ and QKTC$_m$ extend first-order logic, by the semantical completeness of first-order logic we have that

$$\vdash \pi_\phi \tag{24}$$

(i) Notice that, by the previous remark, $\vdash \pi_\phi$ also for $\pi_\phi = \bigvee_{\{\mathfrak{P} | \mathfrak{P} \text{ is a } \epsilon\text{-point for } \phi\}} \beta_\mathfrak{P}$. Moreover, $\phi$ is consistent and by (24) also $\pi_\phi \wedge \phi$ is consistent. Therefore, there is a disjunct $\beta_\mathfrak{P}$ of $\pi_\phi$ such that $\beta_\mathfrak{P} \wedge \phi$ is consistent. So, $\phi \in \mathfrak{t}$ for $\mathfrak{P} = \langle \mathfrak{C}, \mathfrak{t} \rangle$.

(a) By (24) and *Nec* we have $\vdash \bigcirc \pi_\phi$. So, $\beta_\mathfrak{P} \wedge \bigcirc \pi_\phi$ is consistent and there must be a $\iota$-point $\mathfrak{P}'$ such that $\beta_\mathfrak{P} \wedge \bigcirc \beta_{\mathfrak{P}'}$ is also consistent.

(b) The proof is similar to (a).

(c) The proof is by contradiction. Let $U$ be the set of all $\iota$-points $\mathfrak{P}'$ such that there exist $\iota$-points $\mathfrak{P}_j = \langle \mathfrak{C}_j, \mathfrak{t}_j \rangle$ for $j < k$ and $\mathfrak{P} = \mathfrak{P}_0 \Rightarrow \ldots \Rightarrow \mathfrak{P}_k = \mathfrak{P}'$. Let $\theta = \bigvee_{\{\mathfrak{P}'|\mathfrak{P}' \in U\}} \beta_{\mathfrak{P}'}$. We can show that

$$\vdash \theta \to \neg\psi_2 \tag{25}$$

otherwise, we would have a sequence realising $\psi_1 \mathcal{U} \psi_2$. Moreover, by the definition of $U$,

$$\vdash \theta \to \bigcirc\theta \tag{26}$$

From (25) we obtain

$$\vdash G\theta \to G\neg\psi_2$$

and together with (25) and (26) we derive

$$\vdash \theta \to (\neg\psi_2 \wedge G\neg\psi_2) \tag{27}$$

Now consider any $\mathfrak{P}_1 \in U$ such that $\mathfrak{P} \Rightarrow \mathfrak{P}_1$. By (27) we have

$$\begin{aligned}
\vdash & \quad \beta_{\mathfrak{P}_1} \to (\neg\psi_2 \wedge G\neg\psi_2) \\
\vdash & \quad \bigcirc\beta_{\mathfrak{P}_1} \to G\neg\psi_2 \\
\vdash & \quad (\beta_{\mathfrak{P}} \wedge \bigcirc\beta_{\mathfrak{P}_1}) \to G\neg\psi_2
\end{aligned} \tag{28}$$

On the other hand, since $\psi_1 \mathcal{U} \psi_2 \in \mathfrak{t}$ we have

$$\vdash (\beta_{\mathfrak{P}} \wedge \bigcirc\beta_{\mathfrak{P}_1}) \to F\psi_2 \tag{29}$$

but (28) and (29) contradict the fact that $\mathfrak{P} \Rightarrow \mathfrak{P}_1$.

(d) The proof is similar to (c).

(e) First we remark that $\beta_{\mathfrak{P}} \wedge \bar{K}_i(\pi_\phi \wedge \neg\psi)$ is consistent. Thus, there exists a $\iota$-point $\mathfrak{P}' = \langle \mathfrak{C}', \mathfrak{t}' \rangle$ such that $\beta_{\mathfrak{P}} \wedge \bar{K}_i(\beta_{\mathfrak{P}'} \wedge \neg\psi)$ is consistent. Hence, $\mathfrak{P} \approx_i \mathfrak{P}'$ and $\neg\psi \in \mathfrak{t}'$.

(f) The proof is similar to (e).

(g) The proof is by contradiction. Let $V$ be the minimal set of $\iota$-points $\mathfrak{D}$ such that (i) $\mathfrak{P} \in V$; (ii) if $\mathfrak{D} \in V$ and $\mathfrak{D} \approx_i \mathfrak{D}'$ for some $i \in Ag$, then $\mathfrak{D}' \in V$. Let $\theta = \bigvee_{\{\mathfrak{D}|\mathfrak{D} \in V\}} \beta_{\mathfrak{D}}$. We can show

$$\vdash \theta \to \psi \tag{30}$$

If (30) did not hold, we would have a sequence as specified in the lemma. Moreover, by the definition of $V$,

$$\vdash \theta \to K_i\theta \tag{31}$$

for all $i \in Ag$. From (30) and (31) we obtain

$$\vdash \theta \to (\psi \wedge E\theta)$$

and by axiom C2,

$$\vdash \theta \to C\psi \tag{32}$$

but by definition of $\mathfrak{P}$,

$$\vdash \beta_{\mathfrak{P}} \to \neg C\psi$$

which contradicts (32).

(h) The proof is similar to (g).  □

By the following result it is always possible to extend the $\bigcirc$-suitability relation $\Rightarrow$ between types to $\bigcirc$-suitability between points.

**Lemma 7.** *Suppose that* $\mathfrak{t}$ *and* $\mathfrak{t}'$ *are* $\iota$-*types such that* $\mathfrak{t} \Rightarrow \mathfrak{t}'$, *then there are* $\iota$-*points* $\mathfrak{P} = \langle \mathfrak{C}, \mathfrak{t} \rangle$ *and* $\mathfrak{P}' = \langle \mathfrak{C}', \mathfrak{t}' \rangle$ *such that* $\mathfrak{P} \Rightarrow \mathfrak{P}'$. *In particular, for any* $c \in con\phi$, *there are* $\iota$-*points* $\mathfrak{P} = \langle \mathfrak{C}, \mathfrak{t} \rangle$ *and* $\mathfrak{P}' = \langle \mathfrak{C}', \mathfrak{t}' \rangle$ *such that* $\mathfrak{P} \Rightarrow^c \mathfrak{P}'$.

**Proof.** By Lemma 6 we have that $\vdash \pi_\phi$ and $\vdash \bigcirc \pi_\phi$ for $\pi_\phi = \bigvee_{\{\mathfrak{P} | \mathfrak{P} \text{ is a } \iota\text{-point for } \phi\}} \beta_\mathfrak{P}$. Since $\mathfrak{t} \Rightarrow \mathfrak{t}'$, then $\pi_\phi \wedge \mathfrak{t} \wedge \bigcirc(\pi_\phi \wedge \mathfrak{t}')$ is consistent. Thus, there must be $\iota$-points $\mathfrak{P}$ and $\mathfrak{P}'$ such that $\beta_\mathfrak{P} \wedge \mathfrak{t} \wedge \bigcirc(\beta_{\mathfrak{P}'} \wedge \mathfrak{t}')$ is consistent. Then, it is the case that $\mathfrak{P} = \langle \mathfrak{C}, \mathfrak{t} \rangle$ and $\mathfrak{P}' = \langle \mathfrak{C}', \mathfrak{t}' \rangle$ for some $\iota$-state candidates $\mathfrak{C}$ and $\mathfrak{C}'$. As a result, $\mathfrak{P} \Rightarrow \mathfrak{P}'$. The second part of the lemma is proved similarly to the first by observing that if $\mathfrak{t} \Rightarrow \mathfrak{t}'$ then $\mathfrak{t}[x/c] \wedge \bigcirc \mathfrak{t}[x/c]$ is consistent. Hence, also $\pi_\phi \wedge \mathfrak{t}[x/c] \wedge \bigcirc(\pi_\phi \wedge \mathfrak{t}'[x/c])$ is consistent. Thus, there must be $\iota$-points $\mathfrak{P}$ and $\mathfrak{P}'$ such that $\beta_\mathfrak{P} \wedge \mathfrak{t}[x/c] \wedge \bigcirc(\beta_{\mathfrak{P}'} \wedge \mathfrak{t}'[x/c])$ is consistent. So, $\mathfrak{t}^c \in T^{con}$ and $\mathfrak{t}'^c \in T'^{con}$, that is, $\mathfrak{P} \Rightarrow^c \mathfrak{P}'$.  □

According to Lemma 7 we can always extend a possibly infinite sequence of $\iota$-types $\mathfrak{t}_0 \Rightarrow \mathfrak{t}_1 \Rightarrow \ldots$ to a possibly infinite sequence of $\iota$-points $\mathfrak{P}_0 \Rightarrow \mathfrak{P}_1 \Rightarrow \ldots$ such that $\mathfrak{P}_k = \langle \mathfrak{C}_k, \mathfrak{t}_k \rangle$.

**Definition 20.** *Let a* $\Rightarrow$-*sequence be a possibly infinite sequence* $\mathfrak{C}_0 \Rightarrow \mathfrak{C}_1 \Rightarrow \ldots$ *of* $\iota$-*state candidates. A* $\Rightarrow$-*sequence is* acceptable *if for all* $k \geq 0$,

(i) *if* $\psi_1 \mathcal{U} \psi_2 \in \mathfrak{t}_k$, *for* $\mathfrak{t}_k \in \mathfrak{C}_k$, *then* $\psi_1 \mathcal{U} \psi_2$ *is realised in a sequence of* $\iota$-*points* $\mathfrak{P}_j = \langle \mathfrak{C}_j, \mathfrak{t}_j \rangle$ *for* $k \leq j \leq n$;

(ii) *if* $\psi_1 \mathcal{U} \psi_2 \in \mathfrak{t}_k^c$, *for* $\mathfrak{t}_k^c \in \mathfrak{C}_k$, *then* $\psi_1 \mathcal{U} \psi_2$ *is c-realised in a sequence of* $\iota$-*points* $\mathfrak{P}_j = \langle \mathfrak{C}_j, \mathfrak{t}_j \rangle$ *for* $k \leq j \leq n$.

The following lemma entails the completeness result.

**Lemma 8.** *Every finite* $\Rightarrow$-*sequence of* $\iota$-*state candidates can be extended to an infinite acceptable* $\Rightarrow$-*sequence.*

**Proof.** Assume that $\mathfrak{C}_0 \Rightarrow \ldots \Rightarrow \mathfrak{C}_n$ is a finite $\Rightarrow$-sequence $\Sigma$ and $\psi_1 \mathcal{U} \psi_2 \in \mathfrak{t}_k \in \mathfrak{C}_k$ for some $k \leq n$. Either $\psi_1 \mathcal{U} \psi_2$ is realised in $\mathfrak{C}_0 \Rightarrow \ldots \Rightarrow \mathfrak{C}_n$, or by Lemma 6(ii)(c) we can extend $\Sigma$ to a $\Rightarrow$-sequence $\Sigma'$ that realises $\psi_1 \mathcal{U} \psi_2$. This procedure is repeated for all formulas of the form $\psi_1 \mathcal{U} \psi_2$ appearing at some point in the $\Rightarrow$-sequence. Thus, we obtain a (possibly infinite) $\Rightarrow$-sequence $\mathfrak{C}_0 \Rightarrow \mathfrak{C}_1 \Rightarrow \ldots$ such that property (i) in Definition 20 is satisfied. To also satisfy property (ii) we reason similarly by using Lemma 6(ii)(d) instead.  □

Now let $X$ be a new object, a sequence $X, \ldots, X, \mathfrak{C}_n, \mathfrak{C}_{n+1}, \ldots$ is *acceptable from* $n$ if it starts with $n$ copies of $X$ and $\mathfrak{C}_n, \mathfrak{C}_{n+1}, \ldots$ is an acceptable $\Rightarrow$-sequence. We can now consider the completeness proof for each single class of QIS.

## 5.1 The Classes $\mathcal{QIS}_m$, $\mathcal{QIS}_m^{sync}$, $\mathcal{QIS}_m^{uis}$ and $\mathcal{QIS}_m^{sync,uis}$

We start with the completeness proof for the systems $QKT_m$ and $QKTC_m$, where there is no interaction between temporal and epistemic operators.

If a monodic formula $\phi$ is consistent, then by Lemma 6(i) there is a consistent $\epsilon$-state candidate $\mathfrak{C} = \langle T, T^{con} \rangle$ such that $\phi \in \mathfrak{t}$ for some type $\mathfrak{t} \in T$. Also, by Lemma 8 we can extend $\mathfrak{C}$ to an infinite acceptable $\Rightarrow$-sequence. So, the set of infinite acceptable $\Rightarrow$-sequences is non-empty. Let $\mathcal{R}$ be the set of all $\Rightarrow$-sequences acceptable from $n$, for some $n \in \mathbb{N}$. For $r \in \mathcal{R}$, $k \in \mathbb{N}$, define the partial function $\mathfrak{f}$ on $\mathcal{R} \times \mathbb{N}$ as $\mathfrak{f}(r, k) = \mathfrak{C}_k$ if $r$ is the $\Rightarrow$-sequence $X, \dots, X, \mathfrak{C}_n, \mathfrak{C}_{n+1}, \dots$ acceptable from $n$ and $k \geq n$, undefined otherwise. Finally, let $\mathcal{O}$ be the set of all functions $\rho$ associating every $(r, n) \in Dom(\mathfrak{f})$ to a type $\rho(r, n) \in T_{r,n}$ such that

(A) $\rho(r, n) \Rightarrow \rho(r, n+1)$;

(B) $\chi \mathcal{U} \psi \in \rho(r, n)$ iff there is $n' \geq n$ such that $\psi \in \rho(r, n')$ and $\chi \in \rho(r, n'')$ for all $n \leq n'' < n'$;

(C) if $\rho(r, n) \approx_i \mathfrak{t}$ are $\iota$-types, then for some $(r', n)$, $\rho(r', n) = \mathfrak{t}$;

(D) if $\neg C\psi \in \rho(r, n)$ then there exists a point $(r', n)$ and a sequence of $\iota$-points $\mathfrak{P}_j = \langle \mathfrak{C}_j, \mathfrak{t}_j \rangle$ for $j \leq k$, such that $\langle \mathfrak{f}(r, n), \rho(r, n) \rangle = \mathfrak{P}_0 \approx_{i_0} \dots \approx_{i_{k-1}} \mathfrak{P}_k$, $\neg \psi \in \mathfrak{t}_k$, $\mathfrak{f}(r', n) = \mathfrak{C}_k$ and $\rho(r', n) = \mathfrak{t}_k$.

We show that $\mathcal{O}$ is non-empty. Condition (A) is guaranteed by Lemma 6(ii)(a), and condition (B) by the fact that $r$ is an acceptable $\Rightarrow$-sequence. As regards (C) we remark that if $\rho(r, n) \approx_i \mathfrak{t}$ then we can find a consistent $\iota$-point $\mathfrak{P} = \langle \mathfrak{C}, \mathfrak{t} \rangle$ by reasoning similarly as in Lemma 6(i), and by Lemma 8, $\mathfrak{C}$ can be extended to a $\Rightarrow$-sequence $r'$ acceptable from $n$. Finally, set $\rho(r', n) = \mathfrak{t}$. As to (D) we observe that if $\neg C\psi \in \rho(r, n)$ then by Lemma 6(ii)(g) there exists a sequence of $\iota$-points $\mathfrak{P}_j = \langle \mathfrak{C}_j, \mathfrak{t}_j \rangle$ for $j \leq k$, such that $\langle \mathfrak{f}(r, n), \rho(r, n) \rangle = \mathfrak{P}_0 \approx_{i_0} \dots \approx_{i_{k-1}} \mathfrak{P}_k$ and $\neg \psi \in \mathfrak{t}_k$. Now, $\mathfrak{C}_k$ can be extended to a $\Rightarrow$-sequence $r'$ acceptable from $n$ such that $\rho(r', n) = \mathfrak{t}_k$. Finally, for $i \in Ag$, $\rho \in \mathcal{O}$, define $(r, n) \sim_{i,\rho} (r', n')$ iff $\rho(r, n) \approx_i \rho(r', n')$ and $n = n'$.

**Lemma 9.** *The tuple* $\langle \mathcal{R}, \mathcal{O}, \{\sim_{i,\rho}\}_{i \in Ag, \rho \in \mathcal{O}}, \mathfrak{f} \rangle$ *is a synchronous frame.*

**Proof.** We have previously shown that $\mathcal{R}$ and $\mathcal{O}$ are non-empty. Also, each $\sim_{i,\rho}$ is an equivalence relation by definition, and $\mathfrak{f}$ satisfies the conditions in Definition 17. Further, the frame is synchronous by definition of $\sim_{i,\rho}$. $\qquad \square$

Now we can prove the main result.

**Lemma 10.** *The tuple* $\langle \mathcal{R}, \mathcal{O}, \{\sim_{i,\rho}\}_{i \in Ag, \rho \in \mathcal{O}}, \mathfrak{f} \rangle$ *is a synchronous quasimodel for* $\phi$ *and it validates the formulas* $K$ *and* $BF$.

**Proof.** By the previous lemma, it remains to prove that the functions in $\mathcal{O}$ are objects. Conditions (1), (3), (4) and (5) on objects are satisfied by remarks (A)-(D) above. Condition (2) is satisfied by the definition of $\sim_{i,\rho}$. Furthermore, conditions (1), (2) and (3) on quasimodels are satisfied by the definitions of $\mathcal{R}$, $\mathfrak{f}$ and $\sim_{i,\rho}$. As regards (4), we can extend the function $\rho(r, n) = \mathfrak{t}$ to all $Dom(\mathfrak{f})$ by using Lemma 6(ii)(a), (c), (e) and (g). As to (5) the function $\rho^c$ such that $\rho^c(r, n) = \mathfrak{t}^c$ is an object by Lemma 6(ii)(b), (d), (f) and (h). Finally, $\mathfrak{Q}$ validates both the formulas $K$ and $BF$, as all $\mathfrak{t} \in \mathfrak{C}$, for all $\mathfrak{C} \in \mathfrak{Q}$, are consistent with $\mathrm{QKT}_m$ (resp. $\mathrm{QKTC}_m$). $\qquad \square$

The completeness of $QKT_m$ and $QKTC_m$ with respect to the classes $\mathcal{QIS}_m$ and $\mathcal{QIS}_m^{sync}$ of quantified interpreted systems directly follows from Lemma 10 together with Theorem 3. Thus, we obtain the following item in Theorem 2.

**Theorem 4** (Completeness). *The system $QKT_m$ (resp. $QKTC_m$) is complete w.r.t. the classes $\mathcal{QIS}_m$ and $\mathcal{QIS}_m^{sync}$ of QIS.*

To prove completeness for $\mathcal{QIS}_m^{uis}$ and $\mathcal{QIS}_m^{sync,uis}$ we use the next result, which is an extension from the propositional case (Halpern et al., 2004).

**Remark 2.** *Suppose $X$ is a subset of $\{pr, sync\}$. If $\phi \in \mathcal{L}_m^1$ (resp. $\mathcal{LC}_m^1$) is satisfiable in $\mathcal{QIS}_m^X$ then it is also satisfiable in $\mathcal{QIS}_m^{X,uis}$.*

Thus, the system $QKT_m$ (resp. $QKTC_m$) is also complete w.r.t. the classes $\mathcal{QIS}_m^{uis}$ and $\mathcal{QIS}_m^{sync,uis}$ of QIS.

## 5.2 The Classes $\mathcal{QIS}_m^{pr}$ and $\mathcal{QIS}_m^{pr,uis}$

We now begin to investigate systems where interactions between time and knowledge are present. The completeness proof for $QKT_m^1$ with respect to $\mathcal{QIS}_m^{pr}$ and $\mathcal{QIS}_m^{pr,uis}$ relies on the following lemma.

**Lemma 11.** *For all $\iota$-points $\mathfrak{P}_1 = \langle \mathfrak{C}_1, \mathfrak{t}_1 \rangle$, $\mathfrak{P}_2 = \langle \mathfrak{C}_2, \mathfrak{t}_2 \rangle$ and $\iota\sharp i$-type $\mathfrak{t}_2'$, if $\mathfrak{P}_1 \Rightarrow \mathfrak{P}_2$ and $\mathfrak{t}_2 \approx_i \mathfrak{t}_2'$ then there is a $\iota\sharp i$-point $\mathfrak{P}_2' = \langle \mathfrak{C}_2', \mathfrak{t}_2' \rangle$ and a $\Rightarrow$-sequence $\mathfrak{S}_1 \Rightarrow \ldots \Rightarrow \mathfrak{S}_n = \mathfrak{P}_2'$ of $\iota\sharp i$-points such that $\mathfrak{S}_k = \langle \mathfrak{D}_k, \mathfrak{s}_k \rangle$, $\mathfrak{s}_1 \approx_i \mathfrak{t}_1$ and $\mathfrak{s}_k \approx_i \mathfrak{t}_2$ for $1 < k \leq n$. Further, if $\mathfrak{P}_1 \Rightarrow^c \mathfrak{P}_2$ then $\mathfrak{s}_k^c \in T_{\mathfrak{D}_k}^{con}$ for $k \leq n$.*

**Proof.** We extend the proof of Halpern et al. (2004, Lemma 5.5) to deal with state candidates and monodic friendly Kripke frames. By the cited result we can prove that if $\mathfrak{t}_1 \Rightarrow \mathfrak{t}_2$ and $\mathfrak{t}_2 \approx_i \mathfrak{t}_2'$ then there is a sequence of $\iota\sharp i$-types $\mathfrak{s}_1 \Rightarrow \ldots \Rightarrow \mathfrak{s}_n = \mathfrak{t}_2'$ such that $\mathfrak{s}_1 \approx_i \mathfrak{t}_1$ and $\mathfrak{s}_k \approx_i \mathfrak{t}_2$ for $1 < k \leq n$. Now by Lemma 7 we can extend this sequence of $\iota\sharp i$-types to a sequence of $\iota\sharp i$-points $\mathfrak{S}_1 \Rightarrow \ldots \Rightarrow \mathfrak{S}_n$ such that $\mathfrak{S}_k = \langle \mathfrak{D}_k, \mathfrak{s}_k \rangle$ and the lemma's statement is satisfied. In particular, if $\mathfrak{P}_1 \Rightarrow^c \mathfrak{P}_2$ also by Lemma 7 we can assume without loss of generality that $\mathfrak{s}_k^c \in T_{\mathfrak{D}_k}^{con}$ for $k \leq n$. $\square$

For any consistent $\phi \in \mathcal{L}_m^1$ we define a quasimodel$^+$ for $\phi$ to establish the completeness of $QKT_m^1$ with respect to $\mathcal{QIS}_m^{pr}$. Let $\mathcal{R}$ be the set of all acceptable $\Rightarrow$-sequences, and define $\mathfrak{f}$ such that $\mathfrak{f}(r, k) = \mathfrak{C}_k$ if $r$ is the $\Rightarrow$-sequence $\mathfrak{C}_0, \mathfrak{C}_1, \ldots$. Finally, let $\mathcal{O}$ be the set of all functions $\rho$ associating every $(r, n) \in Dom(\mathfrak{f})$ to a type $\rho(r, n) \in T_{r,n}$ such that conditions (A) and (B) above are satisfied and

(C') if $\rho(r, n)$ is a $\iota$-type, $\mathfrak{t}$ is a $\iota\sharp i$-type and $\rho(r, n) \approx_i \mathfrak{t}$, then for some $(r', n')$, $\rho(r', n') = \mathfrak{t}$.

(E) if $\rho(r, n) \approx_i \rho(r', n')$ and $n > 0$ then either (a) $\rho(r, n-1) \approx_i \rho(r', n')$ or (b) there is $k < n'$ such that $\rho(r, n-1) \approx_i \rho(r', k)$ and for all $k'$, $k < k' \leq n'$ implies $\rho(r, n) \approx_i \rho(r', k')$.

Finally, for $i \in Ag$, $\rho \in \mathcal{O}$, we define $(r, n) \sim_{i,\rho} (r', n')$ iff $\rho(r, n) \approx_i \rho(r', n')$.

The following lemma shows that the set $\mathcal{O}$ is non-empty. In particular, conditions (C') and (E) are satisfied by the functions in $\mathcal{O}$.

**Lemma 12.** *The set $\mathcal{O}$ of functions that satisfies conditions (A), (B), (C') and (E) is non-empty.*

**Proof.** Conditions (A) and (B) follow respectively from Lemma 6(ii)(a) and the fact that $r$ is an acceptable $\Rightarrow$-sequence. As regards (C') and (E), the proof proceeds by induction on $n$. The result for $n = 0$ is immediate, as we can take $r'$ to be an acceptable $\Rightarrow$-sequence starting from $\mathfrak{C}$ such that $\mathfrak{t} \in \mathfrak{C}$. Further, we define $\rho(r', 0) = \mathfrak{t}$. Thus, $\rho(r', 0) \approx_i \rho(r, 0)$ and both (C') and (E) are satisfied.

Now suppose that $n > 0$ and the result holds for $n - 1$. Since $\mathfrak{f}(r, n - 1) \Rightarrow \mathfrak{f}(r, n)$ and $\rho(r, n) \approx_i \mathfrak{t}$, it follows by Lemma 11 that there is a $\iota\sharp i$-point $\mathfrak{P} = \langle \mathfrak{C}, \mathfrak{t} \rangle$ and a $\Rightarrow$-sequence of $\iota\sharp i$-points $\mathfrak{P}' \Rightarrow \mathfrak{S}_0 \Rightarrow \ldots \Rightarrow \mathfrak{S}_k = \mathfrak{P}$ such that $\mathfrak{S}_{k'} = \langle \mathfrak{D}_{k'}, \mathfrak{s}_{k'} \rangle$ and $\mathfrak{s}_{k'} \approx_i \rho(r, n)$ for $k' \leq k$. By the induction hypothesis, there exists for every $\iota\sharp i$-type $\mathfrak{s}$ such that $\rho(r, n-1) \approx_i \mathfrak{s}$ a point $(r', n')$ and $\rho(r', n') = \mathfrak{s}$. In case (a), we take $\mathfrak{s} = \mathfrak{t}$; then we have that $\rho(r, n-1) \approx_i \rho(r', n')$ for $\rho(r', n') = \mathfrak{t}$. Thus, it is also the case that $\rho(r, n) \approx_i \rho(r', n')$. In case (b), we take $\mathfrak{s} = \mathfrak{t}'$. Hence, $\rho(r, n - 1) \approx_i \rho(r', n')$ for $\rho(r', n') = \mathfrak{t}'$. Now suppose that $r'$ is derived from the acceptable $\Rightarrow$-sequence $v_0, v_1, \ldots$. Let $r''$ be the run derived from an acceptable sequence with initial segment $v_0, \ldots, v_{n'}, \mathfrak{D}_0, \ldots, \mathfrak{D}_k$. Again, such a run exists by Lemma 8. We define $\rho(r'', n' + k + 1) = \mathfrak{s}_k = \mathfrak{t}$. Thus, we have $\rho(r, n) \approx_i \rho(r'', n' + k + 1)$ and both (C') and (E) are satisfied. $\square$

We can now prove the following lemma.

**Lemma 13.** *The tuple $\langle \mathcal{R}, \mathcal{O}, \{\sim_{i,\rho}\}_{i \in Ag, \rho \in \mathcal{O}}, \mathfrak{f} \rangle$ is a frame that satisfies perfect recall.*

**Proof.** By Lemmas 6(i), 8 and 12 the sets $\mathcal{R}$ and $\mathcal{O}$ are non-empty. Also, $\mathfrak{f}$ satisfies the conditions in Definition 17. Finally, each $\sim_{i,\rho}$ is an equivalence relation by definition, and it satisfies perfect recall by definition of the functions $\rho$ in $\mathcal{O}$. $\square$

Finally, we prove the main result in this section.

**Lemma 14.** *The tuple $\langle \mathcal{R}, \mathcal{O}, \{\sim_{i,\rho}\}_{i \in Ag, \rho \in \mathcal{O}}, \mathfrak{f} \rangle$ is a quasimodel$^+$ for $\phi$ with perfect recall and it validates the formulas $K$ and $BF$.*

**Proof.** By the previous lemma $\langle \mathcal{R}, \mathcal{O}, \{\sim_{i,\rho}\}_{i \in Ag, \rho \in \mathcal{O}}, \mathfrak{f} \rangle$ is a frame satisfying perfect recall; so we are left to prove that the functions in $\mathcal{O}$ are objects$^+$. Conditions (1)-(4') on objects$^+$ are satisfied by remarks (A)-(E) and the definition of $\sim_{i,\rho}$. Furthermore, conditions (1), (2) and (3) on quasimodels$^+$ are satisfied by the definitions of $\mathcal{R}$, $\mathfrak{f}$ and $\sim_{i,\rho}$. As regards (4), it follows from Lemma 11. Finally, condition (5) on quasimodels$^+$ holds by Lemma 6(ii)(b), (d), (f) and (h) and Lemma 11. Finally, $\mathfrak{Q}$ validates both the formulas $K$ and $BF$, as all $\mathfrak{t} \in \mathfrak{C}$, for all $\mathfrak{C} \in \mathfrak{Q}$, are consistent with $QKT_m$. $\square$

This completes the proof for $\mathcal{QIS}_m^{pr}$. Thus, we obtain the following item in Theorem 2.

**Theorem 5** (Completeness). *The system $QKT_m^1$ is complete w.r.t. the class $\mathcal{QIS}_m^{pr}$ of QIS.*

The completeness of $QKT_m^1$ with respect to $\mathcal{QIS}_m^{pr,uis}$ follows by Remark 2.

## 5.3 The Classes $\mathcal{QIS}_m^{pr,sync}$ and $\mathcal{QIS}_m^{pr,sync,uis}$

The completeness of $QKT_m^2$ with respect to $\mathcal{QIS}_m^{pr,sync}$ is proved similarly to the previous case by using the following lemma instead of Lemma 11.

**Lemma 15.** *For $\iota$-state candidates $\mathfrak{C}_1$, $\mathfrak{C}_2$ and $\iota\sharp i$-state candidate $\mathfrak{C}_2'$, there is a $\iota\sharp i$-state candidate $\mathfrak{C}_1'$ such that*

- *if $\mathfrak{C}_1 \Rightarrow \mathfrak{C}_2$ and $\mathfrak{C}_2 \approx_i \mathfrak{C}_2'$ then $\mathfrak{C}_1 \approx_i \mathfrak{C}_1'$ and $\mathfrak{C}_1' \Rightarrow \mathfrak{C}_2'$.*

- *for $c \in con\phi$, for $\mathfrak{P}_1 = \langle \mathfrak{C}_1, \mathfrak{t}_1 \rangle$, $\mathfrak{P}_2 = \langle \mathfrak{C}_2, \mathfrak{t}_2 \rangle$ and $\mathfrak{P}_2' = \langle \mathfrak{C}_2', \mathfrak{t}_2' \rangle$, if $\mathfrak{P}_1 \Rightarrow^c \mathfrak{P}_2$ and $\mathfrak{P}_2 \approx_i^c \mathfrak{P}_2'$ then for $\mathfrak{P}_1' = \langle \mathfrak{C}_1', \mathfrak{t}_1' \rangle$, $\mathfrak{P}_1 \approx_i^c \mathfrak{P}_1'$ and $\mathfrak{P}_1' \Rightarrow^c \mathfrak{P}_2'$.*

**Proof.** if $\mathfrak{C}_1 \Rightarrow \mathfrak{C}_2$ and $\mathfrak{C}_2 \approx_i \mathfrak{C}_2'$ then there exist $\mathfrak{t}_1 \in \mathfrak{C}_1$, $\mathfrak{t}_2 \in \mathfrak{C}_2$ and $\mathfrak{t}_2' \in \mathfrak{C}_2'$ such that $\mathfrak{t}_1 \Rightarrow \mathfrak{t}_2$ and $\mathfrak{t}_2 \approx_i \mathfrak{t}_2'$. Moreover, without loss of generality we can assume that for some $c \in con\phi$, $\mathfrak{t}_1^c \in T_1^{con}$, $\mathfrak{t}_2^c \in T_2^{con}$ and $\mathfrak{t}_2'^c \in T_2'^{con}$. Following the proof by Halpern et al. (2004, Lemma 5.8) we can find a $\iota\sharp i$-type $\mathfrak{t}_1'$ such that $\mathfrak{t}_1 \approx_i \mathfrak{t}_1'$ and $\mathfrak{t}_1' \Rightarrow \mathfrak{t}_2'$. Define $T_1'$ as the set of all such $\mathfrak{t}_1'$ and $T_1'^{con}$ as the set of $\mathfrak{t}_1'^c$. We can show that $\mathfrak{C}_1' = \langle T_1', T_1'^{con} \rangle$ is a consistent $\iota\sharp i$-state candidate such that $\mathfrak{C}_1 \approx_i \mathfrak{C}_1'$, $\mathfrak{C}_1' \Rightarrow \mathfrak{C}_2'$, and for $c \in con\phi$, $\mathfrak{P}_1 \approx_i^c \mathfrak{P}_1'$ and $\mathfrak{P}_1' \Rightarrow^c \mathfrak{P}_2'$. $\qquad\square$

For any consistent $\phi \in \mathcal{L}_m^1$ we define a quasimodel$^+$ for $\phi$ to establish the completeness of $\text{QKT}_m^2$ with respect to $\mathcal{QIS}_m^{pr,sync}$. Let $\mathcal{R}$ be the set of all $\Rightarrow$-sequences acceptable from $n$, for some $n \in \mathbb{N}$, and define $\mathfrak{f}$ such that $\mathfrak{f}(r,k) = \mathfrak{C}_k$ if $r$ is the $\Rightarrow$-sequence $X, \ldots, X, \mathfrak{C}_n, \mathfrak{C}_{n+1}, \ldots$ acceptable from $n$ and $k \geq n$, and undefined otherwise. Finally, let $\mathcal{O}$ be the set of all functions $\rho$ associating every $(r,n) \in Dom(\mathfrak{f})$ to a type $\rho(r,n) \in T_{r,n}$ such that conditions (A) and (B) in Section 5.1 are satisfied and

(C") if $\rho(r,n)$ is a $\iota$-type, $\mathfrak{t}$ is a $\iota\sharp i$-type and $\rho(r,n) \approx_i \mathfrak{t}$, then for some $(r',n)$, $\rho(r',n) = \mathfrak{t}$.

(F) if $\rho(r,n) \approx_i \rho(r',n)$ and $n > 0$ then $\rho(r, n-1) \approx_i \rho(r', n-1)$.

Finally, for $i \in Ag$, $\rho \in \mathcal{O}$, we define $(r,n) \sim_{i,\rho} (r',n')$ iff $\rho(r,n) \approx_i \rho(r',n')$ and $n = n'$.

The following remark shows that the set $\mathcal{O}$ is non-empty. In particular, conditions (C") and (F) are satisfied by the functions in $\mathcal{O}$.

**Lemma 16.** *The set $\mathcal{O}$ of functions that satisfies condition (A), (B), (C") and (F) is non-empty.*

**Proof.** Conditions (A) and (B) follow from Lemma 6(ii)(a) and the fact that $r$ is an acceptable $\Rightarrow$-sequence. As regards (C") and (F), assume that $\rho(r,n) \in \mathfrak{f}(r,n)$ is a $\iota$-type, $\mathfrak{t}$ is a $\iota\sharp i$-type and $\rho(r,n) \approx_i \mathfrak{t}$. For each $\mathfrak{s} \in \mathfrak{f}(r,n)$ different from $\rho(r,n)$ consider the set $U = \{\psi \mid K_i\psi \in \mathfrak{s}\}$. We can check that $U$ is consistent and it can be extended to a $\iota\sharp i$-type $\mathfrak{s}'$ such that $\mathfrak{s} \approx_i \mathfrak{s}'$. Now define $T'$ as the collection of all these $\mathfrak{s}'$. Further, for each $\mathfrak{s}^c \in T^{con}$, we set $\mathfrak{s}'^c \in T'^{con}$. Let $\mathfrak{C}' = \langle T', T'^{con} \rangle$. Clearly, $\mathfrak{C} \approx_i \mathfrak{C}'$ and $\langle \mathfrak{C}, \mathfrak{s} \rangle \approx_i^c \langle \mathfrak{C}', \mathfrak{s}' \rangle$. By Lemma 15 we can construct a $\Rightarrow$-sequence $\mathfrak{C}_0 \Rightarrow \ldots \Rightarrow \mathfrak{C}_n$ such that $\mathfrak{C}_n = \mathfrak{C}'$ and $\mathfrak{f}(r,k) \approx_i \mathfrak{C}_k$ for $k \leq n$. By Lemma 8 we can extend this $\Rightarrow$-sequence to an infinite acceptable $\Rightarrow$-sequence $r'$. In particular, the function $\rho$ can be extended so that for $k \leq n$, $\rho(r,k) \approx_i \rho(r',k)$ and $\rho(r',n) = \mathfrak{t}$. Thus, both (C") and (F) are satisfied. $\qquad\square$

We can now show the following lemma.

**Lemma 17.** *The tuple $\langle \mathcal{R}, \mathcal{O}, \{\sim_{i,\rho}\}_{i \in Ag, \rho \in \mathcal{O}}, \mathfrak{f} \rangle$ is a frame that satisfies perfect recall and synchronicity.*

**Proof.** By Lemmas 6(i), 8 and 16 the sets $\mathcal{R}$ and $\mathcal{O}$ are non-empty. Also, $\mathfrak{f}$ satisfies the conditions in Definition 17. Finally, each $\sim_{i,\rho}$ is an equivalence relation by definition, and it satisfies perfect recall and synchronicity by definition of the functions in $\mathcal{O}$. $\quad\square$

Now we prove the main result.

**Lemma 18.** *The tuple $\langle \mathcal{R}, \mathcal{O}, \{\sim_{i,\rho}\}_{i \in Ag, \rho \in \mathcal{O}}, \mathfrak{f} \rangle$ is a quasimodel$^+$ for $\phi$ with perfect recall and synchronicity, and it validates the formulas $K$ and $BF$.*

**Proof.** By the previous lemma $\langle \mathcal{R}, \mathcal{O}, \{\sim_{i,\rho}\}_{i \in Ag, \rho \in \mathcal{O}}, \mathfrak{f} \rangle$ is a frame satisfying perfect recall and synchronicity; so we are left to prove that the functions in $\mathcal{O}$ are objects$^+$. Conditions (1)-(4') on objects$^+$ are sasified by the remarks (A)-(F) and the definition of $\sim_{i,\rho}$. Furthermore, conditions (1), (2) and (3) on quasimodels$^+$ are satisfied by the definitions of $\mathcal{R}$, $\mathfrak{f}$ and $\sim_{i,\rho}$. As regards condition (4), we can make use of Lemma 15 to show that it holds. Additionally, (5) holds by Lemma 6(ii)(b), (d), (f) and (h) and Lemma 15. Finally, $\mathfrak{Q}$ validates both the formulas $K$ and $BF$, as all $\mathfrak{t} \in \mathfrak{C}$, for all $\mathfrak{C} \in \mathfrak{Q}$, are consistent with QKT$_m$. $\quad\square$

This completes the proof for QKT$_m^2$. Thus, we obtain the following item in Theorem 2.

**Theorem 6** (Completeness). *The system $QKT_m^2$ is complete w.r.t. the class $\mathcal{QIS}_m^{pr,sync}$ of QIS.*

The completeness of QKT$_m^2$ with respect to $\mathcal{QIS}_m^{pr,sync,uis}$ follows again by Remark 2.

## 5.4 The Class $\mathcal{QIS}_m^{nl}$

First, we give the following definitions, which will be used in the completeness proof.

**Definition 21.** *If $\mathfrak{t}$ is a $\iota$-type, then $\Phi_{\mathfrak{t},i}$ is the conjunction of all $\iota$-types $\mathfrak{t}'$ such that $\mathfrak{t} \approx_i \mathfrak{t}'$. Similarly, if $\mathfrak{P}$ is a $\iota$-point, then $\Phi_{\mathfrak{P},i}$ is the set of $\iota$-points $\mathfrak{P}'$ such that $\mathfrak{P} \approx_i \mathfrak{P}'$.*

**Definition 22.** *Two sequences of types $\Lambda$ and $\Lambda'$ are $\approx_i$-concordant if there is some $n \in \mathbb{N}$ (or $n$ may be $\infty$) and non-empty consecutive intervals $\Lambda_1, \ldots, \Lambda_n$ of $\Lambda$ and $\Lambda'_1, \ldots, \Lambda'_n$ of $\Lambda'$ such that for all $\mathfrak{s} \in \Lambda_j$ and $\mathfrak{s}' \in \Lambda'_j$ we have $\mathfrak{s} \approx_i \mathfrak{s}'$ for $j \leq n$.*

*Two sequences $\Pi$ and $\Pi'$ of state candidates are $\approx_i$-concordant if for all $\mathfrak{t} \in \mathfrak{C}$, for either $\mathfrak{C} \in \Pi$ or $\mathfrak{C} \in \Pi'$, there are two sequences $\Lambda$ and $\Lambda'$ of types in $\Pi$ and $\Pi'$ respectively that are $\approx_i$-concordant.*

To prove the completeness of QKT$_m^3$ with respect to $\mathcal{QIS}_m^{nl}$ we need the following lemma, which is dual to Lemma 11.

**Lemma 19.** *For all $\iota$-points $\mathfrak{P}_1 = \langle \mathfrak{C}_1, \mathfrak{t}_1 \rangle$, $\mathfrak{P}_2 = \langle \mathfrak{C}_2, \mathfrak{t}_2 \rangle$ and $\iota \sharp i$-type $\mathfrak{t}'_1$, if $\mathfrak{P}_1 \Rightarrow \mathfrak{P}_2$ and $\mathfrak{t}_1 \approx_i \mathfrak{t}'_1$ then there exists a $\iota \sharp i$-point $\mathfrak{P}'_1 = \langle \mathfrak{C}'_1, \mathfrak{t}'_1 \rangle$ and a $\Rightarrow$-sequence $\mathfrak{P}'_1 = \mathfrak{S}_1 \Rightarrow \ldots \Rightarrow \mathfrak{S}_n$ of $\iota \sharp i$-points such that $\mathfrak{S}_k = \langle \mathfrak{D}_k, \mathfrak{s}_k \rangle$, $\mathfrak{s}_k \approx_i \mathfrak{t}_1$ for $k < n$, and $\mathfrak{t}_2 \approx_i \mathfrak{s}_n$. Further, if $\mathfrak{P}_1 \Rightarrow^c \mathfrak{P}_2$ then $\mathfrak{s}_k^c \in T_{\mathfrak{D}_k}^{con}$ for $k \leq n$.*

**Proof.** By adapting the result of Halpern et al. (2004, Lemma 5.11) to types we can prove that if $\mathfrak{t}_1 \Rightarrow \mathfrak{t}_2$ and $\mathfrak{t}_1 \approx_i \mathfrak{t}'_1$ then there is a sequence of $\iota \sharp i$-types $\mathfrak{t}'_1 = \mathfrak{s}_0 \Rightarrow \ldots \Rightarrow \mathfrak{s}_n$ such that $\mathfrak{s}_k \approx_i \mathfrak{t}_1$ for $k < n$ and $\mathfrak{s}_n \approx_i \mathfrak{t}_2$. Now by Lemma 7 we can extend this sequence of $\iota \sharp i$-types to a sequence of $\iota \sharp i$-points $\mathfrak{S}_1 \Rightarrow \ldots \Rightarrow \mathfrak{S}_n$ such that $\mathfrak{S}_k = \langle \mathfrak{D}_k, \mathfrak{s}_k \rangle$. So, the statement of the lemma is satisfied. In particular, if $\mathfrak{P}_1 \Rightarrow^c \mathfrak{P}_2$ then by Lemma 7 we can assume without loss of generality that $\mathfrak{s}_k^c \in T_{\mathfrak{D}_k}^{com}$ for $k \leq n$. $\square$

As pointed out by Halpern et al. (2004), Lemma 19 is not sufficient to construct a quasimodel$^+$ that satisfies no learning. In fact, given a $\Rightarrow$-sequence $\Sigma = \mathfrak{C}_0, \mathfrak{C}_1, \ldots$ of $\iota$-state candidates and a $\iota\sharp i$-type $\mathfrak{t}_0'$ such that $\mathfrak{t}_0 \approx_i \mathfrak{t}_0'$ for $\mathfrak{t}_0 \in \mathfrak{C}_0$, by Lemma 19 we can find a $\Rightarrow$-sequence $\Sigma' = \mathfrak{C}_0', \mathfrak{C}_1', \ldots$ such that $\mathfrak{t}_0' \in \mathfrak{C}_0'$ and no learning is satisfied. However, it does not follow from the acceptability of $\Sigma$ that $\Sigma'$ is also acceptable. So, as in the propositional case, we have to work with trees of state candidates. Hereafter we extend the definitions given by Halpern et al. (2004) to be able to deal with points and monodic friendly Kripke models.

**Definition 23.** *Let $k \leq ad(\phi)$. A $k$-tree of state candidates for $\phi$ is a set $\Pi$ of $\iota$-state candidates for $\phi$ with $|\iota| \leq k$ that contains a unique $\epsilon$-state candidate, i.e., the root, and for every $\iota$-point $\mathfrak{t}$ in some $\mathfrak{C} \in \Pi$,*

- *if $\mathfrak{t}'$ is a $\iota\sharp i$-type such that $\mathfrak{t} \approx_i \mathfrak{t}'$ and $|\iota\sharp i| \leq k$ then there is some $\iota\sharp i$-state candidate $\mathfrak{C}' \in \Pi$ such that $\mathfrak{t}' \in \mathfrak{C}'$;*

- *if $\iota = \iota'\sharp i$ then there is a $\iota'$-state candidate $\mathfrak{C}' \in \Pi$ and a $\iota'$-type $\mathfrak{t}' \in \mathfrak{C}'$ such that $\mathfrak{t} \approx_i \mathfrak{t}'$.*

*Similarly, we define a $k$-tree of points for $\phi$ as a set $\Sigma$ of $\iota$-points for $\phi$ with $|\iota| \leq k$ that contains a unique $\epsilon$-point, and for every $\iota$-point $\mathfrak{P} = \langle \mathfrak{C}, \mathfrak{t} \rangle \in \Sigma$,*

- *if $\mathfrak{t}'$ is a $\iota\sharp i$-type such that $\mathfrak{t} \approx_i \mathfrak{t}'$ and $|\iota\sharp i| \leq k$, then there is some $\iota\sharp i$-point $\mathfrak{P}' = \langle \mathfrak{C}', \mathfrak{t}' \rangle \in \Sigma$;*

- *if $\iota = \iota'\sharp i$ then there is a $\iota'$-point $\mathfrak{P}' = \langle \mathfrak{C}', \mathfrak{t}' \rangle \in \Sigma$ such that $\mathfrak{t} \approx_i \mathfrak{t}'$.*

Intuitively, a $k$-tree is a view of the epistemic state of our quasimodel from a particular type $\mathfrak{t}$, up to $k$ steps from $\mathfrak{t}$. We now extend the $\bigcirc$-suitability relation $\Rightarrow$ to $k$-trees.

**Definition 24.** *Let $\Pi$ and $\Pi'$ be $k$-trees of state candidates for $\phi$. We say that $\Pi \Rightarrow_f \Pi'$ whenever $f$ is a function associating with each $\iota$-state candidate $\mathfrak{C} \in \Pi$ and each $\iota$-type $\mathfrak{t} \in \mathfrak{C}$ finite $\Rightarrow$-sequences of $\iota$-state candidates in $\Pi \cup \Pi'$ and $\iota$-types such that:*

1. *if $f(\mathfrak{C}) = \mathfrak{C}_0 \Rightarrow \ldots \Rightarrow \mathfrak{C}_k$ then (a) $\mathfrak{C} = \mathfrak{C}_0$ and (b) $\mathfrak{C}_j \in \Pi$ for $j < k$ and $\mathfrak{C}_k \in \Pi'$. Similarly, if $f(\mathfrak{t}) = \mathfrak{t}_0 \Rightarrow \ldots \Rightarrow \mathfrak{t}_k$ then (a) $\mathfrak{t} = \mathfrak{t}_0$ and (b) $\mathfrak{t}_j \in \mathfrak{C}_j$ for $j < k$ and $\mathfrak{t}_k \in \mathfrak{C}_k$.*

2. *Let $\mathfrak{t} \in \mathfrak{C}$ and $\mathfrak{t}' \in \mathfrak{C}'$ for some $\mathfrak{C}, \mathfrak{C}' \in \Pi$. If $\mathfrak{t} \approx_i \mathfrak{t}'$ then $f(\mathfrak{t})$ and $f(\mathfrak{t}')$ are $\approx_i$-concordant;*

3. *for at least one $\mathfrak{C} \in \Pi$ the sequence $f(\mathfrak{C})$ has a length of at least 2.*

*Further, let $\Sigma$ and $\Sigma'$ be $k$-trees of points for $\phi$. We say that $\Sigma \Rightarrow_f \Sigma'$ whenever $f$ is a function associating with each $\iota$-point $\mathfrak{P} \in \Sigma$ a finite $\Rightarrow$-sequence of $\iota$-points in $\Sigma \cup \Sigma'$ such that:*

1. *if $f(\mathfrak{P}) = \mathfrak{P}_0 \Rightarrow \ldots \Rightarrow \mathfrak{P}_k$ then (a) $\mathfrak{P} = \mathfrak{P}_0$ and (b) $\mathfrak{P}_j \in \Sigma$ for $j < k$ and $\mathfrak{P}_k \in \Sigma'$;*

2. Let $\mathfrak{P} = \langle \mathfrak{C}, \mathfrak{t} \rangle$ and $\mathfrak{P}' = \langle \mathfrak{C}', \mathfrak{t}' \rangle$ be in $\Sigma$. If $\mathfrak{t} \approx_i \mathfrak{t}'$ then $f(\mathfrak{t})$ and $f(\mathfrak{t}')$ are $\approx_i$-concordant;

3. for at least one $\mathfrak{P} \in \Sigma$ the sequence $f(\mathfrak{P})$ has a length of at least 2.

Finally, for any constant $c \in con\phi$, we say that $\Sigma \Rightarrow_f^c \Sigma'$ whenever $\Sigma \Rightarrow_f \Sigma'$ and $f(\mathfrak{P}) = \mathfrak{P}_0 \Rightarrow^c \ldots \Rightarrow^c \mathfrak{P}_k$.

Notice that given a $k$-tree $\Pi$ of state candidates with root $\mathfrak{C}$ and $\mathfrak{t} \in \mathfrak{C}$, we can obtain a $k$-tree $\Sigma$ of points such that $\mathfrak{P}' = \langle \mathfrak{C}', \mathfrak{t}' \rangle \in \Sigma$ iff $\mathfrak{C}' \in \Pi$. Also, if $\Pi$, $\Pi'$ are $k$-tree of state candidates and $\Pi \Rightarrow_f \Pi'$, then we also have $\Sigma \Rightarrow_f \Sigma'$ where $\Sigma$ and $\Sigma'$ are $k$-trees of points based on $\Pi$ and $\Pi'$ respectively.

We now show how to obtain acceptable sequences of state candidates from sequences of trees. Given two sequences of $\iota$-state candidates $\lambda = \mathfrak{C}_0, \ldots, \mathfrak{C}_k$ and $\mu = \mathfrak{C}'_0, \ldots,$ where $\lambda$ is finite, the *fusion* $\lambda \cdot \mu$ is defined as $\mathfrak{C}_0, \ldots, \mathfrak{C}_{k-1}, \mathfrak{C}'_0, \ldots$ only if $\mathfrak{C}_k = \mathfrak{C}'_0$. Further, given an infinite sequence $\Theta = \Pi_0 \Rightarrow_{f_0} \Pi_1 \Rightarrow_{f_1} \ldots$ of $k$-trees, we say that a sequence $\lambda$ of $\iota$-state candidates is *compatible with* $\Theta$ if there exists some $h \in \mathbb{N}$ and $\iota$-state candidates $\mathfrak{C}_h, \mathfrak{C}_{h+1}, \ldots,$ with $\mathfrak{C}_j \in \Pi_j$ for $j \geq h$, such that $\lambda = f_h(\mathfrak{C}_h) \cdot f_{h+1}(\mathfrak{C}_{h+1}) \cdot \ldots$. The sequence $\Theta$ is *acceptable* if every $\Rightarrow$-sequence compatible with $\Theta$ is infinite and acceptable.

The basic idea of the completeness proof is to define the quasimodel$^+$ starting from an acceptable sequence $\Theta$. Next we introduce some definitions and lemmas that are essential for the completeness proof.

Given a $k$-tree $\Sigma$ and a $\iota$-point $\mathfrak{P} \in \Sigma$ we inductively define the formula $tree_{\Sigma,\mathfrak{P}}$ that describes the $k$-tree $\Sigma$ from the viewpoint of $\mathfrak{P}$.

**Definition 25.** *If $\mathfrak{P}$ is a $\epsilon$-point, then $tree_{\Sigma,\mathfrak{P}} ::= \beta_{\mathfrak{P}}$. If $\mathfrak{P}$ is a $\iota'\sharp i$-point with $\iota' \neq \iota'\sharp i$ then*

$$tree_{\Sigma,\mathfrak{P}} \;=\; \beta_{\mathfrak{P}} \wedge \bigwedge_{\{\iota'-point \; \mathfrak{P}' \in \Sigma | \mathfrak{t}' \approx_i \mathfrak{t}\}} \bar{K}_i tree_{\Sigma,\mathfrak{P}'}$$

If $\Sigma$ and $\Sigma'$ are $k$-trees, $\mathfrak{P} \in \Sigma$ and $\mathfrak{P}' \in \Sigma'$, then we write $(\Sigma, \mathfrak{P}) \Rightarrow^+ (\Sigma', \mathfrak{P}')$ if there is a sequence of $k$-trees $\Sigma_0, \ldots, \Sigma_l$ and functions $f_0, \ldots, f_{l-1}$ such that (a) $\Sigma = \Sigma_0 \Rightarrow_{f_0} \ldots \Rightarrow_{f_{l-1}} \Sigma_l = \Sigma'$; (b) $f_j(\mathfrak{P}) = \mathfrak{P}$ for $j \leq l-2$ and $f_{l-1}(\mathfrak{P}) = (\mathfrak{P}, \mathfrak{P}')$. Similarly, $(\Sigma, \mathfrak{P}) \Rightarrow^{c+} (\Sigma', \mathfrak{P}')$ if $(\Sigma, \mathfrak{P}) \Rightarrow^+ (\Sigma', \mathfrak{P}')$ and (a') $\Sigma = \Sigma_0 \Rightarrow_{f_0}^c \ldots \Rightarrow_{f_{l-1}}^c \Sigma_l = \Sigma'$.

We prove the following lemma, which extends a result by Halpern et al. (2004, Lemma 5.12) to points.

**Lemma 20.** *Suppose $\Sigma$ is a $k$-tree of points and $\mathfrak{P} = \langle \mathfrak{C}, \mathfrak{t} \rangle \in \Sigma$ is a $\iota$-point with $|\iota| = k$,*

(a) *If $\mathfrak{t}'$ is a $\iota$-type and $tree_{\Sigma,\mathfrak{P}} \wedge \bigcirc(\mathfrak{t}' \wedge \xi)$ is consistent, then there is a $k$-tree $\Sigma'$ and a $\iota$-point $\mathfrak{P}' = \langle \mathfrak{C}', \mathfrak{t}' \rangle \in \Sigma'$ such that $(\Sigma, \mathfrak{P}) \Rightarrow^+ (\Sigma', \mathfrak{P}')$ and $tree_{\Sigma',\mathfrak{P}'} \wedge \xi$ is consistent. Further, if $\mathfrak{t}^c \in T^{con}$ then $(\Sigma, \mathfrak{P}) \Rightarrow^{c+} (\Sigma', \mathfrak{P}')$.*

(b) $\vdash tree_{\Sigma,\mathfrak{P}} \to \bigcirc \bigvee_{\{(\Sigma',\mathfrak{P}')|(\Sigma,\mathfrak{P}) \Rightarrow^+ (\Sigma',\mathfrak{P}')\}} tree_{\Sigma',\mathfrak{P}'}$

(c) *if $tree_{\Sigma,\mathfrak{P}} \wedge \psi \mathcal{U} \psi'$ is consistent, then there is a sequence $\Sigma_0, \ldots, \Sigma_l$ of $k$-trees and points $\mathfrak{P}_0, \ldots, \mathfrak{P}_l$ such that (i) $\mathfrak{P}_j \in \Sigma_j$ for $j \leq l$; (ii) $(\Sigma_0, \mathfrak{P}_0) = (\Sigma, \mathfrak{P})$; (iii) $(\Sigma_j, \mathfrak{P}_j) \Rightarrow^+ (\Sigma_{j+1}, \mathfrak{P}_{j+1})$ for $j < l$; (iv) $tree_{\Sigma_j,\mathfrak{P}_j} \wedge \psi$ is consistent for $j < l$; (v) $tree_{\Sigma_l,\mathfrak{P}_l} \wedge \psi'$ is consistent. Further, if $\mathfrak{t}^c \in T^{con}$ then (iii') $(\Sigma_j, \mathfrak{P}_j) \Rightarrow^{c+} (\Sigma_{j+1}, \mathfrak{P}_{j+1})$ for $j < l$.*

**Proof.** We proceed by induction on $k$. The case for $k = 0$ is immediate using standard arguments as $tree_{\Sigma,\mathfrak{P}}$ is just $\beta_{\mathfrak{P}}$.

Assume that $k > 0$ and $\iota = \iota' \sharp i$ for $\iota \neq \iota'$. We first prove part (a) for $\xi = K_i \xi'$, then part (b), then the general case for (a), and finally (c).

As regards part (a) for $\xi = K_i \xi'$, note that $tree_{\Sigma,\mathfrak{P}} \wedge \bigcirc (\mathfrak{t}' \wedge K_i \xi')$ implies that

$$tree_{\Sigma,\mathfrak{P}} \wedge K_i \Phi_{\mathfrak{P},i} \mathcal{U} K_i (\xi' \wedge \Phi_{\mathfrak{t}',i})$$

By the definition of $k$-tree there is a $\iota'$-point $\mathfrak{P}^* \in \Sigma$ such that $\mathfrak{t} \approx_i \mathfrak{t}^*$. Let $\Sigma^*$ be the $(k-1)$-tree consisting of all $\iota^*$-points in $\Sigma$ for $|\iota^*| \leq k - 1$. By the axiom KT3 also $tree_{\Sigma^*,\mathfrak{P}^*} \wedge K_i \Phi_{\mathfrak{P},i} \mathcal{U} K_i (\xi' \wedge \Phi_{\mathfrak{t}',i})$ is consistent, and by part (c) there is a sequence $\Sigma_0, \ldots, \Sigma_l$ of $(k-1)$-trees and points $\mathfrak{P}_0, \ldots, \mathfrak{P}_l$ such that (i) $\mathfrak{P}_j \in \Sigma_j$ for $j \leq l$; (ii) $(\Sigma_0, \mathfrak{P}_0) = (\Sigma^*, \mathfrak{P}^*)$; (iii) $(\Sigma_j, \mathfrak{P}_j) \Rightarrow^+ (\Sigma_{j+1}, \mathfrak{P}_{j+1})$ for $j < l$; (iv) $tree_{\Sigma_j,\mathfrak{P}_j} \wedge K_i \Phi_{\mathfrak{P},i}$ is consistent for $j < l$; (v) $tree_{\Sigma_l,\mathfrak{P}_l} \wedge K_i (\xi' \wedge \Phi_{\mathfrak{t}',i})$ is consistent.

Again, by the definition of the relation $\Rightarrow^+$ there is a sequence of $(k-1)$-trees $\Gamma_0, \ldots, \Gamma_m$ and functions $f_0, \ldots, f_{m-1}$ such that (a) $\Sigma^* = \Sigma_0 = \Gamma_0 \Rightarrow_{f_0} \ldots \Rightarrow_{f_{m-1}} \Gamma_m = \Sigma_l$. Moreover, there are $(k-1)$-points $u_0, \ldots, u_m$ such that $u_0 = \mathfrak{P}^*$, $u_m = \mathfrak{P}_l$, and for $j < m$, $u_j = \mathfrak{P}_{j'}$ for some $j' \leq j$, and if $u_j = u_{j+1}$ then $f_j(u_j) = u_j$, while if $u_j \neq u_{j+1}$ then $f_j(u_j) = (u_j, u_{j+1})$.

We now show how to define the $k$-tree $\Gamma'_j$ extending $\Gamma_j$ for $j < m$. By (iv) above $\beta_{u_j} \wedge K_i \Phi_{\mathfrak{P},i}$ is consistent for $j < m$, and we have that $u_j \approx_i \mathfrak{P}$. So $\mathfrak{P} \in \Gamma'_j$. Similarly, $\beta_{u_m} \wedge K_i \Phi_{\mathfrak{t}',i}$ is consistent; so there exists $\mathfrak{P}' = \langle \mathfrak{C}', \mathfrak{t}' \rangle$ such that $\mathfrak{P}' \in \Gamma'_m$. Further, we can saturate each $\Gamma'_j$ so that the conditions on $k$-trees are satisfied and in particular $\Gamma'_0 = \Sigma$. We now show how to construct $f'_j$ for $j < m$. For each point $\mathfrak{S}' = \langle \mathfrak{D}', \mathfrak{s}' \rangle \in \Gamma'_j \setminus \Gamma_j$ there must exist a point $\mathfrak{S} = \langle \mathfrak{D}, \mathfrak{s} \rangle \in \Gamma_j$ and an agent $j' \in Ag$ such that $\mathfrak{s} \approx_{j'} \mathfrak{s}'$. From Lemma 19 it follows that there exists a sequence $\lambda_{\mathfrak{S}'}$ starting with $\mathfrak{S}'$ that is $\approx_{j'}$-concordant with $f_j(\mathfrak{S})$. Moreover, we can take $\lambda_{\mathfrak{P}_j} = (\mathfrak{P})$ for $j < m - 1$, and $\lambda_{\mathfrak{P}_{m-1}} = (\mathfrak{P}, \mathfrak{P}')$. We define $f'_j$ such that it agrees with $f_j$ on $\Gamma_j$, and for $\mathfrak{S}' \in \Gamma'_j \setminus \Gamma_j$ we have $f'_j(\mathfrak{S}') = \lambda_{\mathfrak{S}'}$.

Notice that $\Gamma'_0 = \Sigma$ by construction. If $m > 0$ it follows immediately from the definition that $(\Sigma, \mathfrak{P}) \Rightarrow^+ (\Gamma_m, \mathfrak{P}')$ and that $tree_{\Gamma_m,\mathfrak{P}'} \wedge K_i \xi'$ is consistent. If $m = 0$ we can easily check that we have $\mathfrak{P}' \in \Sigma$ as $\mathfrak{t}^* \approx_i \mathfrak{t}'$. Since we also have $\mathfrak{t}^* \approx_i \mathfrak{t}$, it follows that $\mathfrak{t} \approx_i \mathfrak{t}'$. We define $f$ so that $f(u) = u$ for every $u \neq \mathfrak{P}$ and $f(\mathfrak{P}) = (\mathfrak{P}, \mathfrak{P}')$. Then $(\Sigma, \mathfrak{P}) \Rightarrow_f (\Sigma, \mathfrak{P}')$. Since also $\mathfrak{P} \Rightarrow \mathfrak{P}'$ we have $(\Sigma, \mathfrak{P}) \Rightarrow^+ (\Sigma, \mathfrak{P}')$.

The second part of (a) follows by a similar line of reasoning.

To prove part (b), by contradiction we assume that

$$\nvdash tree_{\Sigma,\mathfrak{P}} \rightarrow \bigcirc \bigvee_{\{(\Sigma',\mathfrak{P}')|(\Sigma,\mathfrak{P}) \Rightarrow^+ (\Sigma',\mathfrak{P}')\}} tree_{\Sigma',\mathfrak{P}'}$$

Then $tree_{\Sigma,\mathfrak{P}} \wedge \bigcirc \bigwedge_{\{(\Sigma',\mathfrak{P}')|(\Sigma,\mathfrak{P}) \Rightarrow^+ (\Sigma',\mathfrak{P}')\}} \neg tree_{\Sigma',\mathfrak{P}'}$ is consistent. By temporal reasoning there must be some point $u$ such that

$$tree_{\Sigma,\mathfrak{P}} \wedge \bigcirc (\beta_u \wedge \bigwedge_{\{(\Sigma',\mathfrak{P}')|(\Sigma,\mathfrak{P}) \Rightarrow^+ (\Sigma',\mathfrak{P}')\}} \neg tree_{\Sigma',\mathfrak{P}'}) \tag{33}$$

is consistent. Note that $\neg tree_{\Sigma',\mathfrak{P}'}$ is equivalent to $\neg \beta_{\mathfrak{P}'} \vee \bigvee_{\{\iota'-point\ \mathfrak{P}^* \in \Sigma'|\mathfrak{t}^* \approx_i \mathfrak{t}'\}} K_i \neg tree_{\Sigma',\mathfrak{P}^*}$. Thus, the consistency of (33) implies that for each tree $\Sigma'$ such that $(\Sigma, \mathfrak{P}) \Rightarrow^+ (\Sigma', u)$ there exists a $\iota'$-point $\mathfrak{P}_{\Sigma'} = \langle \mathfrak{C}_{\Sigma'}, \mathfrak{t}_{\Sigma'} \rangle$ such that $\mathfrak{t}_{\Sigma'} \approx_i \mathfrak{t}_u$ and

$$tree_{\Sigma, \mathfrak{P}} \wedge \bigcirc (\beta_u \wedge K_i (\bigwedge_{\{\Sigma' | (\Sigma, \mathfrak{P}) \Rightarrow^+ (\Sigma', \mathfrak{P}_{\Sigma'})\}} \neg tree_{\Sigma', \mathfrak{P}_{\Sigma'}})) \tag{34}$$

is consistent. By part (a) there exists a $k$ tree $\Sigma^*$ and $\mathfrak{P}^* \in \Sigma^*$ such that $(\Sigma, \mathfrak{P}) \Rightarrow^+ (\Sigma^*, \mathfrak{P}^*)$ and $tree_{\Sigma^*, \mathfrak{P}^*} \wedge \beta_u \wedge K_i (\bigwedge_{\{\Sigma' | (\Sigma, \mathfrak{P}) \Rightarrow^+ (\Sigma', \mathfrak{P}_{\Sigma'})\}} \neg tree_{\Sigma', \mathfrak{P}_{\Sigma'}})$ is consistent. But this means that $\mathfrak{P}^* = u$. Thus we have a contradiction, since $tree_{\Sigma^*, u} \wedge K_i \neg tree_{\Sigma^*, \mathfrak{P}_{\Sigma^*}}$ is inconsistent.

The general case for (a) follows from (b). Part (c) also follows from (b). $\qquad\square$

The following lemma is the correspondent of Lemma 8 for $k$-trees.

**Lemma 21.** *If $\phi \in \mathcal{L}_m^1$ is consistent with $QKT_m^3$, then there exists an acceptable sequence $\Theta$ of $ad(\phi)$-trees of state candidates such that $\phi$ belongs to the root of the first tree.*

**Proof.** As in Lemma 8 the key part of this proof consists of showing that, given a finite sequence $\Sigma_0 \Rightarrow_{f_0} \ldots \Rightarrow_{f_{l-1}} \Sigma_l$ of $d$-trees of points and a $\iota$-point $\mathfrak{P} = \langle \mathfrak{C}, \mathfrak{t} \rangle \in \Sigma_l$ such that $\psi \mathcal{U} \psi' \in \mathfrak{t}$ (resp. $\bigcirc \psi \in \mathfrak{t}$), by Lemmas 19 and 20 we can extend the sequence of trees to satisfy acceptability. Specifically, suppose that $\psi \mathcal{U} \psi' \in \mathfrak{t}$. Let $\Sigma$ include $\mathfrak{P}$ and the $\iota'$-points $\mathfrak{P}' = \langle \mathfrak{C}', \mathfrak{t}' \rangle \in \Sigma_l$ with $|\iota'| \leq k = |\iota|$. Note that $\Sigma$ is a $k$-tree. Further, by Lemma 20 we can find a sequence $\Theta_0, \ldots, \Theta_n$ of $k$-trees and points $\mathfrak{P}_0, \ldots, \mathfrak{P}_n$ such that (i) $\mathfrak{P}_j \in \Theta_j$ for $j \leq n$; (ii) $(\Theta_0, \mathfrak{P}_0) = (\Sigma, \mathfrak{P})$; (iii) $(\Theta_j, \mathfrak{P}_j) \Rightarrow^+ (\Theta_{j+1}, \mathfrak{P}_{j+1})$ for $j < l$; (iv) $tree_{\Theta_j, \mathfrak{P}_j} \wedge \psi$ is consistent for $j < l$; and (v) $tree_{\Theta_n, \mathfrak{P}_l} \wedge \psi'$ is consistent. By using Lemma 19 we can extend this to a sequence of $ad(\phi)$-trees starting with $\Sigma_l$ that satisfies $\psi \mathcal{U} \psi'$ as in the proof of Lemma 20(a). For $\bigcirc \psi \in \mathfrak{t}$ the argument is similar. Since $\phi$ is consistent, there must be some tree $\Sigma$ with root $\mathfrak{C}$ such that $\psi \in \mathfrak{t}$ for some $\mathfrak{t} \in \mathfrak{C}$; we can then extend $\Sigma$ as above to complete the proof. $\qquad\square$

For any consistent $\phi \in \mathcal{L}_m^1$ we define a quasimodel$^+$ for $\phi$ to establish the completeness of $QKT_m^3$ with respect to $\mathcal{QIS}_m^{nl}$. Let $\mathcal{R}$ consist of all acceptable $\Rightarrow$-sequences compatible with the $ad(\phi)$-tree $\Theta$, while the function $\mathfrak{f}$ is given by $\mathfrak{f}(r, k) = \mathfrak{C}_k$ if $r$ is the acceptable $\Rightarrow$-sequence $\mathfrak{C}_0, \mathfrak{C}_1, \ldots$. Further, let $\mathcal{O}$ be the set of functions $\rho$ associating every $(r, n) \in Dom(\mathfrak{f})$ to a type $\rho(r, n) \in T_{r,n}$ such that conditions (A), (B) and (C') given previously are satisfied and the following holds:

(G) if $\rho(r, n) \approx_i \rho(r', n')$ then either $\rho(r, n+1) \approx_i \rho(r', n')$ or there exists $k > n'$ such that $\rho(r, n+1) \approx_i \rho(r', k)$ and for all $k'$, $k > k' \geq n'$ implies $\rho(r, n) \approx_i \rho(r', k')$.

Finally, for $i \in Ag$, $\rho \in \mathcal{O}$, $(r, n) \sim_{i,\rho} (r', n')$ iff $\rho(r, n) \approx_i \rho(r', n')$.

As in the previous cases we have the following.

**Lemma 22.** *The set $\mathcal{O}$ of functions that satisfies conditions (A), (B), (C') and (G) is non-empty.*

**Proof.** Conditions (A) and (B) are guaranteed by Lemma 6(ii) and by the fact that $r$ is an acceptable $\Rightarrow$-sequence respectively. As regards (C') and (G), assume that $\rho(r, n)$ is a $\iota$-type, $\mathfrak{t}$ is a $\iota\sharp i$-type and $\rho(r, n) \approx_i \mathfrak{t}$. By using the proofs of Lemmas 20 and 19 we can find an acceptable $\Rightarrow$-sequence $r'$ compatible with the $d$-tree $\Theta$ such that $\mathfrak{t} \in \mathfrak{f}(r', 0)$ and (G) is satisfied. $\qquad\square$

We can now show the following lemma.

**Lemma 23.** *The tuple $\langle \mathcal{R}, \mathcal{O}, \{\sim_{i,\rho}\}_{i \in Ag, \rho \in \mathcal{O}}, \mathfrak{f} \rangle$ is a frame that satisfies no learning.*

**Proof.** By Lemmas 6(i), 21 and 22 the sets $\mathcal{R}$ and $\mathcal{O}$ are non-empty. Also, $\mathfrak{f}$ satisfies the conditions in Definition 17. Further, each $\sim_{i,\rho}$ is an equivalence relation by definition. Finally, the no learning condition is satisfied by definition of the functions in $\mathcal{O}$. $\qquad\square$

**Lemma 24.** *The tuple $\langle \mathcal{R}, \mathcal{O}, \{\sim_{i,\rho}\}_{i \in Ag, \rho \in \mathcal{O}}, \mathfrak{f} \rangle$ is a quasimodel$^+$ for $\phi$ that satisfies no learning and validates the formulas $K$ and $BF$.*

**Proof.** By the previous lemma $\langle \mathcal{R}, \mathcal{O}, \{\sim_{i,\rho}\}_{i \in Ag, \rho \in \mathcal{O}}, \mathfrak{f} \rangle$ is a frame satisfying no learning; so we are left to prove that the functions in $\mathcal{O}$ are objects$^+$. Conditions (1)-(4') on objects$^+$ are safisfied by remarks (A)-(G) and the definition of $\sim_{i,\rho}$. Furthermore, conditions (1), (2) and (3) on quasimodels$^+$ are satisfied by the definitions of $\mathcal{R}$, $\mathfrak{f}$ and $\sim_{i,\rho}$. As regards (4) we use Lemma 19 to show that it holds. Finally, (5) holds by Lemma 6(ii)(b), (d), (f) and (h) and Lemma 19. Finally, $\mathfrak{Q}$ validates both $K$ and $BF$, as all $\mathfrak{t} \in \mathfrak{C}$, for all $\mathfrak{C} \in \mathfrak{Q}$, are consistent with $\mathrm{QKT}_m$. $\qquad\square$

This completes the proof for $\mathrm{QKT}_m^3$. Thus, we obtain the following item in Theorem 2.

**Theorem 7** (Completeness). *The system $QKT_m^3$ is complete w.r.t. the class $\mathcal{QIS}_m^{nl}$ of QIS.*

## 5.5 The Class $\mathcal{QIS}_m^{nl,sync}$

To show that $\mathrm{QKT}_m^4$ is a complete axiomatisation for $\mathcal{QIS}_m^{nl,sync}$, analogously to Lemma 15, we need the following.

**Lemma 25.** *For $\iota$-state candidate $\mathfrak{C}_1$, $\mathfrak{C}_2$ and $\iota \sharp i$-state candidate $\mathfrak{C}_1'$ there exists a $\iota \sharp i$-state candidate $\mathfrak{C}_2'$ such that*

- *if $\mathfrak{C}_1 \Rightarrow \mathfrak{C}_2$ and $\mathfrak{C}_1 \approx_i \mathfrak{C}_1'$ then $\mathfrak{C}_1' \Rightarrow \mathfrak{C}_2'$ and $\mathfrak{C}_2 \approx_i \mathfrak{C}_2'$.*

- *for all $c \in con\phi$, for $\mathfrak{P}_1 = \langle \mathfrak{C}_1, \mathfrak{t}_1 \rangle$, $\mathfrak{P}_2 = \langle \mathfrak{C}_2, \mathfrak{t}_2 \rangle$ and $\mathfrak{P}_1' = \langle \mathfrak{C}_1', \mathfrak{t}_1' \rangle$, if $\mathfrak{P}_1 \Rightarrow^c \mathfrak{P}_2$ and $\mathfrak{P}_1 \approx_i^c \mathfrak{P}_1'$ then $\mathfrak{P}_1' \Rightarrow^c \mathfrak{P}_2'$ and $\mathfrak{P}_2 \approx_i^c \mathfrak{P}_2'$.*

**Proof.** The proof is similar to Lemma 15. If $\mathfrak{C}_1 \Rightarrow \mathfrak{C}_2$ and $\mathfrak{C}_1 \approx_i \mathfrak{C}_1'$ then there exist $\mathfrak{t}_1 \in \mathfrak{C}_1$, $\mathfrak{t}_2 \in \mathfrak{C}_2$ and $\mathfrak{t}_1' \in \mathfrak{C}_1'$ such that $\mathfrak{t}_1 \Rightarrow \mathfrak{t}_2$ and $\mathfrak{t}_1 \approx_i \mathfrak{t}_1'$. Moreover, without loss of generality, we can assume that for some $c \in con\phi$, $\mathfrak{t}_1^c \in T_1^{con}$, $\mathfrak{t}_2^c \in T_2^{con}$ and $\mathfrak{t}_1^{\prime c} \in T_1^{\prime con}$. By adapting the proof of Halpern et al. (2004, Lemma 5.18) we can find a $\iota \sharp i$-type $\mathfrak{t}_2'$ such that $\mathfrak{t}_2 \approx_i \mathfrak{t}_2'$ and $\mathfrak{t}_1' \Rightarrow \mathfrak{t}_2'$. We define $T_2'$ as the set of all such $\mathfrak{t}_2'$ and $T_1^{\prime con}$ as the set of $\mathfrak{t}_2^{\prime c}$. Clearly, $\mathfrak{C}_2' = \langle T_2', T_2^{\prime con} \rangle$ is a consistent $\iota \sharp i$-state candidate such that $\mathfrak{C}_2 \approx_i \mathfrak{C}_2'$, $\mathfrak{C}_1' \Rightarrow \mathfrak{C}_2'$, and for $c \in con\phi$, $\mathfrak{P}_2 \approx_i^c \mathfrak{P}_2'$ and $\mathfrak{P}_1' \Rightarrow^c \mathfrak{P}_2'$. $\qquad\square$

For systems including the axiom $\mathrm{KT}_m^4$ we can define a synchronous version of the relation $\Rightarrow$ between $k$-trees.

**Definition 26.** *If $\Pi$ and $\Pi'$ are $k$-trees of state candidates for $\phi$ then $\Pi \Rightarrow_{\mathfrak{f}}^{sync} \Pi'$ iff $\Pi \Rightarrow_{\mathfrak{f}} \Pi'$ and for all $\mathfrak{C} \in \Pi$, $\mathfrak{f}(\mathfrak{C})$ has exactly a length of $2$. Similarly, if $\Sigma$ and $\Sigma'$ are $k$-trees of points for $\phi$ then $\Sigma \Rightarrow_{\mathfrak{f}}^{sync} \Sigma'$ iff $\Sigma \Rightarrow_{\mathfrak{f}} \Sigma'$ and for all $\mathfrak{P} \in \Sigma$, $\mathfrak{f}(\mathfrak{P})$ has exactly a length of $2$.*

For any $c \in con\phi$, the relation $\Rightarrow_f^{c\ sync}$ is defined similarly. We define a *sync*-acceptable sequence of trees as an acceptable sequence where the relation $\Rightarrow$ is substituted by the relation $\Rightarrow^{sync}$, that is, the sequence $\Theta$ is acceptable if every $\Rightarrow^{sync}$-sequence compatible with $\Theta$ is infinite and acceptable. Similarly, given the relations $\Rightarrow^+$ and $\Rightarrow^{c+}$ for $c \in con\phi$, the definitions of $\Rightarrow^{sync,+}$ and $\Rightarrow^{c\ sync,+}$ are straightforward. We now state the following result, which is a simplified version of Lemma 20. The proof is analogous to that of Lemma 20, in which Lemma 25 is used instead of Lemma 19.

**Lemma 26.** *Let $\Sigma$ be a $k$-tree of points and $\mathfrak{P} \in \Sigma$ is a $\iota$-point with $|\iota| = k$,*

(a) *If $\mathfrak{t}'$ is a $\iota$-type and $tree_{\Sigma,\mathfrak{P}} \wedge \bigcirc(\mathfrak{t}' \wedge \xi)$ is consistent, then there exists a $k$-tree $\Sigma'$ and a $\iota$-point $\mathfrak{P}' = \langle \mathfrak{C}', \mathfrak{t}' \rangle \in \Sigma'$ such that $(\Sigma, \mathfrak{P}) \Rightarrow^{sync,+} (\Sigma', \mathfrak{P}')$ and $tree_{\Sigma',\mathfrak{P}'} \wedge \xi$ is consistent. Further, if $\mathfrak{t}^c \in T^{con}$ then $(\Sigma, \mathfrak{P}) \Rightarrow^{c\ sync,+} (\Sigma', \mathfrak{P}')$.*

(b) $\vdash tree_{\Sigma,\mathfrak{P}} \rightarrow \bigcirc \bigvee_{\{(\Sigma',\mathfrak{P}')|(\Sigma,\mathfrak{P})\Rightarrow^{sync,+}(\Sigma',\mathfrak{P}')\}} tree_{\Sigma',\mathfrak{P}'}$

(c) *if $tree_{\Sigma,\mathfrak{P}} \wedge \psi \mathcal{U} \psi'$ is consistent, then there exists a sequence $\Sigma_0, \ldots, \Sigma_l$ of $k$-trees and points $\mathfrak{P}_0, \ldots, \mathfrak{P}_l$ such that (i) $\mathfrak{P}_j \in \Sigma_j$ for $j \leq l$; (ii) $(\Sigma_0, \mathfrak{P}_0) = (\Sigma, \mathfrak{P})$; (iii) $(\Sigma_j, \mathfrak{P}_j) \Rightarrow^{sync,+} (\Sigma_{j+1}, \mathfrak{P}_{j+1})$ for $j < l$; (iv) $tree_{\Sigma_j,\mathfrak{P}_j} \wedge \psi$ is consistent for $j < l$; and (v) $tree_{\Sigma_l,\mathfrak{P}_l} \wedge \psi'$ is consistent. Further, if $\mathfrak{t}^c \in T^{con}$ then (iii') $(\Sigma_j, \mathfrak{P}_j) \Rightarrow^{c\ sync,+} (\Sigma_{j+1}, \mathfrak{P}_{j+1})$ for $j < l$.*

Further, we make use of Lemma 26 to adapt Lemma 21 and obtain the following result.

**Lemma 27.** *If $\phi \in \mathcal{L}_m^1$ is consistent with $QKT_m^4$, then there exists a sync-acceptable sequence $\Theta$ of $ad(\phi)$-trees of state candidates such that $\phi$ belongs to the root of the first tree.*

For any consistent $\phi \in \mathcal{L}_m^1$ we define a quasimodel$^+$ for $\phi$ to establish the completeness of $QKT_m^4$ with respect to $\mathcal{QIS}_m^{nl,sync}$. Let $X$ be a new object, a sequence $X, \ldots, X, \mathfrak{C}_n, \mathfrak{C}_{n+1}, \ldots$ is *sync-acceptable from $n$* if it starts with $n$ copies of $X$ and $\mathfrak{C}_n, \mathfrak{C}_{n+1}, \ldots$ is a *sync*-acceptable $\Rightarrow$-sequence compatible with the $ad(\phi)$-tree $\Theta$. Let $\mathcal{R}$ consist of all $\Rightarrow$-sequences *sync*-acceptable from $n$, for some $n \in \mathbb{N}$. The function $\mathfrak{f}$ is defined as $\mathfrak{f}(r, k) = \mathfrak{C}_k$ if $r$ is the $\Rightarrow$-sequence $X, \ldots, X, \mathfrak{C}_n, \mathfrak{C}_{n+1}, \ldots$ *sync*-acceptable from $n$ and $k \geq n$; $\mathfrak{f}(r, k)$ is undefined otherwise. Further, Let $\mathcal{O}$ be the set of functions $\rho$ associating every $(r, n) \in Dom(\mathfrak{f})$ to a type $\rho(r, n) \in T_{r,n}$ such that conditions (A), (B) and (C") are satisfied and the following holds:

(H) *if $\rho(r, n) \approx_i \rho(r', n')$ then $\rho(r, n+1) \approx_i \rho(r', n'+1)$.*

Finally, for $i \in Ag$, $\rho \in \mathcal{O}$, $(r, n) \sim_{i,\rho} (r', n')$ iff $\rho(r, n) \approx_i \rho(r', n')$ and $n = n'$.

Similarly to Lemma 22, we can show the following.

**Lemma 28.** *The set $\mathcal{O}$ of functions that satisfies conditions (A), (B), (C") and (H) above is non-empty.*

Moreover, the following result follows from Lemmas 6(i), 27 and 28.

**Lemma 29.** *The tuple $\langle \mathcal{R}, \mathcal{O}, \{\sim_{i,\rho}\}_{i \in Ag, \rho \in \mathcal{O}}, \mathfrak{f} \rangle$ is a frame that satisfies no learning and synchronicity.*

Finally, by adapting the proof for Lemma 24 we can state the following result.

**Lemma 30.** *The tuple $\langle \mathcal{R}, \mathcal{O}, \{\sim_{i,\rho}\}_{i \in Ag, \rho \in \mathcal{O}}, \mathfrak{f}\rangle$ is a quasimodel$^+$ for $\phi$ with no learning and synchronicity, and it validates the formulas $K$ and $BF$.*

This completes the proof for $QKT_m^4$. Thus, we obtain the following item in Theorem 2.

**Theorem 8** (Completeness). *The system $QKT_m^4$ is complete w.r.t. the class $\mathcal{QIS}_m^{nl,sync}$ of QIS.*

### 5.6 The Classes $\mathcal{QIS}_m^{nl,pr}$ and $\mathcal{QIS}_1^{nl,pr,uis}$

To obtain the completeness proof for $\mathcal{QIS}_m^{nl,pr}$ we combine the results shown for $\mathcal{QIS}_m^{pr}$ and $\mathcal{QIS}_m^{nl}$.

If $\phi \in \mathcal{L}_m^1$ is consistent with $QKT_m^{2,3}$ then by Lemma 21 there exists an acceptable sequence $\Theta$ of $ad(\phi)$-trees such that $\phi$ belongs to the root of the first tree. Let $\mathcal{R}$ be the set of all acceptable $\Rightarrow$-sequences *that have a suffix* that is compatible with $\Theta$, while the function $\mathfrak{f}$ is defined as in Section 5.2. Further, $\mathcal{O}$ is the set of all functions $\rho$ associating every $(r, n) \in Dom(\mathfrak{f})$ to a type $\rho(r, n) \in T_{r,n}$ that satisfies the conditions (A), (B), (C'), (E) and (G). Finally, for $i \in Ag$, $\rho \in \mathcal{O}$, $(r, n) \sim_{i,\rho} (r', n')$ iff $\rho(r, n) \approx_i \rho(r', n')$.

**Lemma 31.** *The set $\mathcal{O}$ of functions that satisfies conditions (A), (B), (C'), (E) and (G) is non-empty.*

**Proof.** We can show that all conditions but (C') are satisfied similarly to the cases of $\mathcal{QIS}_m^{pr}$ and $\mathcal{QIS}_m^{nl}$. As to (C'), suppose that $\rho(r, n)$ is a $\iota$-type in $\mathfrak{f}(r, n)$ and $\mathfrak{t}$ is $\iota \sharp i$-type. Also, $\Theta$ is the sequence of $ad(\phi)$-trees $\Pi_0 \Rightarrow_{f_0} \Pi_1 \Rightarrow_{f_1} \ldots$ of state candidates. The run $r$ is derived by definition from a $\Rightarrow$-sequence $\mathfrak{C}_0, \mathfrak{C}_1, \ldots$ that has a suffix $\mathfrak{C}_N, \mathfrak{C}_{N+1}, \ldots$ that is compatible with $\Theta$, and $\mathfrak{f}(r, n) = \mathfrak{C}_n$. We consider two cases.

If $n \geq N$, then there exists some $k \in \mathbb{N}$ such that $\mathfrak{C}_n \in \Pi_k$. By Lemma 11 there exists a $\Rightarrow$-sequence $\mathfrak{S}_0 \Rightarrow \ldots \Rightarrow \mathfrak{S}_h$ of $\iota \sharp i$-state candidates such that $\mathfrak{t} \in \mathfrak{S}_h$ and $\mathfrak{S}_0, \ldots, \mathfrak{S}_h$ is $\approx_i$-concordant with $\mathfrak{C}_0, \ldots, \mathfrak{C}_n$. Further, we can assume that $\mathfrak{S}_h \in \Pi_k$ and let $\mathfrak{S}_h, \mathfrak{S}_{h+1}, \ldots$ be the sequence compatible with $\Theta$. Now consider the $\Rightarrow$-sequence $\mathfrak{S}_0 \Rightarrow \mathfrak{S}_1 \Rightarrow \ldots$. By construction the run $r'$ derived from this sequence is in $\mathcal{R}$ and we can assume that $\rho(r', h) = \mathfrak{t}$.

If $n < N$, then by Lemma 11 there exists a $\Rightarrow$-sequence $\mathfrak{S}_0 \Rightarrow \ldots \Rightarrow \mathfrak{S}_h$ of $\iota \sharp i$-state candidates such that $\mathfrak{t} \in \mathfrak{S}_h$ and $\mathfrak{S}_0, \ldots, \mathfrak{S}_h$ is $\approx_i$-concordant with $\mathfrak{C}_0, \ldots, \mathfrak{C}_n$. By Lemma 19 we can extend this sequence to a $\Rightarrow$-sequence $\mathfrak{S}_0 \Rightarrow \ldots \Rightarrow \mathfrak{S}_k$ that is $\approx_i$-concordant with $\mathfrak{C}_0, \ldots, \mathfrak{C}_N$. Since $\mathfrak{C}_N \in \Pi_M$ for some $M \in \mathbb{N}$, we can assume that $\mathfrak{S}_k \in \Pi_M$ as well. Let $\mathfrak{S}_h, \mathfrak{S}_{h+1}, \ldots$ be the sequence compatible with $\Theta$, and consider the $\Rightarrow$-sequence $\mathfrak{S}_0 \Rightarrow \mathfrak{S}_1 \Rightarrow \ldots$. As in the previous case, the run $r'$ derived from this sequence is in $\mathcal{R}$ by construction and we can assume that $\rho(r', h) = \mathfrak{t}$. $\square$

By Lemmas 6(i), 21 and 31 we obtain the next result.

**Lemma 32.** *The tuple $\langle \mathcal{R}, \mathcal{O}, \{\sim_{i,\rho}\}_{i \in Ag, \rho \in \mathcal{O}}, \mathfrak{f}\rangle$ is a frame that satisfies perfect recall and no learning.*

Finally, we state the following lemma, whose proof follows the lines of the corresponding proofs for $\mathcal{QIS}_m^{pr}$ and $\mathcal{QIS}_m^{nl}$ and Lemma 31.

**Lemma 33.** *The tuple $\langle \mathcal{R}, \mathcal{O}, \{\sim_{i,\rho}\}_{i \in Ag, \rho \in \mathcal{O}}, \mathfrak{f} \rangle$ is a quasimodel$^+$ for $\phi$ that satisfies perfect recall and no learning, and validates the formulas K and BF.*

This establishes the completeness of QKT$^{2,3}$. Thus, we obtain the following item in Theorem 2.

**Theorem 9** (Completeness). *The system $QKT_m^{2,3}$ is complete w.r.t. the class $\mathcal{QIS}_m^{nl,pr}$ of QIS.*

The completeness of QKT$_1^{2,3}$ with respect to $\mathcal{QIS}_1^{nl,pr,uis}$ follows from the following remark, whose proof is analogous to the propositional case.

**Remark 3.** *A formula $\phi \in \mathcal{L}_1^1$ is satisfiable in $\mathcal{QIS}_1^{nl,pr}$ (resp. $\mathcal{QIS}_1^{nl,pr,sync}$) iff it is satisfiable in $\mathcal{QIS}_1^{nl,pr,uis}$ (resp. $\mathcal{QIS}_1^{nl,pr,sync,uis}$).*

### 5.7 The Class $\mathcal{QIS}_m^{nl,pr,sync}$

To prove the completeness of QKT$_m^{1,4}$ with respect to $\mathcal{QIS}_m^{nl,pr,sync}$ we combine the results obtained for $\mathcal{QIS}_m^{nl,pr}$ in the previous section with those for $\mathcal{QIS}_m^{nl,sync}$ and $\mathcal{QIS}_m^{pr,sync}$. Specifically, if $\phi \in \mathcal{L}_m^1$ is consistent with QKT$_m^{1,4}$ by Lemma 27 we can construct a *sync*-acceptable sequence $\Theta$ of $ad(\phi)$-trees such that $\phi$ belongs to the root of the first tree. Let $\mathcal{R}$ be the set of all *sync*-acceptable $\Rightarrow$-sequences with suffixes that are compatible with $\Theta$; and the function $\mathfrak{f}$ is defined as in Section 5.2. Further, $\mathcal{O}$ is the set of all functions $\rho$ associating every $(r, n) \in Dom(\mathfrak{f})$ to a type $\rho(r, n) \in T_{r,n}$ that satisfies the conditions (A), (B), (C"), (F) and (H). Finally, for $i \in Ag$, $\rho \in \mathcal{O}$, $(r, n) \sim_{i,\rho} (r', n')$ iff $\rho(r, n) \approx_i \rho(r', n')$ and $n = n'$.

By adapting the proof of Lemma 31 by means of Lemmas 15 and 25 we can show the following result.

**Lemma 34.** *The set $\mathcal{O}$ of functions that satisfies conditions (A), (B), (C"), (F) and (H) is non-empty.*

By Lemmas 6(i), 27 and 34 we obtain the following result.

**Lemma 35.** *The tuple $\langle \mathcal{R}, \mathcal{O}, \{\sim_{i,\rho}\}_{i \in Ag, \rho \in \mathcal{O}}, \mathfrak{f} \rangle$ is a frame that satisfies perfect recall, no learning and synchronicity.*

Finally, we state the following lemma whose proof follows the lines of the corresponding proofs for $\mathcal{QIS}_m^{pr,sync}$, $\mathcal{QIS}_m^{nl,sync}$ and Lemma 34.

**Lemma 36.** *The tuple $\langle \mathcal{R}, \mathcal{O}, \{\sim_{i,\rho}\}_{i \in Ag, \rho \in \mathcal{O}}, \mathfrak{f} \rangle$ is a quasimodel$^+$ for $\phi$ that satisfies perfect recall, no learning and synchronicity, and validates the formulas K and BF.*

This completes the proof for QKT$_m^{1,4}$. Thus, we obtain the following item in Theorem 2.

**Theorem 10** (Completeness). *The system $QKT_m^{1,4}$ is complete w.r.t. the class $\mathcal{QIS}_m^{nl,pr,sync}$ of QIS.*

## 5.8 The Classes $\mathcal{QIS}_m^{nl,sync,uis}$ and $\mathcal{QIS}_m^{nl,pr,sync,uis}$

We now show that the system $\mathrm{QKT}_m^{1,4,5}$ is complete with respect to the classes $\mathcal{QIS}_m^{nl,sync,uis}$ and $\mathcal{QIS}_m^{nl,pr,sync,uis}$. The completeness result follows from next remark.

**Remark 4.** *A formula* $\phi \in \mathcal{L}_m$ *is valid on* $\mathcal{QIS}_m^{nl,sync,uis}$ *iff it is valid in* $\mathcal{QIS}_m^{nl,pr,sync,uis}$.

The proof is a straightforward extension to first-order of a result by Halpern et al. (2004, Proposition 5.22). Given this remark and the axiom KT5 it is sufficient to prove the completeness of $\mathrm{QKT}_1^{1,4}$ with respect to $\mathcal{QIS}_1^{nl,pr,sync,uis}$. By the result in the previous section, $\mathrm{QKT}_1^{1,4}$ is indeed complete with respect to $\mathcal{QIS}_1^{nl,pr,sync}$. The desired result follows by Remark 3. Thus, we obtain the following item in Theorem 2.

**Theorem 11** (Completeness)**.** *The system* $QKT_m^{1,4}$ *is complete w.r.t. the classes* $\mathcal{QIS}_m^{nl,sync,uis}$ *and* $\mathcal{QIS}_m^{nl,pr,sync,uis}$ *of QIS.*

# 6. Conclusions and Further Work

In this paper we investigated interaction axioms in the context of monodic first-order temporal-epistemic logic. Specifically, we explored classes of quantified interpreted systems satisfying conditions such as synchronicity, no learning, perfect recall, and having a unique initial state. The contribution of the article concerns the provably complete axiomatisation of these classes.

The results presented extend previous contributions on first-order epistemic and temporal logic with no interactions (e.g., see Belardinelli & Lomuscio, 2011, Sturm et al., 2000, Wolter & Zakharyaschev, 2002), in a direction that was previously only explored at the propositional level (Halpern et al., 2004). Our findings show that the characterisation axioms considered at the propositional level can be extended to the first-order monodic setting.

While temporal-epistemic logic in a first-order context has so far mostly attracted theoretical contributions, there is evidence in the literature of it being increasingly embraced in applications. For instance, there is an active interest in verifying artifact-centric systems against first-order modal specifications (Belardinelli, Lomuscio, & Patrizi, 2011a, 2011b; Deutsch, Hull, Patrizi, & Vianu, 2009; Deutsch, Sui, & Vianu, 2007; Calvanese, Giacomo, Lenzerini, & Rosati, 2012; Hariri, Calvanese, Giacomo, Masellis, & Felli, 2011).

Given this, it remains of importance to investigate the questions pertaining to computational aspects of the formalisms introduced, including their decidability and the computational complexity of the satisfiability and model checking problems. Work so far (including Belardinelli & Lomuscio, 2011; Hodkinson at al., 2000; Wolter & Zakharyaschev 2001) has focused on fragments where no interaction is present, but we know from the literature (Halpern et al., 2004) that interactions can make these problems harder. We leave this for further work, particularly in connection with the addition of other epistemic modalities (e.g., explicit and algorithmic knowledge, see Halpern & Pucella, 2005), or branching-time modalities. Epistemic variants of branching-time CTL are well understood at the propositional level (Meyden & Wong, 2003) but their first-order extensions have not yet been explored.

## Acknowledgments

The research presented was supported by the European Commission through the Marie Curie Fellowship "FoMMAS" (grant n. 235329) and the STREP Project "ACSI" (grant n. 257593), and by the UK Engineering and Physical Sciences Research Council Leadership Fellowship "Trusted Autonomous Systems" (grant n. EP/I00520X/1).

We would like to thank the anonymous reviewers and Mr. Andrew V. Jones for valuable comments on the paper.

## References

Belardinelli, F., & Lomuscio, A. (2009). Quantified epistemic logics for reasoning about knowledge in multi-agent systems. *Artificial Intelligence*, *173*(9-10), 982–1013.

Belardinelli, F., & Lomuscio, A. (2011). First-order linear-time epistemic logic with group knowledge: An axiomatisation of the monodic fragment. *Fundamenta Informaticae*, *106*(2-4), 175–90.

Belardinelli, F., & Lomuscio, A. (2008). A complete quantified epistemic logic for reasoning about message passing systems. In *Computational Logic in Multi-Agent Systems, 8th International Workshop, CLIMA VIII. Revised Selected and Invited Papers*, Vol. 5056 of *Lecture Notes in Computer Science*, pp. 248–267. Springer.

Belardinelli, F., & Lomuscio, A. (2010). Interactions between time and knowledge in a first-order logic for multi-agent systems. In *Principles of Knowledge Representation and Reasoning: Proceedings of the 12th International Conference, KR 2010*. AAAI Press.

Belardinelli, F., Lomuscio, A., & Patrizi, F. (2011a). A computationally-grounded semantics for artifact-centric systems and abstraction results. In *Proceedings of the 22nd International Joint Conference on Artificial Intelligence, IJCAI 2011*, pp. 738–743. AAAI Press.

Belardinelli, F., Lomuscio, A., & Patrizi, F. (2011b). Verification of deployed artifact systems via data abstraction. In *Service-Oriented Computing: Proceedings of the 9th International Conference, ICSOC 2011*, Vol. 7084 of *Lecture Notes in Computer Science*, pp. 142–156. Springer.

Calvanese, D., Giacomo, G. D., Lenzerini, M., & Rosati, R. (2012). View-based query answering in description logics: Semantics and complexity. *Journal of Computer and System Sciences*, *78*(1), 26–46.

Cohen, P., & Levesque, H. (1995). Communicative actions for artificial agents. In *Proceedings of the 1st International Conference on Multi-Agent Systems, ICMAS 1995*, pp. 65–72. AAAI Press.

Degtyarev, A., Fisher, M., & Konev, B. (2003). Monodic temporal resolution. In *Automated Deduction: Proceedings of the 19th International Conference on Automated Deduction, CADE-19*, Vol. 2741 of *Lecture Notes in Computer Science*, pp. 397–411. Springer.

Degtyarev, A., Fisher, M., & Lisitsa, A. (2002). Equality and monodic first-order temporal logic. *Studia Logica*, *72*(2), 147–156.

Dennett, D. (1987). *The Intentional Stance*. MIT Press.

Deutsch, A., Hull, R., Patrizi, F., & Vianu, V. (2009). Automatic verification of data-centric business processes. In *Database Theory: Proceedings of the 12th International Conference, ICDT 2009*, Vol. 361 of *ACM International Conference Proceeding Series*, pp. 252–267. ACM Press.

Deutsch, A., Sui, L., & Vianu, V. (2007). Specification and verification of data-driven web applications. *Journal of Computer and System Sciences*, *73*(3), 442–474.

Fagin, R., Halpern, J. Y., Moses, Y., & Vardi, M. Y. (1995). *Reasoning about Knowledge*. MIT Press.

Fagin, R., Halpern, J. Y., & Vardi, M. Y. (1992). What can machines know? On the properties of knowledge in distributed systems. *Journal of the ACM*, *39*(2), 328–376.

Gabbay, D., Kurucz, A., Wolter, F., & Zakharyaschev, M. (2003). *Many-Dimensional Modal Logics: Theory and Applications*, Vol. 148 of *Studies in Logic*. Elsevier.

Garson, J. (2001). Quantification in modal logic. In Gabbay, D., & Guenthner, F. (Eds.), *Handbook of Philosophical Logic*, Vol. 3, pp. 267–323. Reidel.

Halpern, J., & Moses, Y. (1992). A guide to completeness and complexity for modal logics of knowledge and belief. *Artificial Intelligence*, *54*, 319–379.

Halpern, J., van der Meyden, R., & Vardi, M. (2004). Complete axiomatizations for reasoning about knowledge and time. *SIAM Journal on Computing*, *33*(3), 674–703.

Halpern, J., & Vardi, M. (1986). The complexity of reasoning about knowledge and time. In *ACM Symposium on Theory of Computing, STOC 1986*, pp. 304–315. ACM Press.

Halpern, J., & Vardi, M. (1989). The complexity of reasoning about knowledge and time 1: lower bounds. *Journal of Computer and System Sciences*, *38*(1), 195–237.

Halpern, J., & Pucella, R. (2005). Probabilistic algorithmic knowledge. *Logical Methods in Computer Science*, *1*(3).

Hariri, B. B., Calvanese, D., Giacomo, G. D., Masellis, R. D., & Felli, P. (2011). Foundations of relational artifacts verification. In *Business Process Management: Proceedings of the 9th International Conference, BPM 2011*, Vol. 6896 of *Lecture Notes in Computer Science*, pp. 379–395. Springer.

Hodkinson, I. (2002). Monodic packed fragment with equality is decidable. *Studia Logica*, *72*, 185–197.

Hodkinson, I. (2006). Complexity of monodic guarded fragments over linear and real time. *Annals of Pure and Applied Logic*, *138*, 94–125.

Hodkinson, I., Kontchakov, R., Kurucz, A., Wolter, F., & Zakharyaschev, M. (2003). On the computational complexity of decidable fragments of first-order linear temporal logics. In *Proceedings of the 10th International Symposium on Temporal Representation and Reasoning / 4th International Conference on Temporal Logic, TIME-ICTL 2003*, pp. 91–98. IEEE Computer Society Press.

Hodkinson, I., Wolter, F., & Zakharyaschev, M. (2000). Decidable fragment of first-order temporal logics. *Annals of Pure and Applied Logic*, *106*(1-3), 85–134.

Hodkinson, I., Wolter, F., & Zakharyaschev, M. (2002). Decidable and undecidable fragments of first-order branching temporal logics. In *Proceedings of the 17th IEEE Symposium on Logic in Computer Science, LICS 2002*, pp. 393–402. IEEE Computer Society Press.

Lomuscio, A., & Ryan, M. (1998). On the relation between interpreted systems and Kripke models. In *Agent and Multi-Agent Systems: Proceedings of the AI97 Workshop on the theoretical and practical foundations of intelligent agents and agent-oriented systems*, Vol. 1441 of *Lecture Notes in Artificial Intelligence*, pp. 46–59. Springer.

McCarthy, J. (1979). Ascribing mental qualities to machines. In Ringle, M. (Ed.), *Philosophical Perspectives in Artificial Intelligence*, pp. 161–195. Harvester Press.

McCarthy, J. (1990). Artificial intelligence, logic and formalizing common sense. In Thomason, R. (Ed.), *Philosophical Logic and Artificial Intelligence*, pp. 161–190. Kluwer Academic.

Meyden, R. (1994). Axioms for knowledge and time in distributed systems with perfect recall. In *Proceedings of the 9th Annual IEEE Symposium on Logic in Computer Science, LICS 1994*, pp. 448–457. IEEE Computer Society Press.

Meyden, R. v., & Wong, K. (2003). Complete axiomatizations for reasoning about knowledge and branching time. *Studia Logica*, *75*(1), 93–123.

Moore, R. C. (1990). A formal theory of knowledge and action. In Allen, J., Hendler, J., & Tate, A. (Eds.), *Readings in Planning*, pp. 480–519. Kaufmann.

Parikh, R., & Ramanujam, R. (1985). Distributed processes and the logic of knowledge. In *Logics of Programs, Conference Proceedings*, Vol. 193 of *Lecture Notes in Computer Science*, pp. 256–268. Springer.

Pnueli, A. (1977). The temporal logic of programs. In *Proceedings of the 18th International Symposium Foundations of Computer Science, FOCS 1977*, pp. 46–57.

Rao, A., & Georgeff, M. (1991). Deliberation and its role in the formation of intentions. In *Proceedings of the 7th Conference on Uncertainty in Artificial Intelligence*, pp. 300–307. Kaufmann.

Sturm, H., Wolter, F., & Zakharyaschev, M. (2000). Monodic epistemic predicate logic. In *Logics in Artificial Intelligence, European Workshop, JELIA 2000*, Vol. 1919 of *Lecture Notes in Computer Science*, pp. 329–344. Springer.

Sturm, H., Wolter, F., & Zakharyaschev, M. (2002). Common knowledge and quantification. *Economic Theory*, *19*, 157–186.

Wolter, F., & Zakharyaschev, M. (2001). Decidable fragments of first-order modal logics. *Journal of Symbolic Logic*, *66*(3), 1415–1438.

Wolter, F., & Zakharyaschev, M. (2002). Axiomatizing the monodic fragment of first-order temporal logic. *Annals of Pure and Applies Logic*, *118*(1-2), 133–145.

Wooldridge, M. (2000a). Computationally grounded theories of agency. In *Proceedings of the International Conference of Multi-Agent Systems, ICMAS 2000*, pp. 13–22. IEEE Computer Society Press.

Wooldridge, M. (2000b). *Reasoning about Rational Agents*. MIT Press.

Wooldridge, M., & Fisher, M. (1992). A first-order branching time logic of multi-agent systems. In *Proceedings of the 10th European Conference on Artificial Intelligence, ECAI 1992*, pp. 234–238. John Wiley and Sons.

Wooldridge, M., Fisher, M., Huget, M., & Parsons, S. (2002). Model checking multi-agent systems with MABLE. In *Proceedings of the 1st International Conference on Autonomous Agents and Multiagent Systems, AAMAS 2002*, pp. 952–959. ACM Press.

Wooldridge, M., Huget, M., Fisher, M., & Parsons, S. (2006). Model checking for multiagent systems: the MABLE language and its applications. *International Journal on Artificial Intelligence Tools*, *15*(2), 195–226.

Wooldridge, M. (1999). Verifying that agents implement a communication language. In *Proceedings of the 16th National Conference on Artificial Intelligence and 11th Conference on Innovative Applications of Artificial Intelligence*, pp. 52–57. AAAI Press.